\documentclass[10pt]{article}

\usepackage{authblk}
\usepackage{natbib}
\usepackage{mathtools}
\usepackage{amsmath}
\usepackage{amsthm}
\usepackage{amssymb}
\usepackage{amsbsy}
\usepackage{amsfonts}
\usepackage{amscd}
\usepackage{mathrsfs}
\usepackage{bm}
\usepackage{breqn}
\usepackage{color, soul}
\usepackage{enumerate}
\usepackage{empheq}
\usepackage{rotating}
\usepackage{ftnxtra}
\usepackage{fnpos}
\usepackage[titletoc,toc,title]{appendix}
\usepackage{euscript}
\usepackage{graphicx}
\usepackage{epsfig}
\usepackage{epstopdf}
\DeclareGraphicsExtensions{.pdf,.png,.jpg,.eps}
\usepackage{pstool}
\usepackage{upgreek}
\usepackage{mathrsfs}
\usepackage{enumitem}
\usepackage[all,cmtip]{xy}
\usepackage{stmaryrd}
\usepackage{psfrag}
\usepackage[dvipsnames]{xcolor}

\numberwithin{equation}{section}

\theoremstyle{plain}	
\newtheorem{thm}{Theorem}[section]

\newtheorem{prob}[thm]{Problem}
\newtheorem*{prop*}{Proposition}
\theoremstyle{definition}	

\newtheorem{remark}[thm]{Remark}

\usepackage[textwidth=7.0in,%
	textheight=9.3in,%
	paper=letterpaper,%
	centering]{geometry}
\usepackage{caption}
\usepackage{subcaption}

\usepackage{hyperref}
\hypersetup{colorlinks=true, linkcolor=blue}
\hypersetup{colorlinks=true,citecolor=blue}

\DeclareMathAlphabet{\mathpzc}{OT1}{pzc}{m}{it}

\usepackage{amsmath, amsthm, amssymb}

\usepackage{cleveref}

\DeclarePairedDelimiter\abs{\lvert}{\rvert}

\makeatletter
\newsavebox{\@brx}
\newcommand{\llangle}[1][]{\savebox{\@brx}{\(\m@th{#1\langle}\)}%
  \mathopen{\copy\@brx\mkern2mu\kern-0.9\wd\@brx\usebox{\@brx}}}
\newcommand{\rrangle}[1][]{\savebox{\@brx}{\(\m@th{#1\rangle}\)}%
  \mathclose{\copy\@brx\mkern2mu\kern-0.9\wd\@brx\usebox{\@brx}}}%
\let\oldabs\abs
\def\abs{\@ifstar{\oldabs}{\oldabs*}}
\makeatother

\usepackage{accents}

{\end{bmatrix}}%
	
\usepackage[cbgreek]{textgreek} 

\usepackage{booktabs}
\usepackage{multirow}

\usepackage{hhline} 

\usepackage{cases} 

\usepackage[retainorgcmds]{IEEEtrantools} 

\DeclareMathAlphabet{\mathpzc}{OT1}{pzc}{m}{it}

\makeatletter
\DeclareFontFamily{OMX}{MnSymbolE}{}
\DeclareSymbolFont{MnLargeSymbols}{OMX}{MnSymbolE}{m}{n}
\SetSymbolFont{MnLargeSymbols}{bold}{OMX}{MnSymbolE}{b}{n}
\DeclareFontShape{OMX}{MnSymbolE}{m}{n}{
    <-6>  MnSymbolE5
   <6-7>  MnSymbolE6
   <7-8>  MnSymbolE7
   <8-9>  MnSymbolE8
   <9-10> MnSymbolE9
  <10-12> MnSymbolE10
  <12->   MnSymbolE12
}{}
\DeclareFontShape{OMX}{MnSymbolE}{b}{n}{
    <-6>  MnSymbolE-Bold5
   <6-7>  MnSymbolE-Bold6
   <7-8>  MnSymbolE-Bold7
   <8-9>  MnSymbolE-Bold8
   <9-10> MnSymbolE-Bold9
  <10-12> MnSymbolE-Bold10
  <12->   MnSymbolE-Bold12
}{}

\let\llangle\@undefined
\let\rrangle\@undefined
\DeclareMathDelimiter{\llangle}{\mathopen}%
                     {MnLargeSymbols}{'164}{MnLargeSymbols}{'164}
\DeclareMathDelimiter{\rrangle}{\mathclose}%
                     {MnLargeSymbols}{'171}{MnLargeSymbols}{'171}
\makeatother

\newcommand{\vek}[1] {\lceil#1\rceil}
\newcommand{\bigVek}[2] {#1\lceil#2 #1\rceil} 

\newcommand{\lm}[2] {\boldsymbol{\mathsf{#1}}^{#2}_{\mathcal{T}} }     
\newcommand{\gm}[2] {\boldsymbol{\mathsf{#1}}^{#2}_{h}}   
\newcommand{\bz}{\boldsymbol{0}}

\allowdisplaybreaks
\begin{document}

\title{\textbf{\vspace{-1in}\\
Optimal Elastostatic Cloaks}}

\author[1]{Fabio Sozio}
\author[2]{Mostafa Faghih Shojaei}
\author[3,4]{Arash Yavari\thanks{Corresponding author, e-mail: arash.yavari@ce.gatech.edu}}
\affil[1]{\small \textit{Solid Mechanics Laboratory, \'Ecole Polytechnique, Palaiseau, France}}
\affil[2]{\small \textit{Department of Mechanical Engineering, University of Michigan, Ann Arbor, MI 48109, USA}}
\affil[3]{\small \textit{School of Civil and Environmental Engineering, Georgia Institute of Technology, Atlanta, GA 30332, USA}}
\affil[4]{\small \textit{The George W. Woodruff School of Mechanical Engineering, Georgia Institute of Technology, Atlanta, GA 30332, USA}}
	
\date{}
\maketitle

\vspace{-.2in}
\begin{abstract} 
An elastic cloak hides a hole (or an inhomogeneity) from elastic fields. In this paper, a formulation of the optimal design of elastic cloaks based on the adjoint state method, in which the balance of linear momentum is enforced as a constraint, is presented. The design parameters are the elastic moduli of the cloak, and the objective function is a measure of the distance between the solutions in the physical and in the virtual bodies. Both the elastic medium and the cloak are assumed to be made of isotropic linear elastic materials. However, the proposed formulation can easily be extended to anisotropic solids. In order to guarantee smooth inhomogeneous elastic moduli within the cloak a penalization term is added to the objective function. Mixed finite elements are used for discretizing the weak formulation of the optimization problem. Several numerical examples of optimal elastic cloaks designed for both single and multiple loads are presented. We consider different geometries and loading types and observe that in some cases the optimal elastic cloaks for cloaking holes (cavities) are made of auxetic materials.
\end{abstract}

\begin{description}
\item[Keywords:] Elastic cloaking, optimal design, mixed finite elements, adjoint state method, auxetic materials.
\end{description}

\small 
\tableofcontents
\normalsize

\section{Introduction}

Cloaking objects from different types of waves has been a problem of interest for decades. In elasticity some original ideas related to cloaking can be found in \citep{Gurney1938,Reissner1949,Mansfield1953} on reinforced holes in elastic sheets, and in \citep{Hashin1962,Hashin1963,Hashin1985,Hashin1964,Benveniste2003} on neutral inhomogeneities.
One approach to cloaking is to use the invariance of the governing equations of a field theory under certain transformations. This has been referred to as \emph{transformation cloaking} in the literature.
In the case of electromagnetism the first works on transformation cloaking are due to \citet{Pendry2006} and \citet{Leonhardt2006}.
In the literature of elastic transformation cloaking there have been many inconsistent formulations in the past fifteen years, which were recently critically reviewed in \citep{Yavari2019} and \citep{Golgoon2021}.
It is now known that exact elastodynamic (and elastostatic) transformation cloaking is not possible; the obstruction to transformation cloaking is the balance of angular momentum. More specifically, the impossibility of elastodynamic (and elastostatic) transformation cloaking has been proved for classical linear elastic solids, gradient solids (both centrosymmetric and non-centrosymmetric), and (generalized) Cosserat solids \citep{Yavari2019, Sozio2021Cloaking}. It turns out that exact transformation cloaking is not possible even for elastic plates \citep{Golgoon2021}. The following are the no-go theorems of elastodynamic transformation cloaking:

\begin{itemize}[topsep=0pt,noitemsep, leftmargin=15pt]
\item  Nonlinear elastodynamic transformation cloaking is not possible regardless of the shape of the hole and the cloak \citep{Yavari2019}. 
\item  Elastodynamic transformation cloaking is not possible in the setting of classical linear elasticity regardless of the shape of the hole and the cloak \citep{Yavari2019}. 
\item In the small-on-large theory, i.e., linearized elasticity with respect to a pre-stressed configuration, elastodynamic transformation cloaking is not possible regardless of the shape of the hole and the cloak \citep{Yavari2019}. 
\item Assuming that the virtual body is isotropic and centro-symmetric, elastodynamic transformation cloaking is not possible for gradient elastic solids in either $2$D or $3$D for a hole (cavity) of any shape \citep{Yavari2019}.
\item Assuming that the virtual body is isotropic and non-centrosymmetric, elastodynamic transformation cloaking is not possible for any cylindrical hole (not necessarily circular) \citep{Sozio2021Cloaking}.
\item Elastodynamic transformation cloaking is not possible for linear (generalized) Cosserat elastic solids in dimension two \citep{Yavari2019}.
\item Elastodynamic transformation cloaking is not possible for a spherical cavity using a spherical cloak in linear (generalized) Cosserat elastic solids \citep{Yavari2019}.
\end{itemize}

\noindent These no-go theorems imply that one cannot use transformation methods to design cloaks that can work for all possible loadings. 
The engineering solution for elastic cloaking applications is to resort to approximate cloaking formulated as an optimal design problem. This is what is done in the present paper.

There have been recent systematic studies of the optimal design of acoustic cloaks in the literature \citep{Sanders2018,Chen2021,Cominelli2022}. There are no such systematic formulations in the case of elasticity.
\citet{Fachinotti2018} approximated a 2D elasticity problem by displacement-based finite elements and formulated the optimal design of an elastic cloak at the level of finite elements. Each finite element has a vector of design parameters that describe the microstructure in that element.
\citet{Sanders2021} used an optimization method to design 2D elastostatic cloaks for lattice materials.
In a recent paper, \citet{wang2022mechanical} proposed a discretized formulation that allowed them to consider different shapes of voids and cloaks, as well as different boundary conditions and loadings. They used a data-driven approach for the design of unit cell tessellations that best fit the results of the optimization.
There have also been recent efforts in the literature in using topology optimization techniques for the design of elastic cloaks \citep{Ota2022}.

In this paper, we formulate the optimal design of elastic cloaks as a partial differential equation (PDE)-constrained optimization problem. The design variables are the elastic properties of the cloak. We derive both the strong and weak governing equations of the optimal design problem. A class of mixed finite elements are used to discretize the weak form of the governing equations.
We will present several examples of optimal elastic cloaks for both single and multiple loads.

This paper is structured as follows. 
In \S\ref{Problem-Definition} the problem of elastic cloaking for a body with holes/inhomogeneities is defined and a simple 1D example is discussed. The PDE-constrained optimization problem of elastic cloaking is formulated in \S\ref{Optimization-Formulation}. Both the strong and weak governing equations of the optimization problem are derived. In \S\ref{Sec:FEM} the optimization problem is discretized using mixed finite elements. Several numerical examples are presented and discussed in \S\ref{Sec:Examples}.
Conclusions are given in \S\ref{Sec:Conclusions}.

\section{Optimal Design of a Static Elastic Cloak} \label{Problem-Definition}

In this section we formulate elastostatic cloaking as an optimal design problem. We introduce the design variables, define an objective function, and discuss a simple axisymmetric example.

\subsection{Design variables and objective functions}

\begin{figure}[t!]
\centering
\vskip 0.1in
\includegraphics[width=3in]{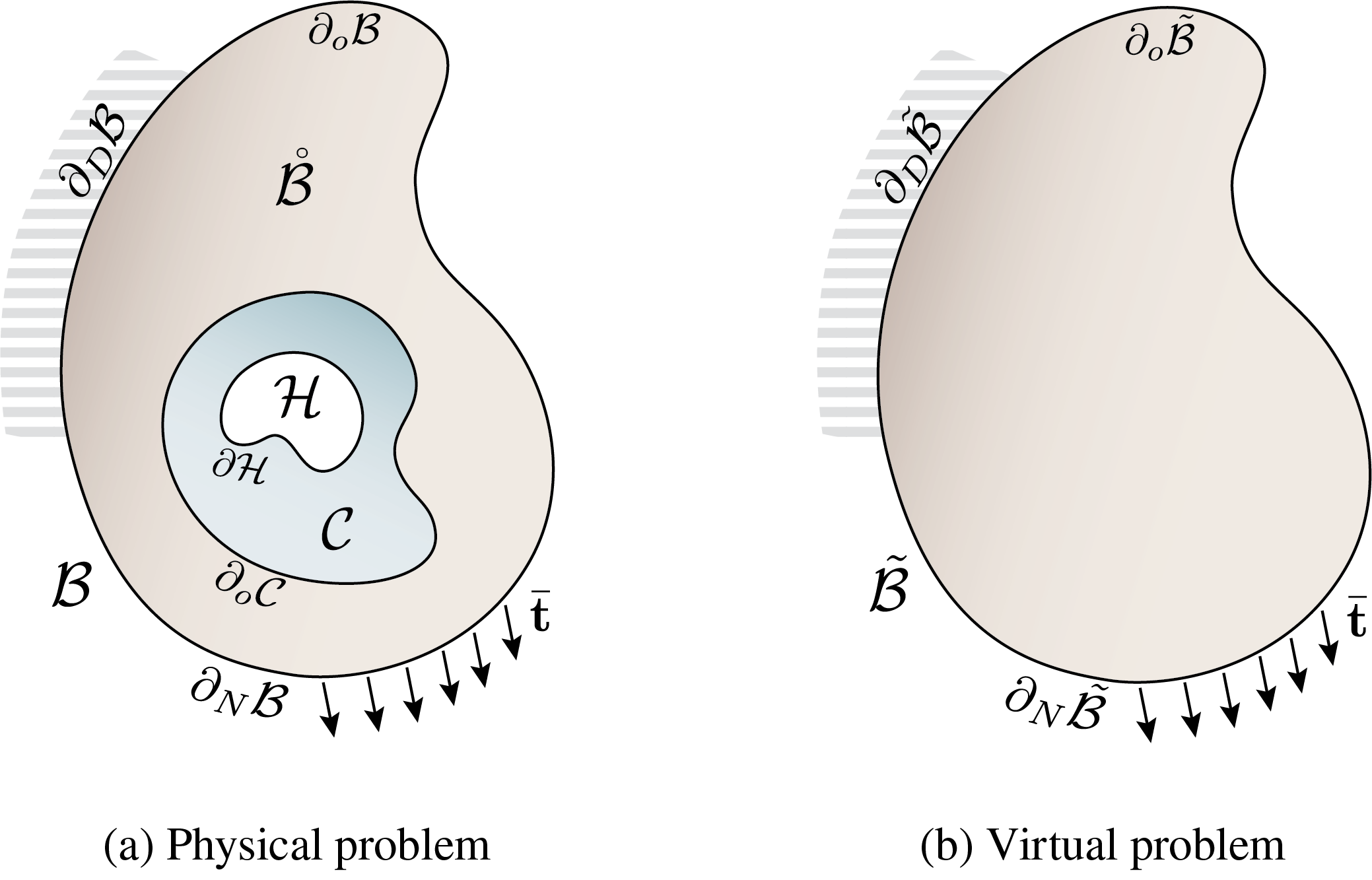}
\vskip 0.1in
\caption{(a) Assuming a fixed cloak, the design parameters are the elastic moduli in the cloak $\mathcal{C}$. (b) The virtual body is without any holes, cavities or inhomogeneities. The optimal elastic constants of the cloak make the response of the physical body in $\mathcal{B}\setminus\mathcal{C}$ as close as possible to that in the same set in the virtual body.}
\label{fig:sets}
\end{figure}

Let us consider a body $\mathcal{B}$ made of a linear elastic solid. We assume that there is an object occupying a region $\mathcal{H}$ in this body. This object can be a hole, a cavity, an inhomogeneity or a combination of the three, while the set $\mathcal{H}$ can be either simply-connected or non-simply-connected. Our goal is to design a cloak that encloses the object and hides it from static loads as much as possible.
Let us denote the cloaking region by $\mathcal{C}\subset\mathcal{B}$. We assume that in the physical body the elastic constants $\mathsf{C}^{abcd}$ and the mass density $\rho_0$ in the exterior domain $\mathring{\mathcal{B}}=\mathcal{B}\setminus\mathcal{C}$ are given and are uniform.
The virtual body has the same uniform mass density and elastic constants everywhere (see Fig.~\ref{fig:sets}). 

We denote the boundary of the cloak by $\partial_o\mathcal{C}$, thus $\partial\mathcal{C}=\partial_o\mathcal{C}\cup\partial\mathcal{H}$.
In the optimal design of a static elastic cloak the goal is to find the optimal elastic moduli inside the cloak. We assume the following traction and displacement boundary conditions
\begin{equation}
\begin{aligned}
	 \boldsymbol{\sigma}\hat{\mathbf{n}} & =\bar{\mathbf{t}},~~~\text{on}~~~\partial_N\mathcal{B}, \\
	\mathbf{u} & =\bar{\mathbf{u}},~~~\text{on}~~~\partial_D\mathcal{B},
\end{aligned}
\end{equation}
where $\partial\mathcal{B}=\partial_N\mathcal{B} \cup\partial_D\mathcal{B}$, and $\partial_N\mathcal{B} \cap\partial_D\mathcal{B}=\emptyset$.
In the case of a hole or a cavity, one can assume traction-free $\partial\mathcal{H}$, while in the case of inhomogeneities, one would enforce the continuity of the traction vector. Other than the external traction forces we assume a given body force $\mathbf{b}$ defined on $\mathring{\mathcal{B}}$.

We define the corresponding virtual body $\tilde{\mathcal{B}}$ to have the geometry of the (physical) body $\mathcal{B}$ but without any holes or inhomogeneities, i.e., $\tilde{\mathcal{B}}=\mathcal{B}\cup\mathcal{H}$. We also assume that the virtual body has the mass density and elastic constants identical to those of the physical body outside the cloak. Outside the cloak, i.e., in $\mathring{\tilde{\mathcal{B}}}=\mathring{\mathcal{B}}$, the virtual body is under the same traction and displacement boundary conditions as the physical body is. Also, the body force distributions in the two problems are identical in the region outside the cloak.

The design parameters are the elastic constants in the cloak, i.e., $\mathsf{C}^{abcd}(x)$, $x\in\mathcal{C}$. Given $\mathbf{b}, \bar{\mathbf{t}}, \bar{\mathbf{u}}$, we would like the mechanical responses of the two bodies to be as close as possible outside the cloak (the exterior domain $\mathring{\mathcal{B}}$ or a subset of it). 
There are several possibilities for the objective functions. Examples are minimizing the difference between displacement fields, stress fields, or energies. 
The objective function that we will be using is 
\begin{equation} \label{objective-function}
	\mathsf{g}= \frac{1}{2}
	\int_{\mathcal{B}\setminus\mathcal{C}} \|\mathbf{u}-\tilde{\mathbf{u}}\|^2\mathrm{d}v \,,
\end{equation}
where $\|\mathbf{u}-\tilde{\mathbf{u}}\|^2=(\mathbf{u}-\tilde{\mathbf{u}})\cdot(\mathbf{u}-\tilde{\mathbf{u}})$.
Note that unlike the classical minimum compliance problem \citep{Jog1994,Bendsoe2013} the objective function in the elastic cloaking problem is not defined on the entire body.
For given boundary conditions, the virtual body has a unique displacement field that can be easily calculated and is independent of the design parameters. The optimization problem is written as
\begin{equation} \label{unconstr-opt}
	\inf_{\boldsymbol{\mathsf{C}} \in \mathsf{Ela}(\mathcal{C})} \mathsf{g}[\boldsymbol{\mathsf{C}}],
\end{equation}
where $\mathsf{Ela}(\mathcal{C})$ is the set of elasticity tensors defined in the cloaking region.
Assuming that the cloak is isotropic\footnote{Extending this analysis to anisotropic cloaks would be straightforward.} but inhomogeneous, in Cartesian coordinates $\mathsf{C}^{abcd}(x)=\lambda(x)\,\delta^{ab}\delta^{cd}+\mu(x)(\delta^{ac}\delta^{bd}+\delta^{ad}\delta^{bc})$,\footnote{In general curvilinear coordinates, $\mathsf{C}^{abcd}(x)=\lambda(x)\,g^{ab}g^{cd}+\mu(x)\big(g^{ac}g^{bd}+g^{ad}g^{bc}\big)$, where $g^{ab}$ are components of the inverse metric tensor of the Euclidean ambient space.} where the Lam\'{e} constants satisfy the constraints $\mu(x)>0$, and $3\kappa(x) = 2\mu(x)+3\lambda(x)>0$.

\subsection{Example of an infinitely-long hollow solid cylinder}

\begin{figure}[t!]
\centering
\includegraphics[width=\textwidth]{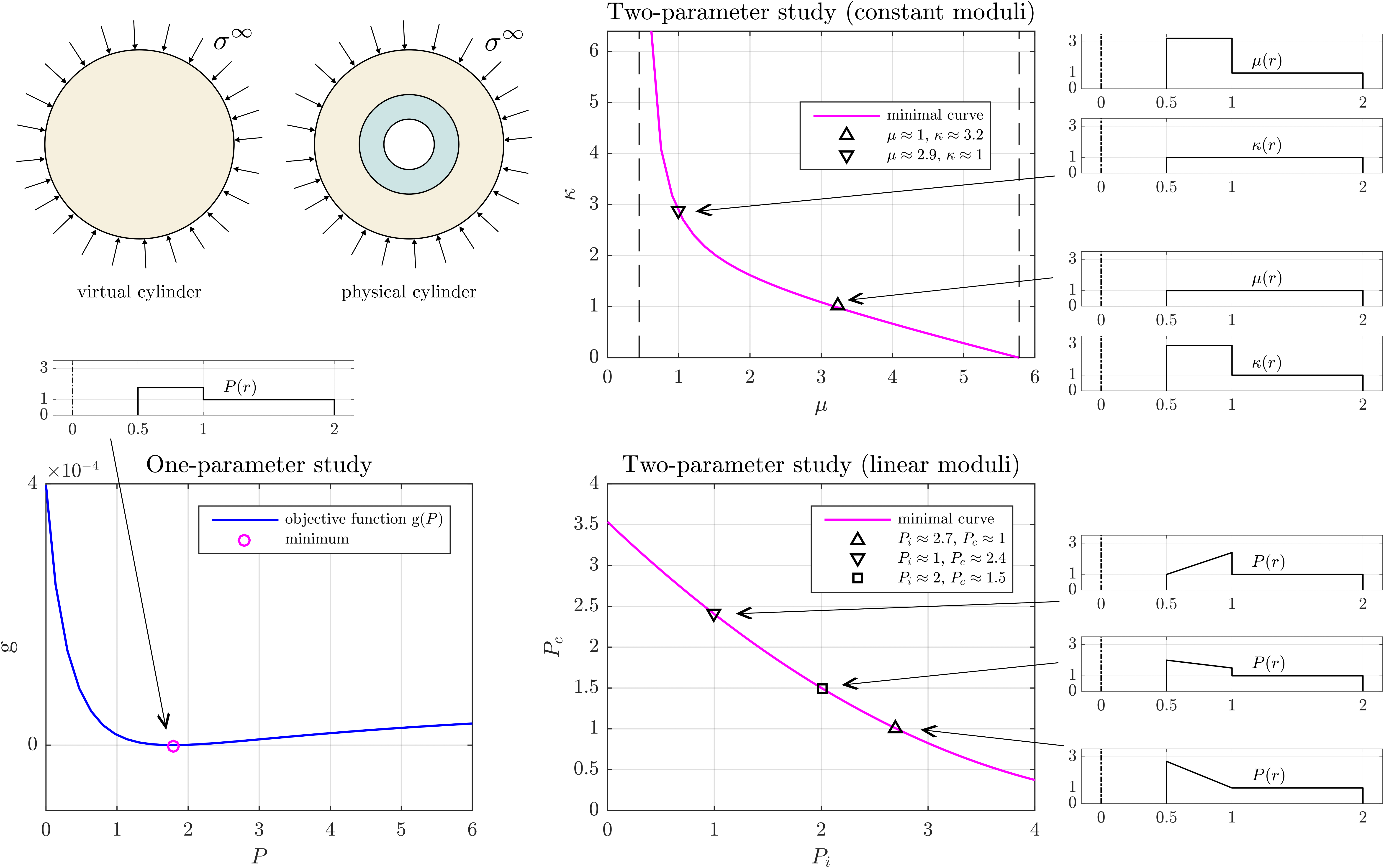}
\caption{Objective function for the cloaking of an infinitely-long hollow solid cylinder, with $r_i = 0.5$, $r_c = 1.0$, $r_o = 2.0$, $\mathring{\mu}/\sigma^{\infty} = 1$, $\mathring{\kappa}/\sigma^{\infty} = 1$.
Bottom left: Objective function $\mathsf{g}=\mathsf{g}(P)$ for the case $\mu(r) \equiv \mathring{\mu} P$ and $\kappa(r) \equiv \mathring{\kappa} P$; $\mathsf{g}=0$ for $P$ as in~\eqref{radial-P}.
Top right: Two uniform independent elastic moduli; the objective function vanishes on the curve~\eqref{radial-kappa-mu} in the design plane $\kappa$-$\mu$; the upper and lower bounds~\eqref{radial-mu-limits} are indicated with dashed lines.
Bottom right: Linear elastic moduli $\kappa(r)= \mathring{\kappa}  P(r)$ and $\mu(r) = \mathring{\mu} P(r)$ \eqref{radial-linear-param}; the objective function vanishes on the curve indicated in the design plane $P_i$-$P_c$.}
\label{fig:radial-plots}
\end{figure}

Let us consider an infinitely-long hollow solid cylinder with inner and outer radii $r_i$ and $r_o$, respectively, as shown in Fig.~\ref{fig:radial-plots} (top left).
The outer radius of the cloak is denoted by $r_c$, with $r_i<r_c<r_o$.
The cylinder is under a far-field traction $\sigma^{\infty}$, i.e., $\sigma^{rr}(r_o)=\sigma^{\infty}$, while the inner boundary is traction-free.
Assuming radial displacements $\mathbf{u}=u(r) \,\hat{\mathbf{e}}_r$, the non-zero strain components are $\epsilon_{rr}(r)=u'(r)$ and $\epsilon_{\theta\theta}(r)=r\,u(r)$.
In cylindrical coordinates $(r,\theta,z)$ the non-zero elastic moduli for an isotropic solid are
\begin{equation} \label{radial-constitutive}
	\mathsf{C}^{rrrr} (r)=\kappa(r)+\frac{4}{3}\mu(r),\quad
	\mathsf{C}^{rr\theta\theta} (r)=\frac{1}{r^2} \left[\kappa(r) -\frac{2}{3}\mu(r) \right],\quad
	\mathsf{C}^{\theta\theta\theta\theta}(r)=\frac{1}{r^4} \left[\kappa(r) + \frac{4}{3}\mu(r)\right],
\end{equation}
while the only non-trivial equilibrium equation is\footnote{The physical components of the Cauchy stress $\hat{\sigma}^{ab}$ are related to the components of the Cauchy stress $\sigma^{ab}$ as $\hat{\sigma}^{ab}=\sqrt{g_{aa}}\sqrt{g_{bb}}\,\sigma^{ab}$ (no summation) \citep{Truesdell1953}. Recall that the non-zero components of the spatial metric in polar coordinates are $g_{rr}=1$ and $g_{\theta\theta}=r^2$. Thus, $\hat{\sigma}^{rr}=\sigma^{rr}$ and $\hat{\sigma}^{\theta\theta}=r^2\sigma^{\theta\theta}$. In terms of the physical components, one recovers the familiar form of the equilibrium equation in the literature: $\frac{d}{dr}\hat{\sigma}^{rr}+\frac{1}{r}(\hat{\sigma}^{rr}-\hat{\sigma}^{\theta\theta})=0$. It should also be noted that in \eqref{radial-constitutive} we are showing the curvilinear components of the elastic constants and not their physical components.}
\begin{equation} \label{radial-balance}
	\frac{d}{dr}\sigma^{rr}+\frac{1}{r}\sigma^{rr}-r\sigma^{\theta\theta}=0
	\,.
\end{equation}

\paragraph{The virtual problem.}
The virtual body is a solid cylinder with uniform elastic constants $\mathring{\kappa}$ and $\mathring{\mu}$.
In this case the equilibrium equation~\eqref{radial-balance} simplifies to read
\begin{equation} \label{radial-balance-uniform}
	r^2 \tilde{u}''(r)+r \tilde{u}'(r)-\tilde{u}(r)=0
	\,.
\end{equation}
Solutions of~\eqref{radial-balance-uniform} have the form $\tilde{u}(r)=\tilde{C}_1r+\tilde{C}_2/r$.
Knowing that the displacement is bounded as $r\rightarrow 0$, we conclude that $\tilde{C}_2=0$, while the traction boundary condition $\sigma^{rr}(r_o)=\sigma^{\infty}$ gives
\begin{equation} \label{radial-virtual}
	\tilde{C}_1=\frac{3\sigma^{\infty}}{2(3\mathring{\kappa}+\mathring{\mu})} \,,\quad
	\tilde{u}(r)=\frac{3\sigma^{\infty} r }{2(3\mathring{\kappa}+\mathring{\mu})} \,,\quad
	\tilde{\sigma}^{rr}(r)\equiv \sigma^{\infty}
	\,.
\end{equation}

\paragraph{The physical problem.}
The physical body is assumed to have the uniform elastic constants $\mathring{\mu}$ and $\mathring{\kappa}$ for $r_c<r<r_o$.
In this region, Eq.~\eqref{radial-balance-uniform} holds, and hence one has $u(r)=C_1 r+C_2/r $.
The traction boundary condition at $r=r_o$ implies that
\begin{equation} \label{C1-C2}
	C_2= \frac{3\mathring{\kappa}+\mathring{\mu}}{\mathring{\mu}} r_o^2 C_1
	-\frac{\sigma^{\infty} }{2\mathring{\mu}} r_o^2
	\,.
\end{equation}
For $r_i<r<r_c$, the elastic moduli $\mu(r)$ and $\kappa(r)$ are the design parameters.
In this region, Eq.~\eqref{radial-balance-uniform} is no longer valid, as~\eqref{radial-balance} is written as
\begin{equation} \label{radial-balance-nonuni}
	u''(r)
	+\left[ \frac{3\kappa' (r)+ 4\mu' (r)}{3\kappa (r)+ 4\mu(r)} + \frac{1}{r} \right] u'(r)
	+\frac{1}{r} \left[ \frac{3\kappa' (r)-2\mu' (r)}{3\kappa (r)+ 4\mu(r)} - \frac{1}{r} \right] u(r)
	=0\,.
\end{equation}

\paragraph{The objective function.}
We consider an objective function as defined in~\eqref{objective-function}, which in this case reads $\mathsf{g}=\mathsf{g}[\mu(r),\kappa(r)]$.
It should be noted that $\mathsf{g}$ is obtained by integrating $\Vert u(r)-\tilde{u}(r)\Vert^2$ over $[r_c,r_o]$.
Therefore, Eq.~\eqref{C1-C2} allows one to write the objective function in the form $\mathsf{g}=\mathsf{g}(C_1)$.
Using~\eqref{radial-virtual} one obtains
\begin{equation}
	\mathsf{g}(C_1)=\pi \left(r_o^2-r_c^2\right) 
	\frac{ 3 \mathring{\mu} \, r_c^2+(3\mathring{\kappa} +\mathring{\mu})\,r_o^2}
	{2 \mathring{\mu} (3\mathring{\kappa} +\mathring{\mu}) \,r_c^2}
	\left[\sigma^{\infty} -\frac{2}{3} (3\mathring{\kappa} +\mathring{\mu}) \,C_1\right]^2 ,
\end{equation}
which needs to be minimized with respect to $C_1$.
It is seen that $\mathsf{g}$ vanishes for the following value of $C_1$:
\begin{equation} \label{radial-min-C1}
	C_1 = \frac{3\sigma^{\infty}}{2(3\mathring{\kappa}+\mathring{\mu})}
	\,.
\end{equation}
The vanishing of $\mathsf{g}$ implies that the physical and virtual solutions are identical, which can be checked by plugging~\eqref{radial-min-C1} into~\eqref{C1-C2} and comparing it with~\eqref{radial-virtual}.
Moreover, the value given in~\eqref{radial-min-C1} does not depend on the choice of the norm that is used to define the objective function, as $\Vert \mathbf{u}-\tilde{\mathbf{u}} \Vert = 0$ if and only if $\mathbf{u}=\tilde{\mathbf{u}}$.
Therefore, any $\mu(r)$ and $\kappa(r)$, $r_i\leq r\leq r_c$, that satisfy the equilibrium equation~\eqref{radial-balance-nonuni}, together with the boundary conditions
\begin{equation}
	u(r_c) =\frac{3\sigma^{\infty}}{2(3\mathring{\kappa}+\mathring{\mu})}r_c \,,\quad
	\sigma^{rr}(r_c) = \sigma^{\infty} \,,\quad
	\sigma^{rr}(r_i)=0 \,,
\end{equation}
generate a solution that is indistinguishable from the virtual problem, and hence, constitute a ``perfect cloak".
This might suggest that the solution to the cloaking problem is not unique, as we see next.

\paragraph{The design variables.}
First we assume elastic moduli in the cloak of the type $\mu(r) \equiv \mathring{\mu} P$ and $\kappa(r) \equiv \mathring{\kappa} P$, for a parameter $P>0$.
This allows us to study the objective function as a function of a single variable, i.e., $\mathsf{g}=\mathsf{g}(P)$, as shown in Fig.~\ref{fig:radial-plots} (bottom left).
The displacement field $u(r)$ for $r_i<r<r_c$ satisfies~\eqref{radial-balance-uniform} and has the form $u(r)=D_1 r+D_2/r $.
It is seen that $\mathsf{g}(P)$ vanishes for
\begin{equation} \label{radial-P}
	 P=\frac{3r_i^2 \mathring{\kappa} + (3r_c^2+r_i^2)\mathring{\mu} }{3(r_c^2-r_i^2)\mathring{\mu}}
	\,.
\end{equation}
%
Next, we consider general uniform elastic moduli $\mu(r) \equiv \mu$ and $\kappa(r) \equiv \kappa$ for $r_i<r<r_c$ .
In this case, the objective function has the form $\mathsf{g}=\mathsf{g}(\kappa,\mu)$, and hence, the displacement field $u(r)$ for $r_i<r<r_c$ still satisfies~\eqref{radial-balance-uniform}.
It can be seen that $\mathsf{g}$ vanishes on the curve
\begin{equation} \label{radial-kappa-mu}
	 \kappa=\mu \frac{-\mu(r_c^2+r_i^2)+(\mathring{\lambda}+\mathring{\mu})(r_c^2+r_i^2)}
	{\mu(r_c^2-r_i^2)-r_i^2(\mathring{\lambda}+\mathring{\mu})}
	\,,
\end{equation}
as long as $\mu>0$ and $\kappa>0$, i.e., when
\begin{equation} \label{radial-mu-limits}
	\frac{r_i^2 (3 \mathring{\kappa}+\mathring{\mu}) }{3 (r_c^2-r_i^2)}
	<\mu<
	\frac{(3 \mathring{\kappa} + \mathring{\mu}) \left(3 r_c^2+r_i^2\right)}{3 \left(r_c^2-r_i^2\right)}
	\,,
\end{equation}
see Fig.~\ref{fig:radial-plots} (top right).
Note that both solutions~\eqref{radial-P} and~\eqref{radial-kappa-mu} for the elastic constants are independent of the far-field load $\sigma^{\infty}$.
Lastly, we assume that the elastic moduli are linear in $r$, and can be written as $\mu(r) = \mathring{\mu} \,P(r)$ and $\kappa(r)= \mathring{\kappa}  \,P(r)$, for
\begin{equation} \label{radial-linear-param}
	P(r)=\frac{(r_c-r)P_i+(r-r_i)P_c}{r_c-r_i} \,.
\end{equation}
This implies that $\mu(r_i)=\mathring{\mu} \,P_i$, $\mu(r_c)=\mathring{\mu} \,P_c$, $\kappa(r_i)=\mathring{\kappa} \,P_i$, and $\kappa(r_c)=\mathring{\kappa} \,P_c$.
In this case, the objective function can be written as $\mathsf{g}=\mathsf{g}(P_i,P_c)$.
Since the elastic moduli are not uniform, Eq.~\eqref{radial-balance-uniform} is no longer valid, and one must solve~\eqref{radial-balance-nonuni} numerically, see Fig.~\ref{fig:radial-plots} (bottom right).

\begin{remark}
\citet{Yavari2019} showed that, for radial deformations, exact transformation cloaking is possible in both nonlinear\footnote{An underlying assumption is that radial deformations in the absence of body forces are permitted for the given energy function. Recall that radial deformations are not universal for compressible solids \citep{Ericksen1955}.}
and linearized elasticity.
In transformation cloaking the reference configuration of the physical body is mapped to the reference configuration of a virtual body, that in this case is a hollow cylinder with inner and outer radii $\epsilon>0$ and $r_o$, respectively.
This is done using a cloaking map $\xi(r,\theta,z)=(f(r),\theta,z)$, which is such that $f(r_i)=\epsilon, f(r_c)=r_c$, $f'(r_c)=1$, while for $r>r_c$ one has $f(r)=r$.
In the case of radial deformations in a cylindrically-symmetric body, one obtains the following elastic moduli \citep{Yavari2019}
\begin{equation}\label{radial-tranformation}	
	\mathsf{C}^{rrrr} (r)= \frac{\left[ 3 \kappa(r)+4 \mu(r) \right] f(r)}{3 r f'(r)}  \,,\quad
	\mathsf{C}^{rr\theta\theta} (r)=\frac{3\kappa(r) -2 \mu(r)}{3 r^2} \,,\quad
	\mathsf{C}^{\theta\theta\theta\theta}(r)=\frac{\left[ 3 \kappa(r) + 4\mu(r)\right]  f(r)}{3 r^3 f'(r)} \,,
\end{equation}
that is to be compared with~\eqref{radial-constitutive}.
However, it can be shown that~\eqref{radial-tranformation} violates the isotropy assumption.
\end{remark}

\section{Elastic Cloaking Optimization Formulation} \label{Optimization-Formulation}

In this section, we obtain the strong form of the governing equations associated with the optimal design problem.
Instead of working with~\eqref{unconstr-opt}, we choose to minimize the objective function in the space of all possible design variables $\boldsymbol{\mathsf{C}}$ and displacements $\mathbf{u}$, while equilibrium on the entire body is enforced as a constraint.
We use the method of adjoint state. This method provides a physical interpretation of the structure of the governing equations in the context of the method of Lagrange multipliers \citep{plessix2006review}.

\subsection{The augmented objective function}

Assuming that the physical body is made of a compressible linear elastic solid, the design of an optimal elastic cloak is written as the following PDE-constrained optimization problem
\begin{equation} \label{minimization-problem} 
	\inf_{\mathbf{u},\boldsymbol{\mathsf{C}}} \mathsf{g}(\mathbf{u},\boldsymbol{\mathsf{C}})  \quad
	\begin{aligned}[t]
		\text{subject to}    & & \operatorname{div}(\boldsymbol{\mathsf{C}}\nabla\mathbf{u}) 
		+\mathbf{b} =\mathbf{0} &~~ \text{in~}\mathcal{B} \,, \\
	         & & \bar{\mathbf{t}} - (\boldsymbol{\mathsf{C}}\nabla\mathbf{u})\hat{\mathbf{n}}=\mathbf{0}
	         &~~ \text{on~}\partial_N\mathcal{B} 
	         \,, 
	\end{aligned}
\end{equation}
where
\begin{equation} \label{obj-func}
	\mathsf{g}= \frac{1}{2} \|\mathbf{u}-\tilde{\mathbf{u}}\|_{L^2}^2  
	= \frac{1}{2}\int_{\mathring{\mathcal{B}}} \|\mathbf{u}-\tilde{\mathbf{u}}\|^2 \,\mathrm{d}v	\,,
\end{equation}
and the displacements are restricted to those that satisfy $\mathbf{u} =\bar{\mathbf{u}}$ on $\partial_D\mathcal{B}$.
In the strong form of the governing equations, we assume as much smoothness as needed. We will be more specific when writing the weak form of the governing equations.
It should be emphasized that in this optimization problem the displacement field is unknown on the entire body $\mathcal{B}$, while $\boldsymbol{\mathsf{C}}$ is unknown only inside the cloak; outside the cloak $\boldsymbol{\mathsf{C}}=\mathring{\boldsymbol{\mathsf{C}}}$, and inside the inhomogeneities (if there are any) $\boldsymbol{\mathsf{C}}$ is given.
It should also be mentioned that the domain over which the integral in~\eqref{obj-func} is evaluated does not need to be the entire set $\mathring{\mathcal{B}}$; it can be reduced to a subset $\mathring{\mathcal{B}}^{M}\subset\mathring{\mathcal{B}}$ in which the effects of the cloak are to be measured.
In order to solve~\eqref{minimization-problem}, we define the following objective function
\begin{equation} \label{modified-objective-0}
	\mathsf{f}=
	 \frac{1}{2}\int_{\mathring{\mathcal{B}}} \|\mathbf{u}-\tilde{\mathbf{u}}\|^2 \,\mathrm{d}v
	+ \int_{\mathcal{B}} \boldsymbol{\gamma} \cdot 
	\left[ \operatorname{div}(\boldsymbol{\mathsf{C}} \nabla\mathbf{u} )
	+\mathbf{b}\right] \,\mathrm{d}v \\
	+ \int_{\partial_N\mathcal{B}} \boldsymbol{\gamma} \cdot 
	\left[  \bar{\mathbf{t}} - (\boldsymbol{\mathsf{C}} \nabla\mathbf{u}) \hat{\mathbf{n}}\right]  \,\mathrm{d}a	\,,
\end{equation}
where the Lagrange multiplier $\boldsymbol{\gamma}$, which is associated with the equilibrium equations, is an adjoint displacement variable, see \S\ref{App:Lag-mult}.
Later in this section we will prove that it satisfies the boundary condition $\boldsymbol{\gamma}|_{\partial_D\mathcal{B}}=\mathbf{0}$.

The way the functional \eqref{modified-objective-0} is written there is no control over how much the elastic constants of the cloak deviate from those of the virtual body. In order to ensure that the cloak is not much softer or stiffer than the outside medium we would need to add an extra term that penalizes large deviations from the elastic constants of the outside medium.\footnote{This extra term is standard in optimal control \citep{Troltzsch2010}.}
We would also like to avoid abrupt changes of the elastic moduli of the cloak by adding a penalizing term involving some norm of $\nabla\boldsymbol{\mathsf{C}}$.
Suppose $d_{\boldsymbol{\mathsf{C}}}(\boldsymbol{\mathsf{C}},\mathring{\boldsymbol{\mathsf{C}}})$ is the distance between the elasticity tensor of the cloak and that of the virtual body, where $d_{\boldsymbol{\mathsf{C}}}(.,.)$ is a Sobolev metric that will be specified for isotropic cloaks in the following. 
The modified objective function is defined as
\begin{equation} \label{modified-objective}
\begin{aligned}
	\mathsf{f} = &
	\frac{k}{2}\int_{\mathring{\mathcal{B}}} \|\mathbf{u}-\tilde{\mathbf{u}}\|^2 \,\mathrm{d}v
	+\frac{1}{2}  d^2_{\boldsymbol{\mathsf{C}}}(\boldsymbol{\mathsf{C}},
	\mathring{\boldsymbol{\mathsf{C}}}) + \int_{\mathcal{B}} \boldsymbol{\gamma} \cdot 
	\left[ \operatorname{div}(\boldsymbol{\mathsf{C}} \nabla\mathbf{u} )+\mathbf{b}\right] \,\mathrm{d}v 
	+ \int_{\partial_N\mathcal{B}} \boldsymbol{\gamma} \cdot 
	\left[  \bar{\mathbf{t}} - (\boldsymbol{\mathsf{C}} \nabla\mathbf{u}) \hat{\mathbf{n}}\right]  \,\mathrm{d}a \,,
\end{aligned}
\end{equation}
where the first term is multiplied by a constant $k>0$ to have a comparable value to $d^2_{\boldsymbol{\mathsf{C}}}(\boldsymbol{\mathsf{C}}, \mathring{\boldsymbol{\mathsf{C}}})$. We later use $k$ as a control parameter in the minimization scheme, see \S\ref{Sec:Examples}.
The minimization problem~\eqref{minimization-problem} is now rewritten as
\begin{equation} \label{minimization-problem-lag}
	\inf_{\boldsymbol{\mathsf{C}}, \mathbf{u}, \boldsymbol{\gamma}}
	\mathsf{f}(\boldsymbol{\mathsf{C}}, \mathbf{u},\boldsymbol{\gamma})   \,.
\end{equation}

\subsection{The design space}

We assume that both the virtual body and the cloak are made of isotropic solids.
Positive-definiteness of $\boldsymbol{\mathsf{C}}$ is equivalent to the following inequalities in the cloak:
\begin{equation} \label{Stability}
	\mu(x)>0\,,~\text{and}\quad  \kappa(x)>0 \,.
\end{equation}
Similarly, outside the cloak one has $\mathring{\mu}>0$, and $\mathring{\kappa}>0$.
It is possible to eliminate the inequality constraints \eqref{Stability} by using a change of variables for the unknown elastic constants. Using an idea similar to what was used in designing approximate acoustic cloaks \citep{Chen2021,Cominelli2022}, let us assume that
\begin{equation} \label{xi-eta-lambda-mu}
	\mu(x)=\mathring{\mu}\,e^{-\xi(x)} \,,\quad
	\kappa(x)=\mathring{\kappa}\,e^{-\eta(x)}
	\,.
\end{equation}
Knowing that $\mathring{\mu}>0$ and $\mathring{\kappa}>0$, one has $\mu(x)>0$ and $\kappa(x)>0$ for any functions $\xi(x)$ and $\eta(x)$.
Instead of using $(\mu(x),\kappa(x))$ as design parameters with two inequality constraints, one can use $(\xi(x),\eta(x))$ without any constraints.\footnote{Another choice for the change of variables is $\mu(x)=\mathring{\mu}\,\xi^2(x)$, and $\kappa(x)=\mathring{\kappa}\,\eta^2(x)$.}
Hence, one can uniquely identify each isotropic elasticity tensor $\boldsymbol{\mathsf{C}}(x)$ with a vector field $\mathbf{v}(x) = (\xi(x), \eta(x))$ with values in $\mathbb{R}^2$.
Notice that a constant elasticity tensor $\mathring{\boldsymbol{\mathsf{C}}}$ in the outside region $\mathring{\mathcal{B}}$ corresponds to $\mathbf{v}\equiv \mathbf{0}$.
Clearly, penalizing large deviations from the elastic constants of the outside medium is equivalent to penalizing the norm of $\mathbf{v}$.
Penalizing sharp gradients of the Lam\'e  constants is equivalent to penalizing sharp gradients of $\mathbf{v}$, as
\begin{equation} 
	\nabla\mu(x)=-\mathring{\mu}\,e^{-\xi(x)}\nabla\, \xi(x) \,,\quad
	\nabla\kappa(x)=-\mathring{\kappa}\,e^{-\eta(x)}\,\nabla\eta(x)
	\,.
\end{equation}
Therefore, we define the following metric in the design space
\begin{equation}
	d^2_{\boldsymbol{\mathsf{C}}} ( \boldsymbol{\mathsf{C}}_1, \boldsymbol{\mathsf{C}}_2) =
	\Vert \mathbf{v}_1 - \mathbf{v}_2 \Vert_{H^1}^2 \,,
\end{equation}
where $\Vert  \cdot \Vert_{H^1}$ is any $H^1$-equivalent Sobolev norm.
For two isotropic elasticity tensors $\boldsymbol{\mathsf{C}}_1$ and $\boldsymbol{\mathsf{C}}_2$ corresponding to the pairs $(\xi_1,\eta_1)$ and $(\xi_2,\eta_2)$, we choose the following metric 
\begin{equation} \label{metric-H2}
	d^2_{\boldsymbol{\mathsf{C}}}(\boldsymbol{\mathsf{C}}_1,\boldsymbol{\mathsf{C}}_2)
	= \int_{\mathcal{C}} \Bigl[ m_1(\xi_1-\xi_2 )^2+m_2(\eta_1-\eta_2)^2 \Bigr]\mathrm d v
	+ \int_{\mathcal{C}} \Bigl[  \alpha_1\Vert \nabla (\xi_1-\xi_2) \Vert^2+
	\alpha_2\Vert\nabla(\eta_1-\eta_2) \Vert^2 \Bigr] \mathrm d v	\,,
\end{equation}
where $m_1$, $m_2$, $\alpha_1$, and $\alpha_2$ are some positive constants.
In~\S\ref{App:Metric} we prove that $d_{\boldsymbol{\mathsf{C}}}$ is indeed a metric.
Hence, the second term in~\eqref{modified-objective} reads
\begin{equation} \label{metric-H2-CCo}
	d^2_{\boldsymbol{\mathsf{C}}}(\boldsymbol{\mathsf{C}},\mathring{\boldsymbol{\mathsf{C}}})
	= \int_{\mathcal{C}} \Bigl[ m_1 \xi^2+m_2 \eta^2 \Bigr]\mathrm d v
	+ \int_{\mathcal{C}} \Bigl[  \alpha_1\Vert \nabla \xi \Vert^2+
	\alpha_2\Vert\nabla \eta \Vert^2 \Bigr] \mathrm d v	\,.
\end{equation}

\subsection{The strong form of the governing equations}

We solve~\eqref{minimization-problem-lag} by using the methods of calculus of variations.
We start by taking $\boldsymbol{\mathsf{C}}$-variations of the objective function. 

\paragraph{$\boldsymbol{\mathsf{C}}$-variations.}
Plugging~\eqref{metric-H2-CCo} in~\eqref{modified-objective} and taking variations with respect to $\xi$ and $\eta$, one can write
\begin{equation} \label{C-var}
	\delta_{\boldsymbol{\mathsf{C}}} \mathsf{f}  =
	\int_{\mathcal{C}} \left( m_1\xi \,\delta\xi+m_2 \eta  \,\delta\eta  \right)\mathrm{d}v
	+\int_{\mathcal{C}} \left[\alpha_1 \nabla\xi\cdot \nabla \delta\xi+\alpha_2 \nabla\eta 
	\cdot \nabla \delta\eta \right] \mathrm{d}v 
	+\int_{\mathcal{B}} \boldsymbol{\gamma} \cdot \operatorname{div}
	(\delta\boldsymbol{\mathsf{C}} \nabla\mathbf{u})\,\mathrm{d}v
	- \int_{\partial_N\mathcal{B}} \boldsymbol{\gamma} \cdot 
	(\delta\boldsymbol{\mathsf{C}} \nabla\mathbf{u}) \hat{\mathbf{n}}\,\mathrm{d}a
	\,.
\end{equation}
Note that the second term in~\eqref{C-var} can be written as
\begin{equation} \label{div-grad-pen}
	\int_{\mathcal{C}} \left[\alpha_1 \nabla\xi\cdot \nabla \delta\xi+\alpha_2 \nabla\eta 
	\cdot \nabla \delta\eta \right] \mathrm{d}v 
	=\int_{\partial\mathcal{C}} \left(\alpha_1\nabla\xi\cdot\hat{\mathbf{n}}\,\delta\xi
	+\alpha_2\nabla\eta \cdot\hat{\mathbf{n}}\,\delta\eta \right)\,\mathrm{d}a 
	-\int_{\mathcal{C}} \left( \alpha_1 \nabla^2\xi\,\delta\xi+\alpha_2 \nabla^2\eta\,
	\delta\eta \right)\mathrm{d}v \,.
\end{equation}
For the last two terms in~\eqref{C-var} one has
\begin{equation} \label{C-var-bal}
\begin{aligned}
	& \int_{\mathcal{B}} \boldsymbol{\gamma} \cdot \operatorname{div}
	(\delta\boldsymbol{\mathsf{C}} \nabla\mathbf{u})\,\mathrm{d}v
	- \int_{\partial_N\mathcal{B}}  \boldsymbol{\gamma}
	\cdot (\delta\boldsymbol{\mathsf{C}} \nabla\mathbf{u})  \hat{\mathbf{n}}\,\mathrm{d}a  \\
	& \quad= \int_{\mathcal{B}} \operatorname{div}[ (\delta\boldsymbol{\mathsf{C}} 
	\nabla\mathbf{u} ) \boldsymbol{\gamma}]\,\mathrm{d}v
	- \int_{\mathcal{B}} \nabla\boldsymbol{\gamma} : \delta\boldsymbol{\mathsf{C}} 
	\nabla\mathbf{u}\,\mathrm{d}v
	- \int_{\partial_N\mathcal{B}} \boldsymbol{\gamma} \cdot (\delta\boldsymbol{\mathsf{C}} 
	\nabla\mathbf{u}) \hat{\mathbf{n}}\,\mathrm{d}a  \\
	& \quad = \int_{\partial_N\mathcal{B}}  (\delta\boldsymbol{\mathsf{C}} \nabla\mathbf{u} ) 
	\boldsymbol{\gamma}\cdot\hat{\mathbf{n}}\,\mathrm{d}a
	- \int_{\mathcal{C}} \nabla\boldsymbol{\gamma} : \delta\boldsymbol{\mathsf{C}} 
	\nabla\mathbf{u}\,\mathrm{d}v
	- \int_{\partial_N\mathcal{B}} \boldsymbol{\gamma} \cdot (\delta\boldsymbol{\mathsf{C}} 
	\nabla\mathbf{u}) \hat{\mathbf{n}}\,\mathrm{d}a  \\
	& \quad = - \int_{\mathcal{C}} \nabla\boldsymbol{\gamma} : \delta\boldsymbol{\mathsf{C}} 
	\nabla\mathbf{u}\,\mathrm{d}v \,,	
\end{aligned}
\end{equation}
where use was made of the fact that $\delta\boldsymbol{\mathsf{C}}|_{\mathcal{B}\setminus\mathcal{C}}=\boldsymbol{0}$,
and $\boldsymbol\gamma = \boldsymbol{0}$ on $\partial_D\mathcal{B}$ (as is shown later in this section).
Note that for isotropic solids one has
\begin{equation}
\begin{split}
	 \nabla\boldsymbol{\gamma} : \delta \boldsymbol{\mathsf{C}} \nabla\mathbf{u} &= 
	 \delta\lambda (\operatorname{div}\boldsymbol{\gamma})(\operatorname{div}\mathbf{u})
	 +2\delta\mu \left( \mathsf{sym}\nabla\boldsymbol{\gamma}  : \mathsf{sym}\nabla\mathbf{u} \right)
	 \\&= \left(\delta \kappa - \frac{2}{3}\delta \mu \right)
	 (\operatorname{div}\boldsymbol{\gamma})(\operatorname{div}\mathbf{u})
	 +2\delta\mu \left( \mathsf{sym}\nabla\boldsymbol{\gamma}  : \mathsf{sym}\nabla\mathbf{u} \right) \,,
\end{split}
\end{equation}
as $\mathsf{C}^{abcd}=\lambda\,\delta^{ab}\delta^{cd}+\mu\big(\delta^{ac}\delta^{bd}+\delta^{ad}\delta^{bc}\big)$ in Cartesian coordinates.
Therefore, using  $\delta\mu=-\mathring{\mu}\,e^{-\xi(x)}\,\delta \xi$ and $\delta\kappa=-\mathring{\kappa}\,e^{-\eta(x)}\,\delta\eta$, one obtains
\begin{equation} \label{C-Variation}
\begin{aligned}
	- \int_{\mathcal{C}} \nabla\boldsymbol{\gamma} : \delta\boldsymbol{\mathsf{C}} 
	\nabla\mathbf{u}\, \mathrm{d}v
	& =\int_{\mathcal{C}} \left[ - \frac{2}{3}\mathring{\mu}\,
	e^{-\xi}(\operatorname{div}\boldsymbol{\gamma})(\operatorname{div}\mathbf{u})
	 +2\mathring{\mu}\,e^{-\xi}\left(\mathsf{sym}\nabla\boldsymbol{\gamma} 
	 : \mathsf{sym}\nabla\mathbf{u} \right) \right] \delta\xi \,\mathrm{d}v \\
	  & \quad+ \int_{\mathcal{C}} \mathring{\kappa} \,e^{-\eta} 
	(\operatorname{div}\boldsymbol{\gamma})(\operatorname{div}\mathbf{u})\,\delta\eta \, \mathrm{d}v  \,,
\end{aligned}
\end{equation}
where $\mathsf{sym}$ indicates the symmetric part of a second-order tensor, i.e., $(\mathsf{sym}\,A)_{ab}=\frac{1}{2}(A_{ab}+A_{ba})$.
We are looking for extremal points of $\mathsf{f}$, i.e., fields $\boldsymbol{\mathsf{C}}, \mathbf{u}, \boldsymbol{\gamma}$ such that $\delta_{\boldsymbol{\mathsf{C}}} \mathsf{f}=0$.
Using the arbitrariness of both $\delta\xi$ and $\delta\eta$, from~\eqref{div-grad-pen} and~\eqref{C-Variation} one obtains the strong form of the governing equations associated with $\boldsymbol{\mathsf{C}}$-variations:
\begin{equation} \label{C-eq-isotropic}
\begin{aligned}
	-\frac{2}{3}\mathring{\mu}\,
	e^{-\xi}(\operatorname{div}\boldsymbol{\gamma})(\operatorname{div}\mathbf{u})
	 +2\mathring{\mu}\,e^{-\xi}\left(\mathsf{sym}\nabla\boldsymbol{\gamma} 
	 : \mathsf{sym}\nabla\mathbf{u} \right)+m_1\xi -\alpha_1 \nabla^2\xi & = 0 \,,\\
	\mathring{\kappa} \,e^{-\eta}
	(\operatorname{div}\boldsymbol{\gamma})(\operatorname{div}\mathbf{u})+m_2\eta
	-\alpha_2 \nabla^2\eta &= 0 \,,
\end{aligned}
\end{equation}
while on $\partial\mathcal{C}$ we have the Neumann boundary conditions $\nabla\xi\cdot\hat{\mathbf{n}}=0$ and $\nabla\eta \cdot\hat{\mathbf{n}}=0$.

\paragraph{$\mathbf{u}$-variations.}Next, we take the $\mathbf{u}$-variation of the objective function, which reads
\begin{equation} \label{u-variation}
	\delta_{\mathbf{u}} \mathsf{f} =
	\int_{\mathring{\mathcal{B}}} k(\mathbf{u}-\tilde{\mathbf{u}})\cdot \delta\mathbf{u}\,\mathrm{d}v
	+\int_{\mathcal{B}} \boldsymbol{\gamma} \cdot \operatorname{div}(\boldsymbol{\mathsf{C}}\nabla\delta\mathbf{u})\,
	\mathrm{d}v
	- \int_{\partial_N\mathcal{B}} \boldsymbol{\gamma} \cdot (\boldsymbol{\mathsf{C}} 
	\nabla\delta\mathbf{u})\hat{\mathbf{n}}\,\mathrm{d}a
	\,.
\end{equation}
The second term on the right-hand side of \eqref{u-variation} can be simplified as
\begin{equation}  \label{u-variation2}
\begin{aligned}
	\int_{\mathcal{B}} \boldsymbol{\gamma} \cdot 
	\operatorname{div}(\boldsymbol{\mathsf{C}}\nabla\delta\mathbf{u})\,\mathrm{d}v
	& =\int_{\mathcal{B}} 
	\left[\operatorname{div}(\boldsymbol{\mathsf{C}}\nabla\delta\mathbf{u}\cdot\boldsymbol{\gamma})
	-\nabla\boldsymbol{\gamma} : \boldsymbol{\mathsf{C}}\nabla\delta\mathbf{u}\right]\,\mathrm{d}v \\
	& =-\int_{\mathcal{B}} \boldsymbol{\mathsf{C}}\nabla\boldsymbol{\gamma}
	:\nabla\delta\mathbf{u}\,\mathrm{d}v
	+ \int_{\partial\mathcal{B}} \boldsymbol{\gamma} \cdot (\boldsymbol{\mathsf{C}} 
	\nabla\delta\mathbf{u})\hat{\mathbf{n}}\,\mathrm{d}a \\
	& =-\int_{\mathcal{B}} \Big\{ \operatorname{div}\left[ (\boldsymbol{\mathsf{C}}\nabla\boldsymbol{\gamma})
	\delta\mathbf{u}\right]-\operatorname{div}(\boldsymbol{\mathsf{C}}\nabla\boldsymbol{\gamma})
	\cdot\delta\mathbf{u} \Big\}\mathrm{d}v
	+ \int_{\partial\mathcal{B}} \boldsymbol{\gamma} \cdot ( \boldsymbol{\mathsf{C}} 
	\nabla\delta\mathbf{u})\hat{\mathbf{n}}\,\mathrm{d}a \\
	& =\int_{\mathcal{B}} \operatorname{div}(\boldsymbol{\mathsf{C}}\nabla\boldsymbol{\gamma})
	\cdot\delta\mathbf{u}\,\mathrm{d}v
	-\int_{\partial_N\mathcal{B}} (\boldsymbol{\mathsf{C}}\nabla\boldsymbol{\gamma})
	\hat{\mathbf{n}}\cdot\delta\mathbf{u} \,\mathrm{d}a
	+ \int_{\partial\mathcal{B}} \boldsymbol{\gamma} \cdot (\boldsymbol{\mathsf{C}} 
	\nabla\delta\mathbf{u})\hat{\mathbf{n}}\,\mathrm{d}a 
	\,.
\end{aligned}
\end{equation}
The boundary condition $\mathbf{u} =\bar{\mathbf{u}}$ on $\partial_D\mathcal{B}$ implies that $\delta\mathbf{u}=\mathbf{0}$ on $\partial_D\mathcal{B}$.
Knowing that $\delta\mathbf{u}=\mathbf{0}$ on $\partial_D\mathcal B$ is not enough to imply the vanishing of $\nabla \delta\mathbf{u}$ on $\partial_D\mathcal B$, as $\nabla_{\hat{\mathbf{n}}} \delta\mathbf{u}$ is not specified.
The last term on the right-hand side of \eqref{u-variation2} can be written as
\begin{equation} 	
	\int_{\partial\mathcal{B}} \boldsymbol{\gamma} \cdot ( \boldsymbol{\mathsf{C}} 
	\nabla\delta\mathbf{u})\hat{\mathbf{n}}\,\mathrm{d}a =
	\int_{\partial_N\mathcal{B}} \boldsymbol{\gamma} \cdot (\boldsymbol{\mathsf{C}} 
	\nabla\delta\mathbf{u})\hat{\mathbf{n}}\,\mathrm{d}a +
	\int_{\partial_D\mathcal{B}}
	\nabla_{\hat{\mathbf{n}}} \delta\mathbf{u}
	\cdot \left[ \boldsymbol{\mathsf{C}} (\boldsymbol{\gamma} \otimes \hat{\mathbf{n}}) \right] \hat{\mathbf{n}}\,\mathrm{d}a \,.
\end{equation}
The $\mathbf{u}$-variation \eqref{u-variation} now reads
\begin{equation}
\begin{aligned}
	\delta_{\mathbf{u}} \mathsf{f} & =
	\int_{\mathring{\mathcal{B}}} k(\mathbf{u}-\tilde{\mathbf{u}})\cdot \delta\mathbf{u}\,\mathrm{d}v+
	\int_{\mathcal{B}} \operatorname{div}(\boldsymbol{\mathsf{C}}\nabla\boldsymbol{\gamma})
	\cdot\delta\mathbf{u}\,\mathrm{d}v
	-\int_{\partial_N\mathcal{B}}(\boldsymbol{\mathsf{C}}\nabla\boldsymbol{\gamma})
	\hat{\mathbf{n}}\cdot\delta\mathbf{u} 
	\,\mathrm{d}a \\
	&\quad+
	\int_{\partial_D\mathcal{B}}
	\nabla_{\hat{\mathbf{n}}} \delta\mathbf{u}
	\cdot \left[ \boldsymbol{\mathsf{C}} (\boldsymbol{\gamma} \otimes \hat{\mathbf{n}}) \right] 
	\hat{\mathbf{n}}\,\mathrm{d}a
	=0
	\,.
\end{aligned}
\end{equation}
Arbitrariness of $\nabla_{\hat{\mathbf{n}}} \delta\mathbf{u}$ on $\partial_D\mathcal{B}$ implies that
\begin{equation} 
	\left[\boldsymbol{\mathsf{C}} (\boldsymbol{\gamma} \otimes \hat{\mathbf{n}}) \right] \hat{\mathbf{n}} 
	= \mathbf{0}	\qquad\text{on}~\partial_D\mathcal{B},
\end{equation}
or in components $\mathsf{C}^{abcd}\,n_b\, \gamma_c\, n_d = 0$.
Using an argument similar to that used in \S\ref{App:Lag-mult} one can show that $\boldsymbol{\gamma}=\mathbf{0}$ on $\partial_D\mathcal{B}$.
Therefore, the $\mathbf{u}$-variation \eqref{u-variation} is simplified to read
\begin{equation}
\begin{aligned}
	\delta_{\mathbf{u}} \mathsf{f} =
	\int_{\mathring{\mathcal{B}}} k(\mathbf{u}-\tilde{\mathbf{u}})\cdot \delta\mathbf{u}\,\mathrm{d}v+
	\int_{\mathcal{B}} \operatorname{div}(\boldsymbol{\mathsf{C}}\nabla\boldsymbol{\gamma})
	\cdot\delta\mathbf{u}\,\mathrm{d}v
	-\int_{\partial_N\mathcal{B}}(\boldsymbol{\mathsf{C}}\nabla\boldsymbol{\gamma})
	\hat{\mathbf{n}}\cdot\delta\mathbf{u} 
	\,\mathrm{d}a
	\,.
\end{aligned}
\end{equation}
Thus, the strong form of the governing equations associated with $\mathbf{u}$-variations is
\begin{equation} \label{u-eq-anisotropic}
\begin{aligned}
    && \operatorname{div}(\boldsymbol{\mathsf{C}}\nabla\boldsymbol{\gamma}) = \mathbf{0} &&& \text{in~}\mathcal C \,, \\
    && \operatorname{div}(\mathring{\boldsymbol{\mathsf{C}}}\nabla\boldsymbol{\gamma})
    +k(\mathbf{u}-\tilde{\mathbf{u}}) = \mathbf{0} 
    &&& \text{in~}\mathring{\mathcal{B}} \,, \\
    && (\boldsymbol{\mathsf{C}}\nabla\boldsymbol{\gamma})\hat{\mathbf{n}} =\mathbf{0} &&& \text{on}~\partial_N\mathcal{B} \,, \\
    && \boldsymbol{\gamma}=\mathbf{0}  &&& \text{on}~\partial_D\mathcal{B} \,.
\end{aligned}
\end{equation}
Eqs.~\eqref{u-eq-anisotropic} represent the \emph{adjoint} elasticity problem, in which
$\boldsymbol\gamma$ is a displacement field on $\mathcal B$.
In this adjoint problem, the body forces are discontinuous, being $k(\mathbf{u}-\tilde{\mathbf{u}})$ in $\mathring{\mathcal{B}}$, while they vanish in $\mathcal C$.
As for the adjoint boundary conditions, $\boldsymbol\gamma$ is fixed on $\partial_D\mathcal{B}$, while $\partial_N\mathcal{B}$ is traction-free.

\paragraph{The governing equations of the optimization problem.}
Notice that $\boldsymbol\gamma$-variations give the balance of linear momentum with its associated boundary conditions.
Therefore, it is now possible to write the complete set of governing equations.
First, we define the following operators:\footnote{The sharp operator $\sharp$ raises indices. $\mathbf{g}$ is the spatial metric and has components $g_{ab}$ in the curvilinear coordinates $\{x^a\}$. $\mathbf{g}^{\sharp}$ is the inverse of the spatial metric and has components $g^{ab}$ such that $g^{ac}\,g_{cb}=\delta^a_b$.}
\begin{equation}  \label{w-eq}
\begin{aligned}
	W_1(\xi,\mathbf{u},\boldsymbol{\gamma})  &= 
	-\frac{2}{3}\mathring{\mu}\,e^{-\xi}
         (\operatorname{div}\boldsymbol\gamma )(\operatorname{div}\mathbf{u} )\mathbf{g}^{\sharp}
	+2 \mathring{\mu}\,e^{-\xi} (\operatorname{sym}\nabla\boldsymbol\gamma 
	: \operatorname{sym}\nabla \mathbf{u}) \,, \\
	W_2(\eta,\mathbf{u},\boldsymbol{\gamma})  &= 
	\mathring{\kappa} \,e^{-\eta} (\operatorname{div}\boldsymbol{\gamma})(\operatorname{div}\mathbf{u})  \,,
\end{aligned}
\end{equation}
representing the work done by the stress on the adjoint displacements $\boldsymbol{\gamma}$ (or vice versa, the work done by the adjoint stress on the standard displacements $\mathbf{u}$ ).
In particular, $W_1$ is associated with shear deformations, while $W_2$ is associated with changes in volume.
The total work is given by $W_1+W_2$.
We also define the stress operator as
\begin{equation} \label{sigma-eq}
	\boldsymbol\sigma( \xi,\eta, \mathbf{y}) = 
	\left( -\frac{2}{3}\mathring{\mu}\,e^{-\xi}+\mathring{\kappa} \,e^{-\eta} \right)
       (\operatorname{div} \mathbf{y}) \mathbf{g}^{\sharp}
	+2\mathring{\mu}\,e^{-\xi} (\operatorname{sym}\nabla \mathbf{y})^{\sharp} \,.
\end{equation}
In summary, the complete optimization BVP reads:
\begin{equation} \label{strong-form-isotropic}
\begin{aligned}
    && W_1(\xi,\mathbf{u},\boldsymbol{\gamma})  \,e^{-\xi} +m_1\xi -\alpha_1 \nabla^2\xi= 0 
  &&&  \text{in~}\mathcal C \,,\\
    && W_2(\eta,\mathbf{u},\boldsymbol{\gamma}) \,e^{-\eta}  
	+m_2\eta-\alpha_2 \nabla^2\eta= 0 &&&  \text{in~}\mathcal C \,, \\
	&& \nabla\xi\cdot\hat{\mathbf{n}}= 0 &&&  \text{on~}\partial\mathcal C \,, \\
	&& \nabla\eta \cdot\hat{\mathbf{n}}=0= 0 &&&  \text{on~}\partial\mathcal C \,, \\[4pt]
    &&  \operatorname{div}  \boldsymbol\sigma( \xi,\eta, \boldsymbol\gamma )= \mathbf{0} &&& \text{in~}\mathcal C \,, \\
    &&  \operatorname{div}\boldsymbol\sigma(  0,0, \boldsymbol\gamma) +k(\mathbf{u}-\tilde{\mathbf{u}}) = \mathbf{0}
	&&& \text{in~}\mathring{\mathcal{B}} \,, \\
    && \boldsymbol\sigma( \xi,\eta, \boldsymbol\gamma) \hat{\mathbf{n}} =\mathbf{0} &&& \text{on}~\partial_N\mathcal{B} \,, \\
    && \boldsymbol{\gamma}=\mathbf{0}  &&& \text{on}~\partial_D\mathcal{B} \,, \\[4pt]
        && \operatorname{div}  \boldsymbol\sigma(\xi,\eta, \mathbf{u})  +\mathbf{b} = \mathbf{0}&&& \text{in~}\mathcal{B} \,, \\
    && \bar{\mathbf{t}} - \boldsymbol\sigma(\xi,\eta, \mathbf{u} ) \hat{\mathbf{n}}
	=\mathbf{0} &&&\text{on~}\partial_N\mathcal{B} \,, \\
    && \mathbf{u} - \bar{\mathbf{u}} = \mathbf{0} &&& \text{on~}\partial_D\mathcal{B} \,.
\end{aligned}
\end{equation}

\begin{remark} \label{Rem:linearity}
Under a proper rescaling of the penalty factor $k$, the design of a cloak is not affected by the intensity of the load it is optimized for.
Moreover, the performance of a given cloak is not affected by the intensity of the external loads.
In order to show this, let us assume that $\bar{\mathbf{u}}$, $\bar{\mathbf{t}}$ and $\mathbf{b}$ are replaced by $\bar{\mathbf{u}}' = c\,\bar{\mathbf{u}}$, $\bar{\mathbf{t}}' =c\,\bar{\mathbf{t}}$ and $\mathbf{b}' = c\,\mathbf{b}$, for $c\in\mathbb{R}$. 
Since the virtual problem is linear, its solution is $c\,\tilde{\mathbf{u}}$.
Then, if the penalty factor is rescaled as $k' = k / c^2$, it is straightforward to see that $c\,\mathbf{u}$, $\boldsymbol{\gamma}/ c$, $\xi$, $\eta$ are solutions of~\eqref{strong-form-isotropic} for the loads and boundary conditions $\bar{\mathbf{u}}' $, $\bar{\mathbf{t}}' $ and $\mathbf{b}'$. Moreover, since we measure the performance of a cloak using a normalized metric, one has
\begin{equation}
	\frac{\Vert c\,\mathbf{u} - c\,\tilde{\mathbf{u}} \Vert_{L^2}}{\Vert c\, \tilde{\mathbf{u}} \Vert_{L^2} }=
	\frac{\Vert \mathbf{u} - \tilde{\mathbf{u}} \Vert_{L^2}}{\Vert \tilde{\mathbf{u}} \Vert_{L^2}} \,.
\end{equation}
Note that the rescaling of the penalty factor can be avoided by normalizing it with respect to $\Vert \tilde{\mathbf{u}} \Vert_{L^2}$.
The same rescaling property does not hold when one replaces  $\bar{\mathbf{u}}$, $\bar{\mathbf{t}}$ and $\mathbf{b}$ with linear combinations 
\begin{equation}
	\bar{\mathbf{u}}' = \sum_{i} c_i \bar{\mathbf{u}}_i \,,\qquad
	\bar{\mathbf{t}}' = \sum_{i} c_i \bar{\mathbf{t}} \,,\qquad
	\mathbf{b}' = \sum_{i} c_i \mathbf{b} \,.
\end{equation}
This is due to the fact that, although the $L^2$ norm satisfies absolute homogeneity $\Vert c\, \mathbf{x}\Vert^2_{L^2}= c^2 \Vert \mathbf{x}\Vert$, in the case of sums only the triangular inequality holds.
Therefore, while the intensity is not a factor, the placement and direction of loads affect both the design of a cloak and its performance.
For this reason, it is crucial to consider optimal design for multiple loads as well.
\end{remark}

\subsection{Cloaking under multiple loads}

Consider $N$ different loads with their corresponding boundary tractions $\bar{\mathbf{t}}_{(i)}$, and body forces $\mathbf{b}_{(i)}$, $i=1,2,...,N$. We denote the corresponding physical and virtual displacement fields by $\mathbf{u}_{(i)}$ and $\tilde{\mathbf{u}}_{(i)}$ respectively. We assign a weight $w_i\geq 0$ to each load such that $\sum_{i=1}^{N}w_i=1$.
Each combination of traction and displacement boundary conditions is associated with a partition of $\partial\mathcal{B}$ into $\partial_{D_{(i)}}\mathcal{B}$ and $\partial_{N_{(i)}}\mathcal{B}$, $i=1,2,...,N$.
Recalling~\eqref{modified-objective} and~\eqref{metric-H2}, the objective function is defined as
\begin{equation} \label{objective-multiple}
\begin{split}
	\mathsf{f} &=
	\frac{k}{2} \, \sum_{i=1}^{N} w_i \int_{\mathring{\mathcal{B}}}  \|\mathbf{u}_{(i)}
	-\tilde{\mathbf{u}}_{(i)}\|^2 \,\mathrm{d}v
	+ \int_{\mathcal{C}} \Bigl[ m_1\xi^2+m_2\eta^2 \Bigr] \mathrm d v
	+ \int_{\mathcal{C}} \Bigl[  \alpha_1\Vert \nabla\xi \Vert^2+
	\alpha_2\Vert \nabla\eta \Vert^2 \Bigr] \mathrm d v
	\\
	& \quad+ \sum_{i=1}^{N} w_i \int_{\mathcal{B}} \boldsymbol{\gamma}_{(i)} \cdot 
	\left[ \operatorname{div} \left(\boldsymbol{\mathsf{C}} \nabla\mathbf{u}_{(i)} \right)
	+\mathbf{b}_{(i)}\right] \mathrm{d}v  
	+\sum_{i=1}^{N}  w_i \int_{\partial_{N_{(i)}}\mathcal{B}} \boldsymbol{\gamma}_{(i)} \cdot 
	\left[  \bar{\mathbf{t}}_{(i)} - \left(\boldsymbol{\mathsf{C}} \nabla\mathbf{u}_{(i)} \right) 
	\hat{\mathbf{n}}\right]  \mathrm{d}a \,,
\end{split}
\end{equation}
where $\boldsymbol{\gamma}_{(i)}$ is the Lagrange multiplier field enforcing equilibrium equations for the $i$-th loading.
The minimization problem for the optimal design of an elastic cloak under multiple loads is rewritten as
\begin{equation} \label{minimization-problem-multiple}
	\inf_{\substack{\boldsymbol{\mathsf{C}} \\\mathbf{u}_{(1)},\hdots,\mathbf{u}_{(N)}\\ 
	\boldsymbol{\gamma}_{(1)},\hdots,\boldsymbol{\gamma}_{(N)}}}
	\mathsf{f}(\boldsymbol{\mathsf{C}},\mathbf{u}_{(1)},\hdots,\mathbf{u}_{(N)},
	\boldsymbol{\gamma}_{(1)},\hdots,\boldsymbol{\gamma}
	_{(N)})   \,.
\end{equation}
In the case of isotropic solids, we take variations of $\xi$, $\eta$, $\mathbf{u}_{(i)}$, $\boldsymbol\gamma_{(i)}$.
Following the same calculations as in the single-load case, one obtains the strong form of the governing equations as
\begin{equation} \label{strong-form-isotropic-multi}
\begin{aligned}
    && e^{-\xi} \sum\nolimits_{j=1}^{N} w_i  W_1\left( \xi, \mathbf{u}_{(j)},\boldsymbol{\gamma}_{(j)} \right)  +m_1\xi -\alpha_1 \nabla^2\xi= 0 
  &&&  \text{in~}\mathcal C \,,\\
    && e^{-\eta}   \sum\nolimits_{j=1}^{N} w_i  W_2 \left(\eta, \mathbf{u}_{(j)},\boldsymbol{\gamma}_{(j)}\right) 
	+m_2\eta-\alpha_2 \nabla^2\eta= 0 &&&  \text{in~}\mathcal C \,, \\
	&& \nabla\xi\cdot\hat{\mathbf{n}}= 0 &&&  \text{on~}\partial\mathcal C \,, \\
	&& \nabla\eta \cdot\hat{\mathbf{n}}=0= 0 &&&  \text{on~}\partial\mathcal C \,, \\[4pt]
    &&  \operatorname{div}  \boldsymbol\sigma( \xi,\eta, \boldsymbol\gamma_{(i)} )= \mathbf{0} &&& \text{in~}\mathcal C \,, \\
    &&  \operatorname{div}\boldsymbol\sigma( 0,0, \boldsymbol\gamma_{(i)} ) +k (\mathbf{u}_{(i)}-\tilde{\mathbf{u}}_{(i)}) = \mathbf{0}
	&&& \text{in~}\mathring{\mathcal{B}} \,, \\
    && \boldsymbol\sigma( \xi,\eta,  \boldsymbol\gamma_{(i)}) \hat{\mathbf{n}} =\mathbf{0} &&& \text{on}~\partial_{N_{(i)}}\mathcal{B} \,, \\
    && \boldsymbol{\gamma}_{(i)}=\mathbf{0}  &&& \text{on}~\partial_{D_{(i)}}\mathcal{B} \,, \\[4pt]
        && \operatorname{div}  \boldsymbol\sigma( \xi,\eta, \mathbf{u}_{(i)} )  +\mathbf{b}_{(i)} = \mathbf{0}&&& \text{in~}\mathcal{B} \, \\
    && \bar{\mathbf{t}}_{(i)} - \boldsymbol\sigma( \xi,\eta, \mathbf{u}_{(i)} ) \hat{\mathbf{n}}
	=\mathbf{0} &&&\text{on~}\partial_{N_{(i)}}\mathcal{B} \,, \\
    && \mathbf{u}_{(i)} - \bar{\mathbf{u}}_{(i)} = \mathbf{0} &&& \text{on~}\partial_{D_{(i)}}\mathcal{B} \,,
\end{aligned}
\end{equation}
for $i=1,2,\hdots,N$.

\section{Finite Element Discretization of the Optimization Problem} \label{Sec:FEM}

In this section we propose a weak formulation of the cloaking optimization problem and discuss a mixed finite element discretization of the weak governing equations.

\subsection{The weak form of the governing equations}
Let $L^2(\mathcal{B})$, $L^{2}(T\mathcal{B})$, and $L^{2}(\otimes^2 T\mathcal{B})$ be the spaces of square integrable scalar fields, vector fields, and $\binom{2}{0}$-tensor fields in $\mathcal{B}$, respectively. We also define the same spaces for $\mathcal{C},$ and $\mathring{\mathcal{B}} \subset \mathcal{B}$ and their boundaries. Let us also define 
\begin{equation}
\begin{aligned}
	H^{1}(T\mathcal{B}) & :=\left\{ \mathbf{u}\in L^{2}(T\mathcal{B}):  \nabla\mathbf{u} 
	\in L^{2}(\otimes^2 T\mathcal{B}) \right\} \,,\\
	H^{1}(T\mathcal{B},\partial_D\mathcal{B},\bar{\mathbf{u}}) &:=\left\{\mathbf{u}\in H^{1}(T\mathcal{B}): 
	\mathbf{u}|_{\partial_D\mathcal{B}}=\bar{\mathbf{u}} \right\}\,, \\
	H^{1}(T\mathcal{B},\partial_D\mathcal{B}) &:= H^{1}(T\mathcal{B},\partial_D\mathcal{B},\boldsymbol{0})\,, 
	\\
	H^{1}(\mathcal{C}) & :=\left\{ f \in L^{2}(\mathcal{C}):  \nabla f	\in L^{2}(T\mathcal{C}) \right\}
	\,,
\end{aligned}
\end{equation}
and $H^{1/2}(T\partial\mathcal{B}) := \mathrm{tr}\left(H^{1}(T\mathcal{B})\right) $, where \mbox{$\mathrm{tr}: H^{1}(T\mathcal{B}) \rightarrow L^{2}(T\partial\mathcal{B})$} is the trace operator \citep{Evans2010}.

\begin{prob}[Weak form of the governing equations] \label{prob-weak-form}
Let $\boldsymbol{b}_{(i)}$ be a body force of $L^{2}$-class, $\overline{\mathbf{u}}_{(i)}$ a boundary displacement on $\partial_{D_{(i)}}\mathcal{B}$ of $H^{1/2}$-class, and $\overline{\mathbf{t}}_{(i)}$ a boundary traction on $\partial_{N_{(i)}}\mathcal{B}$ of $L^{2}$-class for $i = 1,2,...,N$. Let the displacement field \mbox{$\tilde{\mathbf{u}}_{(i)} \in H^{1}(T\tilde{\mathcal{B}},\partial_{D_{(i)}}\tilde{\mathcal{B}},\bar{\mathbf{u}}_{(i)})$} be the solution of the virtual problem for the same given $\boldsymbol{b}_{(i)}$, $\overline{\mathbf{u}}_{(i)}$, $\overline{\mathbf{t}}_{(i)}$, for $i = 1,2,...,N$.\footnote{Note that $\tilde{\mathbf{u}}_{(i)}$ is the solution of the following problem
\begin{equation}
	\int_{\tilde{\mathcal{B}}} \left[\mathring{\lambda}(\operatorname{div}\tilde{\mathbf{u}}_{(i)})
	\mathbf{g}^{\sharp}
	+2\mathring{\mu}(\operatorname{sym}\nabla\tilde{\mathbf{u}}_{(i)})^{\sharp})\right]
	\!:\!\nabla\mathbf{w}_{(i)} \, \mathrm{d}v
	=\int_{\tilde{\mathcal{B}}} \mathbf{b}_{(i)}\cdot \mathbf{w}_{(i)}\, \mathrm{d}v
	+\int_{\partial_{N_{(i)}}\tilde{\mathcal{B}}} \bar{\mathbf{t}}_{(i)}\cdot \mathbf{w}_{(i)}\, \mathrm{d}a, 
	\quad \forall \mathbf{w}_{(i)}\in H^{1}(T\tilde{\mathcal{B}},\partial_{D_{(i)}}\tilde{\mathcal{B}})
	\,.
\end{equation}
} 
Find $(\xi,\eta,\mathbf{u}_{(1)},...,\mathbf{u}_{(N)},\boldsymbol{\gamma}_{(1)},...,\mathbf{u}_{(N)}) \in H^{1}(\mathcal{C}) \times H^{1}(\mathcal{C}) \times   H^{1}(T\mathcal{B},\partial_{D_{(1)}}\mathcal{B},\bar{\mathbf{u}}_{(1)}) \times ... \times
 H^{1}(T\mathcal{B},\partial_{D_{(N)}}\mathcal{B},\bar{\mathbf{u}}_{(N)}) \times H^{1}(T\mathcal{B},\partial_{D_{(1)}}\mathcal{B}) \times ... \times H^{1}(T\mathcal{B},\partial_{D_{(N)}}\mathcal{B})$ such that, for $i=1,2,...,N$,
\begin{equation} \label{weak-form}
\begin{alignedat}{3} 
		\sum_{j=1}^N w_j  \int_{\mathcal{C}} W_1(\xi, \mathbf{u}_{(i)},\boldsymbol{\gamma}_{(j)}) \, \delta\xi 
		\,\mathrm{d}v  +\int_{\mathcal{C}}   m_1\xi \, \delta\xi \,\mathrm{d}v  
		+ \int_{\mathcal{C}}  \alpha_1 \nabla\xi\cdot \nabla \delta\xi \mathrm{d}v  
		& = 0, & &\forall \delta\xi\in H^1(\mathcal{C}) \,,\\
		\sum_{j=1}^N w_j  \int_{\mathcal{C}} W_2(\eta, \mathbf{u}_{(j)},\boldsymbol{\gamma}_{(j)}) 
		\, \delta\eta\, \mathrm{d}v 
		+ \int_{\mathcal{C}} m_2\eta\, \delta\eta \, \mathrm{d}v 
		+  \int_{\mathcal{C}} \alpha_2 \nabla\eta \cdot \nabla \delta\eta \, \mathrm{d}v  
		& = 0, & &\forall\delta\eta\in H^1(\mathcal{C})\,, \\
		k \int_{\mathring{\mathcal{B}}} (\mathbf{u}_{(i)}-\tilde{\mathbf{u}}_{(i)})\cdot 
		\delta\mathbf{u}_{(i)}\,\mathrm{d}v
		 -\int_{\mathcal{B}}   \boldsymbol{\sigma}(\xi,\eta,\boldsymbol{\gamma}_{(i)}) 
		 : \nabla\delta\mathbf{u}_{(i)} \,\mathrm{d}v
		& = 0,
		& &\forall \delta\mathbf{u}_{(i)}\in H^{1}(T\mathcal{B},\partial_{D_{(i)}}\mathcal{B})\,,\\
		-\int_{\mathcal{B}} \boldsymbol{\sigma}(\xi,\eta,\mathbf{u}_{(i)}) 
		: \nabla\delta\boldsymbol{\gamma} \,,\mathrm{d}v
		+\int_{\mathcal{B}} \mathbf{b}_{(i)}\cdot \delta\boldsymbol{\gamma}_{(i)}\, \mathrm{d}v
		+\int_{\partial_N\mathcal{B}} \bar{\mathbf{t}}_{(i)}\cdot \delta\boldsymbol{\gamma}_{(i)}\, 
		\mathrm{d}a 
		& =0,
		&\quad &\forall \delta\boldsymbol{\gamma}_{(i)}\in H^{1}(T\mathcal{B},\partial_{D_{(i)}}\mathcal{B})\,.
\end{alignedat}
\end{equation}
\end{prob}

Let $\llangle,\rrangle_{\mathcal{A}}$ denote the $L^{2}$-inner products of scalar, vector, and tensor fields on a set $\mathcal{A}$, which are defined as $\llangle f,g \rrangle_{\mathcal{A}}:=\int_{\mathcal{A}}fg\, \mathrm{d}a$, $\llangle\boldsymbol{Y},\boldsymbol{Z}\rrangle_{\mathcal{A}}:=\int_{\mathcal{A}}Y^{I}Z^{I}dA$, and $\llangle\boldsymbol{S},\boldsymbol{T}\rrangle_{\mathcal{A}}:=\int_{\mathcal{A}}S^{IJ}T^{IJ}dA$, respectively. 
The weak form of the governing equations is written more compactly as
\begin{equation} \label{Cloaking-Optimization-Problem}
\begin{alignedat}{3}
	\sum\nolimits_{j=1}^N  w_j \left \llangle  W_1(\xi, \mathbf{u}_{(j)},\boldsymbol{\gamma}_{(j)})  + m_1\xi,
	\delta\xi \right\rrangle_{\mathcal{C}}
	+\left \llangle \alpha_1 \nabla \xi , \nabla \delta\xi \right\rrangle_{\mathcal{C}}
	& = 0\,, & &\forall \delta\xi\in H^1(\mathcal{C}) \,,\\
	\sum\nolimits_{j=1}^N w_j  \left \llangle W_2(\eta,\mathbf{u}_{(j)},\boldsymbol{\gamma}_{(j)}) + m_2\eta,
	\delta\eta \right\rrangle_{\mathcal{C}}
	+\left \llangle \alpha_2 \nabla \eta , \nabla \delta\eta \right\rrangle_{\mathcal{C}}
	& = 0\,, & &\forall \delta\eta\in H^1(\mathcal{C}) \,,\\
	k \llangle (\mathbf{u}_{(i)}-\tilde{\mathbf{u}}_{(i)}), \delta\mathbf{u}_{(i)} 
	 \rrangle_{\mathring{\mathcal{B}}}
	- \llangle \boldsymbol{\sigma}(\xi,\eta,\boldsymbol{\gamma}_{(i)}), 
	\nabla\delta\mathbf{u}_{(i)}	\rrangle_{\mathcal{B}}& =0\,,
	& &\forall \delta\mathbf{u}_{(i)} \in H^{1}(T\mathcal{B},\partial_{D_{(i)}}\mathcal{B})\,,\\
	-\llangle \boldsymbol{\sigma}(\xi,\eta,\mathbf{u}_{(i)}), 
	\nabla\delta\boldsymbol{\gamma}_{(i)} \rrangle_{\mathcal{B}}
	+\llangle \mathbf{b}_{(i)}, \delta\boldsymbol{\gamma}_{(i)}\rrangle_{\mathcal{B}}
	+  \llangle \bar{\mathbf{t}}_{(i)}, \delta\boldsymbol{\gamma}_{(i)}\rrangle_{\partial_{N_{(i)}}\mathcal{B}}
	& =0\,,
	&\quad &\forall \delta\boldsymbol{\gamma}_{(i)}\in H^{1}(T\mathcal{B},\partial_{D_{(i)}}\mathcal{B}) \,,\\
\end{alignedat}
\end{equation}
for $i = 1,2,\hdots,N$.
Eq.~\eqref{Cloaking-Optimization-Problem} represents a system of $2N+2$ equations,\footnote{In the anisotropic case, one would obtain $2N+M$ equations, where $M$ is the number of independent elastic moduli.} all coupled by the effect of the elastic constants $\xi(x)$ and $\eta(x)$.
Each load case independently contributes to the system with the balance of the standard and adjoint linear momenta~\eqref{Cloaking-Optimization-Problem}$_3$ and~\eqref{Cloaking-Optimization-Problem}$_4$.
Moreover, all the load cases appear in the $\boldsymbol{\mathsf{C}}$-equations~\eqref{Cloaking-Optimization-Problem}$_1$ and~\eqref{Cloaking-Optimization-Problem}$_2$ through the mixed energies $W_1(\xi, \mathbf{u}_{(j)},\boldsymbol{\gamma}_{(j)})$ and $W_2(\eta,\mathbf{u}_{(j)},\boldsymbol{\gamma}_{(j)})$, $i = 1,2,...,N$.

As for the second variations of the objective function, they can be arranged in a matrix with the following structure
\begin{equation} \label{2nd-variations}
\begin{bmatrix}
	\delta_{\xi\xi} \mathsf{f} & \text{sym}  & \text{sym}  & \text{sym}  \\
	\delta_{\xi\eta} \mathsf{f}  & \delta_{\eta\eta} \mathsf{f}  & \text{sym} & \text{sym} \\
	  [\delta_{\xi\mathbf{u}_{(i)}} \mathsf{f} ]^{N \times 1}  & [\delta_{\eta\mathbf{u}_{(i)}} \mathsf{f} ]^{N \times 1}   & [ \delta_{\mathbf{u}_{(i)}\mathbf{u}_{(i)}} \mathsf{f} ]^{N \times N} & \text{sym} \\
	 [\delta_{\xi \boldsymbol{\gamma}_{(i)}} \mathsf{f} ]^{N \times 1} & [ \delta_{\eta\boldsymbol{\gamma}_{(i)} } \mathsf{f} ]^{N \times 1} & [\delta_{\mathbf{u}_{(i)}\boldsymbol{\gamma}_{(i)}}  \mathsf{f} ]^{N \times N} & [\delta_{\boldsymbol{\gamma}_{(i)}\boldsymbol{\gamma}_{(i)}}   \mathsf{f} ]^{N \times N}
\end{bmatrix} \,.
\end{equation}
Recalling the definitions~\eqref{w-eq} and~\eqref{sigma-eq}, in the single-load case the matrix~\eqref{2nd-variations} is reduced to the following $4\times 4$ matrix:
{\footnotesize
\begin{equation}
\def\arraystretch{2.2}
\left[
\begin{array}{c}
  \left \llangle  -W_1( \xi, \mathbf{u},\boldsymbol{\gamma})  + m_1 ,
	\delta\xi \delta\xi  \right\rrangle_{\mathcal{C}}
	+\left \llangle \alpha_1 \nabla \delta\xi , \nabla \delta\xi \right\rrangle_{\mathcal{C}} \\
   0 \\
   \llangle W_1(\xi, \delta\mathbf{u},\boldsymbol{\gamma}) , \delta\xi \rrangle_{\mathcal{C}} \\
   \llangle  W_1(\xi,\mathbf{u},\delta\boldsymbol{\gamma}) , \delta\xi  \rrangle_{\mathcal{C}}
\end{array}
\hspace{-0.6in}
\begin{array}{c}
  \text{sym}\\
  \left \llangle -W_2(\eta,\mathbf{u},\boldsymbol{\gamma}) + m_2 ,
	\delta\eta \delta\eta \right\rrangle_{\mathcal{C}}
	+\left \llangle \alpha_2 \nabla \delta\eta , \nabla \delta\eta \right\rrangle_{\mathcal{C}} \\
   \llangle W_2(\eta, \delta\mathbf{u},\boldsymbol{\gamma}) , \delta\eta \rrangle_{\mathcal{C}} \\
  \llangle W_2(\eta, \mathbf{u},\delta\boldsymbol{\gamma}) , \delta\eta \rrangle_{\mathcal{C}}
\end{array}
\hspace{-0.5in}
\begin{array}{c}
   \text{sym} \\
    \text{sym} \\
  k \llangle \delta\mathbf{u}, \delta\mathbf{u} \rrangle_{\mathring{\mathcal{B}}} \\
  -\llangle W_1(\xi,\delta\mathbf{u},\delta\boldsymbol{\gamma})  + W_2(\eta,\delta\mathbf{u},\delta\boldsymbol{\gamma})  ,1 \rrangle_{\mathcal{B}}  \\
\end{array}
\begin{array}{c}
  \text{sym} \\
  \text{sym} \\
  \text{sym} \\
  0 \\
\end{array}
\right] .
\end{equation}
}

\subsection{Mixed finite elements}

Let $\mathcal{B}_{h}$ denote an arbitrary triangulation (or simply a mesh) of the reference configuration $\mathcal{B}$, where $h:=\mathsf{max}~\mathsf{diam}\,\mathcal{T}$ for all triangles ${\forall \mathcal{T}\in\mathcal{B}_{h}}$. Also, $\mathcal{C}_{h}$ and $\mathring{\mathcal{B}}_h$ are the triangulation of the cloaking region $\mathcal{C}$ and its complement $\mathring{\mathcal{B}}$, respectively.
We define the following finite element space:
\begin{equation}
\begin{aligned}
	V_{h}(T\mathcal{B}_h) &:= \left\{\boldsymbol{V}_h \in L^2(T\mathcal{B}_h) 
	: \forall\mathcal{T}\in\mathcal{B}_h,~ \boldsymbol{V}_h|_{\mathcal{T}}\in\mathcal{P}_{1}(T\mathcal{T}),
	~ \forall\EuScript{E}\in \EuScript{E}_{\mathcal{B}_h}^{i},~\llbracket\boldsymbol{V}_h\rrbracket_{\EuScript{E}}=\boldsymbol{0}\right\},\\
	V_{h}(\mathcal{C}_h) &:= \left\{f_h \in L^2(\mathcal{C}_h) : \forall\mathcal{T}\in\mathcal{C}_h,
	~ f_{h}|_{\mathcal{T}}\in\mathcal{P}_{1}(\mathcal{T}),
	~\forall\EuScript{E}\in \EuScript{E}_{\mathcal{C}_h}^{i},~\llbracket f_h \rrbracket_{\EuScript{E}}=\boldsymbol{0}\right\}
	\,,
\end{aligned}
\end{equation}
where $\EuScript{E}_{\mathcal{B}_h}^{i}$ and $\EuScript{E}_{\mathcal{C}_h}^{i}$ are the sets of interior edges of $\mathcal{B}_h$ and $\mathcal{C}_h$, respectively. Note that $\mathcal{P}_{1}(\mathcal{T})$ and $\mathcal{P}_{1}(T\mathcal{T})$ are first-order (linear) scalar-valued and vector-valued polynomial spaces in a triangle (element) $\mathcal{T}$.
Note that $V_{h}(T\mathcal{B}_h) \subset H^{1}(T\mathcal{B}_h)$ and $V_{h}(\mathcal{C}_h) \subset H^{1}(\mathcal{C}_h)$.
Let us also define
\begin{equation}
\begin{aligned}
	V_{h}(T\mathcal{B}_h,\partial_D\mathcal{B}_h,\bar{\mathbf{u}}) &:= \left\{\boldsymbol{V}_h \in 
	V_{h}(T\mathcal{B}_h),~\forall\EuScript{E}\in \EuScript{E}_h^D,
	~\boldsymbol{V}_h|_{\EuScript{E}}=\mathrm{I}_{\EuScript{E}}(\bar{\mathbf{u}}) \right\},\\
	V_{h}(T\mathcal{B}_h,\partial_D\mathcal{B}_h) &:= V_{h}(T\mathcal{B}_h,\partial_D\mathcal{B}_h,\mathbf{0})
	\,,
\end{aligned}
\end{equation}
where $\EuScript{E}_h^D$ is the set of edges in $\partial_D\mathcal{B}_h$, and $\mathrm{I}_{\EuScript{E}}$ is a linear interpolation operator over the edge $\EuScript{E}$ with the property $\mathrm{I}_{\EuScript{E}}(\boldsymbol{0})=\boldsymbol{0}$. Using the above approximation spaces the mixed finite element form of the cloaking optimization problem \eqref{Cloaking-Optimization-Problem} is written as:

\begin{prob}[Finite elements equations]\label{fem-problem}
Let $\boldsymbol{b}_{(i)}$ be a body force of $L^{2}$-class, $\overline{\mathbf{u}}_{(i)}$ a boundary displacement on $\partial_{D_{(i)}}\mathcal{B}$ of $H^{1/2}$-class, and $\overline{\mathbf{t}}_{(i)}$ a boundary traction on $\partial_{N_{(i)}}\mathcal{B}$ of $L^{2}$-class for $i = 1,2,\hdots,N$. Let the displacement field \mbox{$\tilde{\mathbf{u}}_{h_{(i)}} \in V_h(T\tilde{\mathcal{B}}_h,\partial_{D_{(i)}}\tilde{\mathcal{B}}_h,\bar{\mathbf{u}}_{(i)})$} be the solution of the discretized virtual problem for the same given $\boldsymbol{b}_{(i)}$, $\overline{\mathbf{u}}_{(i)}$, $\overline{\mathbf{t}}_{(i)}$ for $i = 1,2,\hdots,N$. Find $(\xi_h,\eta_h,\mathbf{u}_{h_{(1)}},\hdots,\mathbf{u}_{h_{(N)}},\boldsymbol{\gamma}_{h_{(1)}},\hdots,\mathbf{u}_{h_{(N)}}) \in V_h(\mathcal{C}_h) \times V_h(\mathcal{C}_h) \times V_h(T\mathcal{B}_h,\partial_{D_{(1)}}\mathcal{B}_h,\bar{\mathbf{u}}_{(1)}) \times \hdots \times V_h(T\mathcal{B}_h,\partial_{D_{(N)}}\mathcal{B}_h,\bar{\mathbf{u}}_{(N)}) \times V_h(T\mathcal{B}_h,\partial_{D_{(1)}}\mathcal{B}_h) \times \hdots \times V_h(T\mathcal{B}_h,\partial_{D_{(N)}}\mathcal{B}_h)$ such that, for $i=1,2,\hdots,N$,
\begin{equation}
\begin{alignedat}{3}
	\sum\nolimits_{j=1}^N w_j  \left \llangle  W_{1_h}(\xi_{h}, \mathbf{u}_{h_{(j)}},\boldsymbol{\gamma}_{h_{(j)}})  + m_1\xi_{h},
	\delta\xi_{h} \right\rrangle_{\mathcal{C}_h}
	+\left \llangle \alpha_1 \nabla \delta\xi_{h} , \nabla \delta\xi_{h} \right\rrangle_{\mathcal{C}_h}
	& = 0\,, & &\forall \delta\xi_{h}\in V_{h}(\mathcal{C}_h) \,,\\
	\sum\nolimits_{j=1}^N w_j  \left \llangle W_{2_h}(\eta_{h}, \mathbf{u}_{h_{(j)}},\boldsymbol{\gamma}_{h_{(j)}}) + m_2\eta_{h},
	\delta\eta_{h} \right\rrangle_{\mathcal{C}_h}
	+\left \llangle \alpha_2 \nabla \delta\eta_{h} , \nabla \delta\eta_{h} \right\rrangle_{\mathcal{C}_h}
	& = 0\,, & &\forall \delta\eta_{h}\in V_{h}(\mathcal{C}_h) \,,\\
	k  \llangle \mathbf{u}_{h_{(i)}}-\tilde{\mathbf{u}}_{(i)}, \delta\mathbf{u}_{h_{(i)}} 
	 \rrangle_{\mathring{\mathcal{B}}_h}
	- \llangle \boldsymbol{\sigma}_h(\xi_{h},\eta_{h},\boldsymbol{\gamma}_{h_{(i)}}), 
	\nabla\delta\mathbf{u}_{h_{(i)}}	\rrangle_{\mathcal{B}_h}& =0\,,
	& &\forall \delta\mathbf{u}_{h_{(i)}} \in V_{h}(T\mathcal{B}_h,\partial_{D_{(i)}}\mathcal{B}_h)\,,\\
	-\llangle \boldsymbol{\sigma}_h(\xi_{h},\eta_{h},\mathbf{u}_{h_{(i)}}), 
	\nabla\delta\boldsymbol{\gamma}_{h_{(i)}} \rrangle_{\mathcal{B}_h}
	+\llangle \mathbf{b}_{(i)}, \delta\boldsymbol{\gamma}_{h_{(i)}}\rrangle_{\mathcal{B}_h}
	+  \llangle \bar{\mathbf{t}}_{(i)}, \delta\boldsymbol{\gamma}_{h_{(i)}}\rrangle_{\partial_{N_{(i)}}\mathcal{B}_h}
	& =0\,,
	&~ &\forall \delta\boldsymbol{\gamma}_{h_{(i)}}\in V_{h}(T\mathcal{B}_h,\partial_{D_{(i)}}\mathcal{B}_h)\,.\\
\end{alignedat}
\end{equation}
\end{prob}

\subsection{Matrix formulation}

Next, we discuss a matrix formulation of the finite element discretization. For the sake of simplicity of presentation, we assume $2$D finite elements. However, the formulation can be used for $3$D problems as well. We define the column vector representation of a second-order tensor $\boldsymbol{T}$ by $\vek{\boldsymbol{T}} := \begin{bmatrix} T^{11} &T^{12} &T^{21} &T^{22} \end{bmatrix}^{\mathsf{T}}$.
Note that $\llangle\boldsymbol{Y},\boldsymbol{Z}\rrangle_{\mathcal{A}} = \llangle\vek{\boldsymbol{Y}},\vek{\boldsymbol{Z}}\rrangle_{\mathcal{A}} = \int_{\mathcal{A}}\vek{\boldsymbol{Y}}^{\mathsf{T}}\vek{\boldsymbol{Z}}\mathrm{d}v = \int_{\mathcal{A}}\vek{\boldsymbol{Z}}^{\mathsf{T}}\vek{\boldsymbol{Y}} \mathrm{d}v$.
Using the Lagrange basis functions of $\mathcal{P}_{1}(\mathcal{T})$ and $\mathcal{P}_{1}(T\mathcal{T})$, one can approximate the field variables of the weak formulation in an element $\mathcal{T}$ using the following matrix relations: 
\begin{equation}\label{interpolated}
  \begin{alignedat}{5}
 	\xi_{\mathcal{T}} &= \boldsymbol{\mathsf{b}}_{\mathcal{T}} \,\lm{q}{\xi},	&\quad
    \nabla\xi_{\mathcal{T}} &= \boldsymbol{\mathsf{G}}_{\mathcal{T}} \,\lm{q}{\xi}\,,\\
 	\eta_{\mathcal{T}} &= \boldsymbol{\mathsf{b}}_{\mathcal{T}} \,\lm{q}{\eta},	&\quad
    \nabla\eta_{\mathcal{T}} &= \boldsymbol{\mathsf{G}}_{\mathcal{T}} \,\lm{q}{\eta}\,,\\
    \mathbf{u}_{\mathcal{T}_{(i)}} &= \boldsymbol{\mathsf{B}}_{\mathcal{T}} \,\lm{q}{u_{(i)}},   &\quad
    \operatorname{div}\mathbf{u}_{\mathcal{T}_{(i)}} &=  \boldsymbol{\mathsf{d}}_{\mathcal{T}}  
    \,\lm{q}{u_{(i)}}\,,       &\quad
   \bigVek{\big}{\operatorname{sym}\nabla \mathbf{u}_{\mathcal{T}_{(i)}}} =  \boldsymbol{\mathsf{S}}_{\mathcal{T}} \,\lm{q}{u_{(i)}}\,,\\
    \boldsymbol{\gamma}_{\mathcal{T}_{(i)}} &= \boldsymbol{\mathsf{B}}_{\mathcal{T}} \,\lm{q}{\gamma_{(i)}},   &\quad
    \operatorname{div}\boldsymbol{\gamma}_{\mathcal{T}_{(i)}} &=  \boldsymbol{\mathsf{d}}_{\mathcal{T}}  \,\lm{q}{\gamma_{(i)}}\,,  &\quad
    \bigVek{\big}{\operatorname{sym}\nabla \boldsymbol{\gamma}_{\mathcal{T}_{(i)}}}  
    =  \boldsymbol{\mathsf{S}}_{\mathcal{T}} \,\lm{q}{\gamma_{(i)}}\,,
  \end{alignedat} 
\end{equation}
where the vectors ${\lm{q}{\xi}}_{3\times 1}$, ${\lm{q}{\eta}}_{3\times 1}$, ${\lm{q}{u_{(i)}}}_{6\times 1}$, and ${\lm{q}{\gamma_{(i)}}}_{6\times 1}$ contain the values of degrees of freedom, i.e., the values of $\xi$, $\eta$, $\mathbf{u}_{(i)}$, and $\boldsymbol{\gamma}_{(i)}$ at the three vertices of $\mathcal{T}$, respectively. The matrices ${\boldsymbol{\mathsf{b}}_{\mathcal{T}}}_{1\times 3}$ and ${\boldsymbol{\mathsf{B}}_{\mathcal{T}}}_{2\times 6}$ contain the Lagrange basis (shape) functions of $\mathcal{T}$ and ${\boldsymbol{\mathsf{G}}_{\mathcal{T}}}_{2\times 3}$, ${\boldsymbol{\mathsf{d}}_{\mathcal{T}}}_{1\times 6}$, and ${\boldsymbol{\mathsf{S}}_{\mathcal{T}}}_{4\times 6}$ consist of their spatial derivatives. One can obtain the variations of \eqref{interpolated} by replacing the vectors of degrees of freedom with the vectors of arbitrary real numbers of the same size, e.g., $\delta\xi_{\mathcal{T}} = \boldsymbol{\mathsf{b}}_{\mathcal{T}}\, \boldsymbol{\mathsf{a}}$, where ${\boldsymbol{\mathsf{a}}}_{3 \times 1}$ is a vector of arbitrary real numbers. Using the above discretized fields, we can approximate \eqref{w-eq} in ${\mathcal{T}}$ in terms of degrees of freedom: 
\begin{equation} 
    \begin{aligned}
	W_{1_{\mathcal{T}}}\left(\lm{q}{\xi}, \lm{q}{u_{(i)}},\lm{q}{\gamma_{(i)}}\right)  
	= {\lm{q}{\gamma_{(i)}}}^{\!\mathsf{T}} \,\boldsymbol{\mathsf{W}}_{1_{\mathcal{T}}}(\lm{q}{\xi}) 
	\,\lm{q}{u_{(i)}}\,, \\
	W_{2_{\mathcal{T}}}\left(\lm{q}{\eta},  \lm{q}{u_{(i)}},\lm{q}{\gamma_{(i)}}\right) 
	= {\lm{q}{\gamma_{(i)}}}^{\!\mathsf{T}} \,\boldsymbol{\mathsf{W}}_{2_{\mathcal{T}}}(\lm{q}{\eta}) 
	\,\lm{q}{u_{(i)}}\,,
\end{aligned}
\end{equation}
where the symmetric matrices ${\boldsymbol{\mathsf{W}}_{1_{\mathcal{T}}}}_{6\times6}$ and ${\boldsymbol{\mathsf{W}}_{2_{\mathcal{T}}}}_{6\times6}$ are defined as 
\begin{equation}
	\begin{alignedat}{3}
		\boldsymbol{\mathsf{W}}_{1_{\mathcal{T}}}(\lm{q}{\xi}) &= -\frac{2}{3}\mathring{\mu}\,
		e^{-(\boldsymbol{\mathsf{b}}_{\mathcal{T}} \,\lm{q}{\xi})} \,\lm{d}{\mathsf{T}} \,\lm{d}{}
	+2 \mathring{\mu}\,e^{-(\boldsymbol{\mathsf{b}}_{\mathcal{T}} \,\lm{q}{\xi})} \,
	\lm{S}{\mathsf{T}} \,\lm{S}{} \,, \\
		\boldsymbol{\mathsf{W}}_{2_{\mathcal{T}}}(\lm{q}{\eta}) &= \mathring{\kappa} 
		\,e^{-(\boldsymbol{\mathsf{b}}_{\mathcal{T}} \,\lm{q}{\eta})} \,\lm{d}{\mathsf{T}} \,\lm{d}{}\,.\\
	\end{alignedat} 
  \end{equation}
 Next, guided by \eqref{2nd-variations}, we define the following matrices in $\mathcal{T}$:
\begin{equation}
	\begin{alignedat}{-1}
	\lm{K}{\xi\xi} &= \int_{\mathcal{T}} \left(m_1\,\lm{b}{\mathsf{T}}\,\lm{b}{} 
	+ \alpha_1 \,\lm{G}{\mathsf{T}} \,\lm{G}{} \right) \mathrm{d}v\,, \\
	\lm{K}{\eta\eta} &= \int_{\mathcal{T}} \left(m_2 \,\lm{b}{\mathsf{T}} \,\lm{b}{} 
	+ \alpha_2 \, \lm{G}{\mathsf{T}} \,\lm{G}{} \right) \mathrm{d}v\,, \\
	\lm{\bar{K}}{\xi\xi}\left( \lm{q}{\xi},\lm{q}{u_{(i)}},\lm{q}{\gamma_{(i)}} \right)
	 &= -\int_{\mathcal{T}} 
		W_{1_{\mathcal{T}}}\left( \lm{q}{\xi},\lm{q}{u_{(i)}},\lm{q}{\gamma_{(i)}} \right) 
		\lm{b}{\mathsf{T}}\,\lm{b}{} \mathrm{d}v\,,  \\
	\lm{\bar{K}}{\eta\eta}\left( \lm{q}{\eta},\lm{q}{u_{(i)}},\lm{q}{\gamma_{(i)}} \right)
	  &= -\int_{\mathcal{T}}
		W_{2_{\mathcal{T}}}\left( \lm{q}{\eta},\lm{q}{u_{(i)}},\lm{q}{\gamma_{(i)}} \right)  
		\lm{b}{\mathsf{T}}\,\lm{b}{} \mathrm{d}v\,,  \\
	\lm{K}{\xi u}\left(\lm{q}{\xi},\lm{q}{\gamma_{(i)}}\right) &= \int_{\mathcal{T}}   
	\lm{b}{\mathsf{T}} \, {\lm{q}{\gamma_{(i)}}}^{\!\mathsf{T}} 
	\,\boldsymbol{\mathsf{W}}_{1_{\mathcal{T}}}(\lm{q}{\xi})  \,\mathrm{d}v \,,\quad
	\lm{K}{\xi\gamma}\left(\lm{q}{\xi},\lm{q}{u_{(i)}}\right)  = \int_{\mathcal{T}}   
	\lm{b}{\mathsf{T}}  \, {\lm{q}{u_{(i)}}}^{\!\mathsf{T}} 
	\, \boldsymbol{\mathsf{W}}_{1_{\mathcal{T}}}(\lm{q}{\xi})  \,\mathrm{d}v \,,\\
    \lm{K}{\eta u}\left(\lm{q}{\eta},\lm{q}{\gamma_{(i)}}\right) &= \int_{\mathcal{T}}   
	\lm{b}{\mathsf{T}}  \, {\lm{q}{\gamma_{(i)}}}^{\!\mathsf{T}} 
	\, \boldsymbol{\mathsf{W}}_{2_{\mathcal{T}}}(\lm{q}{\eta})  \,\mathrm{d}v \,,\quad
	\lm{K}{\eta \gamma}\left(\lm{q}{\eta},\lm{q}{u_{(i)}}\right) = \int_{\mathcal{T}}  
	\lm{b}{\mathsf{T}}  \, {\lm{q}{u_{(i)}}}^{\!\mathsf{T}} 
	\, \boldsymbol{\mathsf{W}}_{2_{\mathcal{T}}}(\lm{q}{\eta})  \,\mathrm{d}v \,,\\
	\lm{K}{u \gamma}\left(\lm{q}{\xi},\lm{q}{\eta}\right) &= - \int_{\mathcal{T}} 
	\left( \boldsymbol{\mathsf{W}}_{1_{\mathcal{T}}}(\lm{q}{\xi}) 
	+ \boldsymbol{\mathsf{W}}_{2_{\mathcal{T}}}(\lm{q}{\eta}) \right)  \mathrm{d}v\,,\\
	\lm{K}{uu} &=\left\{ \begin{array}{l l} 
		k\, \int_{\mathcal{T}}  \,\lm{B}{\mathsf{T}} \,\lm{B}{} \,\mathrm{d}v\,,
		& \text{if } \mathcal{T} \in  \mathring{\mathcal{B}}_h\,, \\
		\bz_{6\times6}, & \text{otherwise}\,,
		\end{array} \right.\\
	\lm{F}{u}\left(\tilde{\mathbf{u}}_{h_{(i)}}\right) &= \left\{ \begin{array}{l l}
		 k\int_{\mathcal{T}} \lm{B}{\mathsf{T}} \,\tilde{\mathbf{u}}_{h_{(i)}} \, \mathrm{d}v, 
		 & \text{if } \mathcal{T} \in  \mathring{\mathcal{B}}_h\,, \\
		\bz_{6\times1}, & \text{otherwise}\,,
		\end{array} \right.\\
	\lm{F}{\gamma}\left(\mathbf{b}_{(i)},\overline{\mathbf{t}}_{(i)}\right) &= -\int_{\mathcal{T}} \lm{B}{\mathsf{T}} 
	\,\mathbf{b}_{(i)} \,\mathrm{d}v 
	- \left\{ \begin{array}{l l}
		\int_{\mathcal{E}} \boldsymbol{\mathsf{B}}^{\mathsf{T}}_{\mathcal{E}} 
		\,\overline{\mathbf{t}}_{(i)} \,\mathrm{d}a, 
		& \text{if } \mathcal{E} \in  \partial_{{N}_{(i)}}\mathcal{B}_h\,, \\
	   \bz_{6\times1}\,, & \text{otherwise}\,.
	   \end{array} \right.\\
	\end{alignedat} 
\end{equation}
We assemble the vectors of degrees of freedom and the above matrices as
\begin{equation}
\begin{aligned}
	\{\gm{Q}{\xi}, \gm{Q}{\eta} \} &=
	\widetilde{\boldsymbol{\mathsf{A}}}_{\mathcal{T}\in\mathcal{C}_{h}}
	\{ \lm{q}{\xi}, \lm{q}{\eta} \}\,,\\
	\{\gm{Q}{u_{(i)}}, \gm{Q}{\gamma_{(i)}} \} &=
	\widetilde{\boldsymbol{\mathsf{A}}}_{\mathcal{T}\in\mathcal{B}_{h}}
	\{ \lm{q}{u_{(i)}}, \lm{q}{\gamma_{(i)}} \}\,,\\ 
	\{ \gm{K}{\xi\xi}, \gm{K}{\eta \eta}, \gm{\bar{K}}{\xi\xi}, \gm{\bar{K}}{\eta \eta} \} &=
	\boldsymbol{\mathsf{A}}_{\mathcal{T}\in\mathcal{C}_{h}}
	\{ \lm{K}{\xi\xi},\lm{K}{\eta \eta}, \lm{\bar{K}}{\xi\xi},\lm{\bar{K}}{\eta \eta} \}\,,\\ 
	\{ \gm{K}{uu,(i)}, \gm{K}{u \gamma}, \gm{F}{u,(i)},\gm{F}{\gamma,(i)} \} &=
	\boldsymbol{\mathsf{A}}_{\mathcal{T}\in\mathcal{B}_{h}}
	\{ \gm{K}{uu,(i)}, \lm{K}{u \gamma}, \lm{F}{u,(i)},\lm{F}{\gamma,(i)} \}\,,\\
	\{ \gm{K}{\xi u},\gm{K}{\xi \gamma},\gm{K}{\eta u},\gm{K}{\eta \gamma}  \} &=
	\overline{\boldsymbol{\mathsf{A}}}_{\mathcal{T}\in{\mathcal{C}_{h},\mathcal{B}_{h}}}
	\{ \lm{K}{\xi u},\lm{K}{\xi \gamma},\lm{K}{\eta u},\lm{K}{\eta \gamma} \} \,.
\end{aligned} 
\end{equation}
We then define the following matrices by considering $N$ load cases:
\begin{equation}
	\begin{alignedat}{3}
		&\mathbb{Q}^{u}_h =
		\begin{bmatrix}
			\gm{Q}{u_{(1)}}\\
			\vdots\\
			\gm{Q}{u_{(N)}}
		\end{bmatrix},\quad
		\mathbb{Q}^{\gamma}_h =
		\begin{bmatrix}
			\gm{Q}{\gamma_{(1)}}\\
			\vdots\\
			\gm{Q}{\gamma_{(N)}}
		\end{bmatrix},\quad
		\mathbb{F}^{u}_h =
		\begin{bmatrix}
			w_1 \gm{F}{u}\left(\tilde{\mathbf{u}}_{h_{(1)}}\right)\\
			\vdots\\
			w_N \gm{F}{u}\left(\tilde{\mathbf{u}}_{h_{(N)}}\right)
		\end{bmatrix},\quad
		\mathbb{F}^{\gamma}_h =
		\begin{bmatrix}
			w_1 \gm{F}{\gamma}\left(\mathbf{b}_{(1)},\mathbf{t}_{(1)}\right)\\
			\vdots\\
			w_N \gm{F}{\gamma}\left(\mathbf{b}_{(N)},\mathbf{t}_{(N)}\right)
		\end{bmatrix},\\
		&\mathbb{K}^{uu}_{h} =
		\begin{bmatrix}
			\begin{array}{ccc}
			w_1\gm{K}{uu} & & \\
			 &\ddots & \\
			& & w_N \gm{K}{uu}
			\end{array}
		\end{bmatrix},\quad
		\mathbb{K}^{\gamma u}_{h} = \mathbb{K}^{u \gamma}_{h} = 
		\begin{bmatrix}
			\begin{array}{ccc}
			w_1\gm{K}{u \gamma}(\gm{Q}{\xi},\gm{Q}{\eta}) & & \\
			 &\ddots & \\
			& & w_N \gm{K}{u \gamma}(\gm{Q}{\xi},\gm{Q}{\eta})
			\end{array}
		\end{bmatrix},\\
		& \mathbb{K}^{\xi u}_{h}(\gm{Q}{\xi},\mathbb{Q}^{\gamma}_h) =
		\begin{bmatrix}
			   w_1 \gm{K}{\xi u}(\gm{Q}{\xi},\gm{Q}{\gamma_{(1)}})
			 & \cdots
			 & w_N \gm{K}{\xi u}(\gm{Q}{\xi},\gm{Q}{\gamma_{(N)}})
		\end{bmatrix}, \\
		& \mathbb{K}^{\xi \gamma}_{h}(\gm{Q}{\xi},\mathbb{Q}^{u}_h) =
		\begin{bmatrix}
			    w_1 \gm{K}{\xi u}(\gm{Q}{\xi},\gm{Q}{u_{(1)}})
			 & \cdots
			 & w_N  \gm{K}{\xi u}(\gm{Q}{\xi},\gm{Q}{u_{(N)}})
		\end{bmatrix},\\
		& \mathbb{K}^{\eta u}_{h}(\gm{Q}{\eta},\mathbb{Q}^{\gamma}_h) =
		\begin{bmatrix}
				w_1\gm{K}{\eta u}(\gm{Q}{\eta},\gm{Q}{\gamma_{(1)}})
			 & \cdots
			 & w_N  \gm{K}{\eta u}(\gm{Q}{\eta},\gm{Q}{\gamma_{(N)}})
		\end{bmatrix}, \\
		& \mathbb{K}^{\eta \gamma}_{h}(\gm{Q}{\eta},\mathbb{Q}^{u}_h) =
		\begin{bmatrix}
				w_1\gm{K}{\xi u}(\gm{Q}{\eta},\gm{Q}{u_{(1)}})
			 & \cdots
			 & w_N  \gm{K}{\xi u}(\gm{Q}{\eta},\gm{Q}{u_{(N)}})
		\end{bmatrix},\\
		& \gm{J}{\xi\xi} (\gm{Q}{\xi},\mathbb{Q}^{u}_h,\mathbb{Q}^{\gamma}_h) = 
		 \sum_{j=1}^N w_j \gm{\bar{K}}{\xi\xi}(\gm{Q}{\xi},\gm{Q}{u_{(j)}},\gm{Q}{\gamma_{(j)}}),\quad
		\gm{J}{\eta\eta} (\gm{Q}{\eta},\mathbb{Q}^{u}_h,\mathbb{Q}^{\gamma}_h) =
		\sum_{j=1}^N w_j \gm{\bar{K}}{\eta\eta}(\gm{Q}{\eta},\gm{Q}{u_{(j)}},\gm{Q}{\gamma_{(j)}}).
	\end{alignedat}
\end{equation}
Note that $\mathbb{K}^{\xi u}_{h}(\gm{Q}{\xi},\mathbb{Q}^{\gamma}_h)\,\mathbb{Q}^{u}_h = \mathbb{K}^{\xi \gamma}_{h}(\gm{Q}{\xi},\mathbb{Q}^{u}_h)\,\mathbb{Q}^{\gamma}_h$, and $\mathbb{K}^{\eta u}_{h}(\gm{Q}{\eta},\mathbb{Q}^{\gamma}_h)\,\mathbb{Q}^{u}_h = \mathbb{K}^{\eta \gamma}_{h}(\gm{Q}{\eta},\mathbb{Q}^{u}_h)\,\mathbb{Q}^{\gamma}_h$. We write the matrix formulation of problem \ref{fem-problem} as:
\begin{prob}[Matrix equation]\label{matrix-problem}
	Find $\mathbb{Q}_h$ such that $\mathbb{K}_h(\mathbb{Q}_h)\mathbb{Q}_h = \mathbb{F}_h$, where
	\begin{equation}
		\mathbb{Q}_{h} =
		\begin{bmatrix}
			\gm{Q}{\xi}  \\
			\gm{Q}{\eta} \\
			\mathbb{Q}^{u}_h                 \\
			\mathbb{Q}^{\gamma}_h
		\end{bmatrix}, \quad
\mathbb{K}_{h}(\mathbb{Q}_h) =
\begin{bmatrix}
	\begin{array}{llll}
		\gm{K}{\xi\xi} & \bz              & \mathbb{K}_{h}^{\xi  u}(\gm{Q}{\xi},\mathbb{Q}^{\gamma}_h)  & \bz                                                    \\
		\bz            & \gm{K}{\eta\eta} & \mathbb{K}_{h}^{\eta u}(\gm{Q}{\eta},\mathbb{Q}^{\gamma}_h) & \bz                                                    \\
		\bz            & \bz              & \mathbb{K}_{h}^{uu}                                         & \mathbb{K}_{h}^{u    \gamma}(\gm{Q}{\xi},\gm{Q}{\eta}) \\
		\bz            & \bz              & \mathbb{K}_{h}^{\gamma u}(\gm{Q}{\xi},\gm{Q}{\eta})         & \bz
	\end{array}
\end{bmatrix}\,, ~ \text{and} \quad
		\mathbb{F}_{h} =
		\begin{bmatrix}
			\boldsymbol{0}   \\
			\boldsymbol{0}   \\
			\mathbb{F}^{u}_h \\
			\mathbb{F}^{\gamma}_h
		\end{bmatrix}\,.
	\end{equation}
\end{prob}
One can show that the symmetric Jacobian matrix 
\begin{equation}
	{[\mathbb{J}_{h}]}_{ij} = \frac{\partial\left[\mathbb{K}_h(\mathbb{Q}_h)\mathbb{Q}_h 
	- \mathbb{F}_h \right]_{i}}{{\partial[\mathbb{Q}_h]}_{j}}\,,
\end{equation}
for the nonlinear problem \ref{matrix-problem} reads
\begin{equation}\label{Jacobian-matrix}
	\mathbb{J}_{h}(\mathbb{Q}_h) =
	\begin{bmatrix}
		\begin{array}{llll}
			\gm{K}{\xi\xi} + \gm{J}{\xi\xi} ( \gm{Q}{\xi},\mathbb{Q}^{u}_h,\mathbb{Q}^{\gamma}_h ) & \bz                                                                                                 & \mathbb{K}_{h}^{\xi u}(\gm{Q}{\xi},\mathbb{Q}^{\gamma}_h)   & \mathbb{K}_{h}^{\xi\gamma}(\gm{Q}{\xi},\mathbb{Q}^{u}_h)    \\
			\bz                                                                                               & \gm{K}{\eta\eta}+\gm{J}{\eta\eta} (\gm{Q}{\eta},\mathbb{Q}^{u}_h,\mathbb{Q}^{\gamma}_h) & \mathbb{K}_{h}^{\eta u}(\gm{Q}{\eta},\mathbb{Q}^{\gamma}_h) & \mathbb{K}_{h}^{\eta \gamma}(\gm{Q}{\eta},\mathbb{Q}^{u}_h) \\
			\mathbb{K}_{h}^{u \xi}(\gm{Q}{\xi},\mathbb{Q}^{\gamma}_h)                                         & \mathbb{K}_{h}^{u \eta}(\gm{Q}{\eta},\mathbb{Q}^{\gamma}_h)                                         & \mathbb{K}_{h}^{uu}                                         & \mathbb{K}_{h}^{u \gamma}(\gm{Q}{\xi},\gm{Q}{\eta})         \\
			\mathbb{K}_{h}^{\gamma\xi}(\gm{Q}{\xi},\mathbb{Q}^{u}_h)                                          & \mathbb{K}_{h}^{\gamma\eta}(\gm{Q}{\eta},\mathbb{Q}^{u}_h)                                          & \mathbb{K}_{h}^{\gamma u}(\gm{Q}{\xi},\gm{Q}{\eta})         & \bz
		\end{array}
	\end{bmatrix} \,,
\end{equation}
where $\mathbb{K}_{h}^{u \xi}=(\mathbb{K}_{h}^{\xi u})^{\mathsf{T}}$, $\mathbb{K}_{h}^{u \eta}=(\mathbb{K}_{h}^{\eta u})^{\mathsf{T}}$, $\mathbb{K}_{h}^{\gamma \xi}=(\mathbb{K}_{h}^{\xi \gamma})^{\mathsf{T}}$, and $\mathbb{K}_{h}^{\gamma \eta}=(\mathbb{K}_{h}^{\eta \gamma})^{\mathsf{T}}$.

\begin{figure}[bp]
\centering
\includegraphics[width = .95\textwidth]{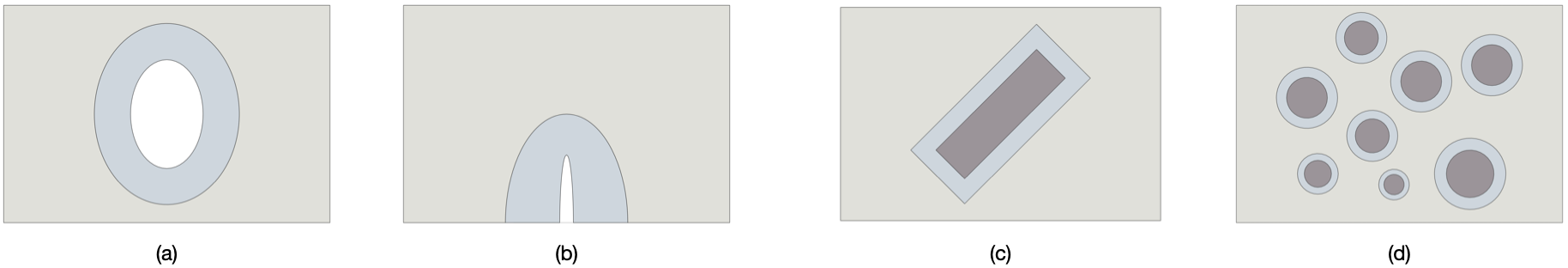}
\vspace{.0in}
\caption{
Cloaking of an elliptic hole (a), an elliptic cut (b), a rectangular inhomogeneity (c), and randomly distributed circular inhomogeneities (d). The blue color shows the cloak $\mathcal{C}$.}
\label{fig:configurations}
\end{figure}

\section{Numerical Results} \label{Sec:Examples}

\begin{figure}[bp]
\centering
\includegraphics[width=.6\textwidth]{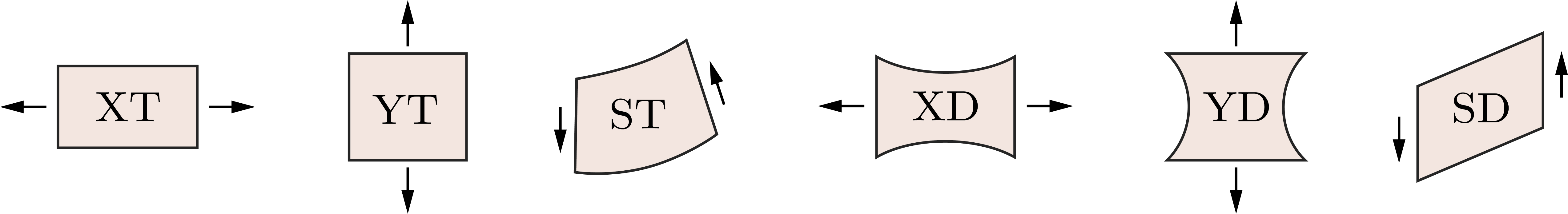}
\vspace{0.1in}
\caption{Symbols for the load cases.}
\label{fig:loads}
\end{figure}

In this section we discuss a few numerical examples of the optimization problem for isotropic cloaks in dimension two.

\subsection{Implementation}

We have implemented a mixed finite element code in MATLAB to construct and solve the nonlinear system of algebraic equations of Problem~\ref{matrix-problem}. 
Given an explicit symmetric Jacobian matrix $\mathbb{J}_h$ \eqref{Jacobian-matrix}, one can efficiently solve Problem~\ref{matrix-problem} using the Newton method. We use MATLAB direct solver for the linear system in each Newton iteration. 
To solve the problem, first some numerical values for the control parameters $m_1$, $m_2$, $\alpha_1$, and $\alpha_2$ are chosen.
Recall that larger values of $m_1$ and $m_2$ penalize large deviations of the elastic moduli from their corresponding elastic moduli in the virtual body, whereas larger values of $\alpha_1$ and $\alpha_2$ are more preventive of sharpe gradients in the material parameters of the cloak.

We use $k$ as a load control parameter, which agrees with what was observed in Eq.~\eqref{strong-form-isotropic}, i.e., the term $k\| \mathbf{u} - \tilde{\mathbf{u}}\|$ represents a body force in the adjoint balance of linear momentum associated with the field $\boldsymbol{\gamma}$ (see also Eq.~\eqref{strong-form-isotropic-multi} for the multiple-load case).
In the load control process, we start with a very small $k$ and the initial guess $\mathbb{Q}_{h} = [\boldsymbol{0},\boldsymbol{0},\mathbb{U}_{h}^{\mathsf{T}},\boldsymbol{0}]^{\mathsf{T}}$ for the Newton method, where $\mathbb{U}_{h}$ is the displacement degrees of freedom linearly solved for a domain without a cloak, which also minimizes the augmented objective function for $k=0$.
Once the Newton method has converged, we gradually increase the value of $k$ and use the solution as the initial guess for the next load-control step. We keep repeating this process until the solution for the desired value of $k$ is obtained. 

The value of $k$ must be carefully chosen based on each load case. For multiple-load optimization, better results can be achieved if one works with $N$ different $k_i$'s to take into account the differences between load cases.
As was mentioned earlier, in the general multiple-load optimal design, each load case participates in the $\boldsymbol{\mathsf{C}}$-equations~\eqref{weak-form}$_{1,2}$ according to the work done by the standard forces on the adjoint displacements, defined in~\eqref{w-eq}.
In particular, the terms $W_1(\xi, \mathbf{u}_{(i)},\boldsymbol{\gamma}_{(i)}) $ and $W_2(\eta, \mathbf{u}_{(i)},\boldsymbol{\gamma}_{(i)}) $ represent the work done on the adjoint shear deformations and the adjoint changes of volume, respectively, for each load case $i = 1,\hdots,N$.
These terms are linear in the $\mathbf{u}_{(i)}$'s and $\boldsymbol{\gamma}_{(i)}$'s, and hence, are
proportional to the external loads (enforced via either tractions or displacement boundary conditions), as well as to the adjoint distributed loads $k ( \mathbf{u}_{(i)} - \tilde{\mathbf{u}}_{(i)} )$.
A reasonable choice would be to normalize $k$ with respect to $\|\tilde{\mathbf{u}}_{(i)}\|^2$ to obtain $N$ different $k_i$'s \citep{Fachinotti2018}.
Another fairly reasonable choice for the normalization of $k$ is to divide it by $\|\mathbf{u}_{(i)} - \tilde{\mathbf{u}}_{(i)}\|^2$ in each step.
Alternatively, one can use combinations of these quantities.
As the normalization is not unique, we choose suitable normalizers for each load case depending on the example we are solving.

\subsection{Examples of optimal elastic cloaks}

We consider a rectangular sheet made of an isotropic linear elastic solid, homogeneous in the virtual setting, and with several distributions of holes or inhomogeneities in the physical problem, see Fig.~\ref{fig:configurations}.
The elastic sheets undergo either extension in the $x$ or $y$ directions or shear in the $xy$ plane.
These deformations can be enforced either via traction or displacement boundary conditions.
No body forces are considered.
In total, we consider six load cases that will be denoted as XT, YT, ST, XD, YD, SD, where the first letter indicates the load direction (X and Y for the extension in the $x$ and $y$ directions, S for shear), and the second letter the type of boundary conditions (T and D for traction or displacement controlled, respectively), see Fig.~\ref{fig:loads}.
External loads will be considered in two different settings: i) the design loads are the ones used in the Optimization Problem \ref{prob-weak-form}, i.e., the loading combination one is optimizing for, and ii) the service loads are the ones used to test each design, and test the efficacy of the cloak in terms of how much it differs from the virtual solution.
Additionally, we will be considering combinations of the traction-controlled load cases XT, YT, ST---denoted as MT---and of the displacement-controlled load cases XD, YD, SD---denoted as MD. It should be noted that the multiple-load combinations MT and MD are only considered in the context of design loads, while the only service loads used to test the designs are the six cases XT, YT, ST, XD, YD, SD.

The performance of a design can be measured using the $L^2$ distance between $\mathbf{u}$ and $\tilde{\mathbf{u}}$ calculated on $\mathring{\mathcal{B}}$, cf.~\eqref{objective-function}.
We introduce the normalized distance between the virtual and the physical displacements outside the cloak as
\begin{equation} \label{performance}
	\hat{\mathsf{g}} = \frac{ \Vert \mathbf{u}-\tilde{\mathbf{u}} \Vert_{L^2} }{ \Vert \tilde{\mathbf{u}} \Vert_{L^2} }
	= \left(
	\frac{ \int_{\mathring{\mathcal{B}} } \Vert \mathbf{u} - \tilde{\mathbf{u}} \Vert^2 \mathrm dv }
	{ \int_{ \mathring{\mathcal{B}} } \Vert\tilde{\mathbf{u}}\Vert^2  \mathrm dv }
	\right)^{\frac{1}{2}}  \,.
\end{equation}
As for multiple-load deigns, we are mainly interested in their performance under a single service load, which can still be measured via Eq.~\eqref{performance}. However, since for the convergence plots a performance measure that takes into account a service load made of a combination of different loads is needed, we define:
\begin{equation} \label{performance-multi}
	\hat{\mathsf{g}}^{\text{M}} = \sum_{j} w_{j}
	\frac{ \Vert \mathbf{u}-\tilde{\mathbf{u}}_{(j)} \Vert_{L^2} }{ \Vert \tilde{\mathbf{u}}_{(j)}  \Vert_{L^2}  \,} .
\end{equation}

All the quantities are dimensionless, and are expressed with respect to a characteristic length $L_o$, and a characteristic stiffness that we take equal to the shear modulus $\mathring{\mu}$ of the homogeneous medium, i.e., of the region $\mathring{\mathcal{B}}$ outside the cloak in the physical problem.
We consider a bulk modulus $\mathring{\kappa} = 2\mathring{\mu}$, and hence obtain a Young's modulus $\mathring{E} = \frac{18}{7}\mathring{\mu} \approx 2.57 \mathring{\mu}$ and a Poisson's ratio $\mathring{\nu} = \frac{2}{7}\approx 0.29$.

Lastly, it should be noticed that in the present formulation the size and the shape of the cloak are not determined as a result of the optimization algorithm, but they are simply given as inputs of the problem.
Moreover, in the following examples the areas of the cloaks are taken to be comparable to that of the heterogeneities. To justify this choice, let us denote with $|\mathcal{C}|$ and $|\mathcal{H}|$ the areas of the sets $\mathcal{C}$ and $\mathcal{H}$, respectively. If $|\mathcal{C}|/|\mathcal{H}| \ll 1$, then the cloak will be either highly inhomogeneous or ineffective. On the other hand, if $|\mathcal{C}|/|\mathcal{H}| \gg 1$ the cloak will occupy a large part of the body and that would not be a desirable design either.

\begin{figure}[tp]
    \centering
    \includegraphics[width=\textwidth]{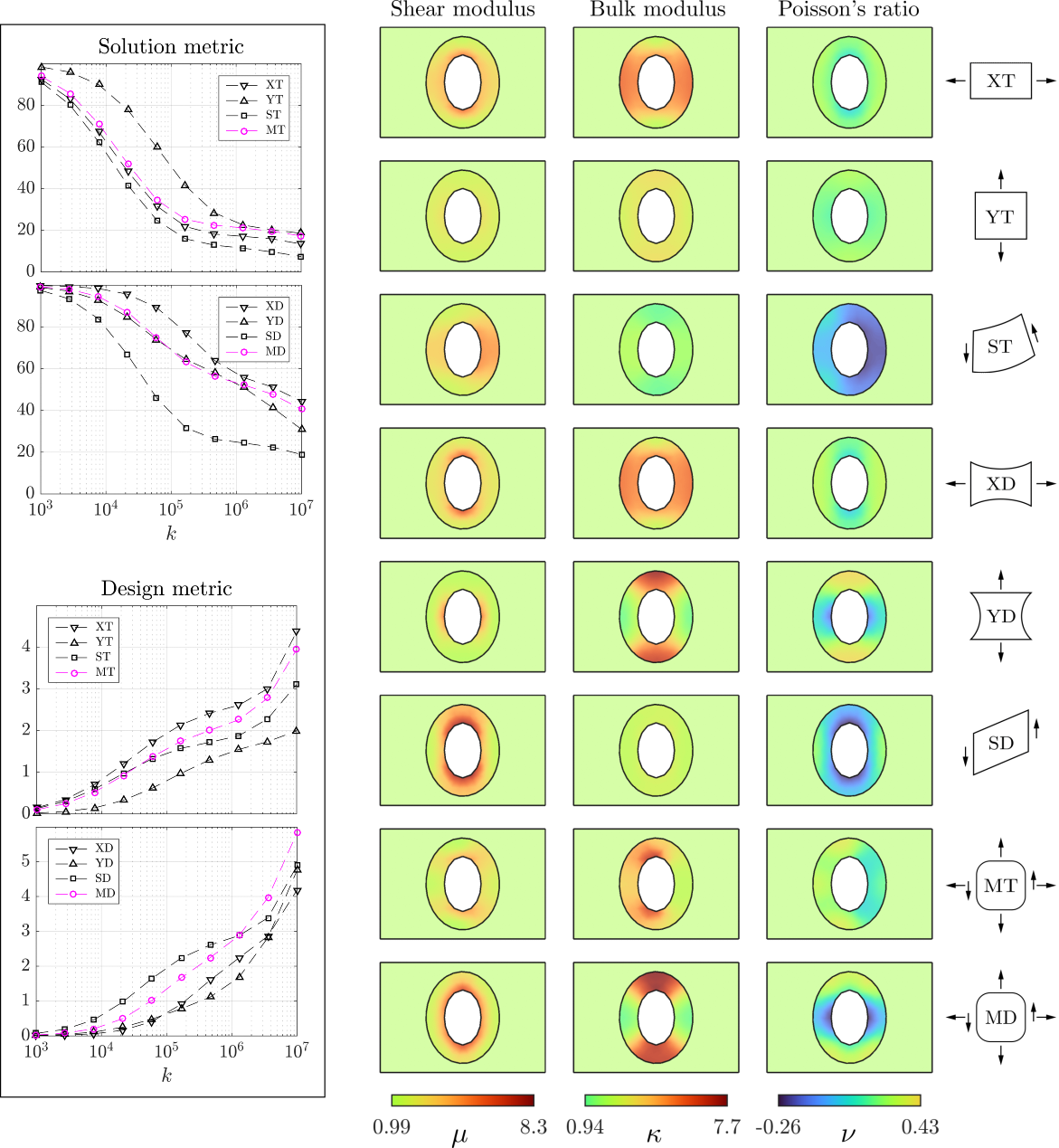}
    \vspace{-0.1in}
    \caption{Design of the optimal cloak for a single elliptic hole.
    Left: Convergence in terms of the solution metric~\eqref{performance},~\eqref{performance-multi}, and of the design $H^1$ metric~\eqref{metric-H2}. The solution metric is normalized with respect to the no-cloak case and is expressed in percentages.
    Right: Distribution of the elastic moduli in the solid for each design load.}
    \label{fig:hole-design}
\end{figure}
\begin{figure}[tp]
\centering
\includegraphics[width=\textwidth]{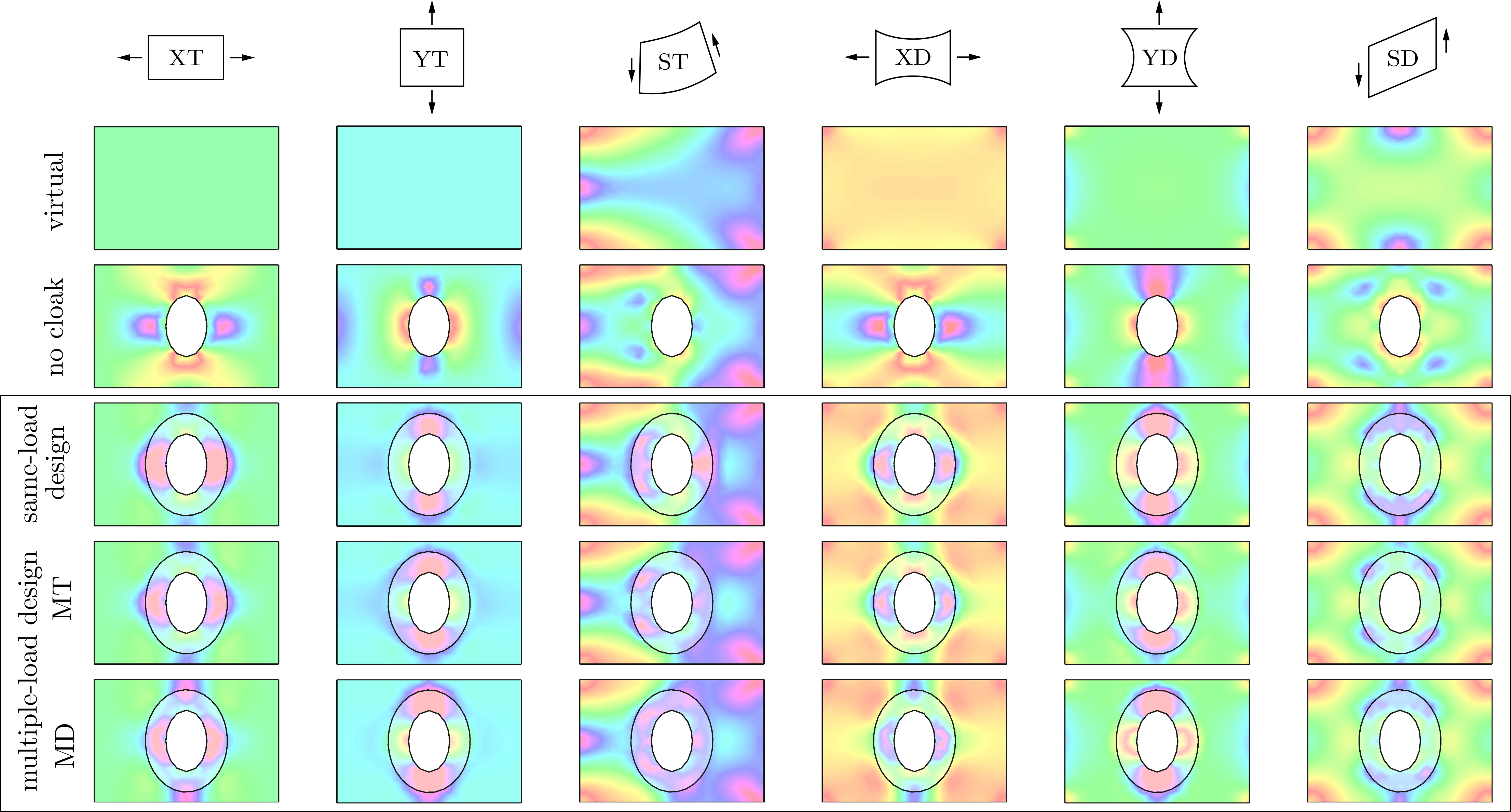}
\vspace{-0.15in}
\caption{Performance of the optimal cloaks under different service loads they are optimized for, in the symmetric case of a single elliptic hole. The performance is given in terms of the Frobenius norm of the stress. Each column represents a loading condition: XT, YT, ST, XD, YD, SD. Each row refers to virtual body, body with no cloak, same-load design, multiple-load design MT, and multiple-load design MD.}
\label{fig:hole-stress}
\end{figure}

\subsubsection{Example 1: Design of a symmetric elastic cloak for a hole} \label{Sec:symmetric-hole}

We consider an elliptic hole at the center of a rectangle of sides $a=6L_o$ and $b=4L_o$, with the cloak consisting of an elliptic annulus, see Fig.~\ref{fig:configurations}(a).
The semi-axes of the hole are $\frac{2}{3}L_o$ and $L_o$, while the semi-axes of the outer rim of the cloak are $\frac{4}{3} L_o$ and $\frac{5}{3} L_o$.
Both combinations MT and MD have weights $w_{\mathrm X}=w_{\mathrm Y}=w_{\mathrm S}=\frac{1}{3}$.
The coefficients in the augmented objective functions are $k = 10^7$, $m_1 = m_2 = \alpha_1 = \alpha_2 = 1$.
The designs are shown in Fig.~\ref{fig:hole-design}, while their efficacy is reported in Table~\ref{table:hole} with respect to each service load.
Fig.~\ref{fig:hole-stress} shows the stress distribution for some of the combinations in Table~\ref{table:hole}.

\begin{table}[h!]
\centering
{\footnotesize
\begin{tabular}{c | c c c c c c | c}  
\toprule
  & XT & YT & ST & XD & YD & SD & average\\
\midrule  
NC & 62.9 & 45.8 & 22.1 & 17.7 &  4.9 & 7.4 & 26.8 \\
 \midrule
XT &  \underline{8.4} & 17.6 &  6.8 &  7.8 &  7.1 & 1.7 &  8.2 \\
YT & 21.0 &  \underline{8.5} &  3.1 & 11.2 &  4.2 & 3.4 &  8.6 \\
ST & 24.3 & 18.8 &  \underline{1.6} & 13.3 & 14.9 & 2.7 & 12.6 \\
XD &  8.9 & 15.6 &  7.0 & \underline{7.8} &  6.4 & 1.7 &  7.9 \\
YD & 24.7 &  9.1 &  5.6 & 11.8 &  \underline{1.5} & 3.9 &  9.4 \\
SD & 11.2 & 18.6 &  8.1 &  9.4 &  6.7 & \underline{1.4} &  9.3 \\
 \midrule
XT & 10.1 &  9.6 &  2.9 &  8.8 &  4.3 & 2.5 &  6.4 \\
MD & 10.3 &  9.3 &  7.4 &  8.7 &  2.1 & 1.4 &  6.5 \\
 \bottomrule
\end{tabular}
}
\caption{Efficacy of an optimal cloak in the symmetric case of a single elliptic hole.
Each row corresponds to one of the six optimization loads (XT, YT, ST, XD, YD, SD) plus the no-cloak case (NC) at the top.
Each column corresponds to a service load, with the last column showing the average.
Excluding the first row and the last column, the main diagonal of the table corresponds to cases in which the optimization and service loads are the same.}
\label{table:hole}
\end{table}

\subsubsection{Example 2: Design of an elastic cloak for an elliptic cut}
\begin{figure}[tp]
    \includegraphics[width=\textwidth]{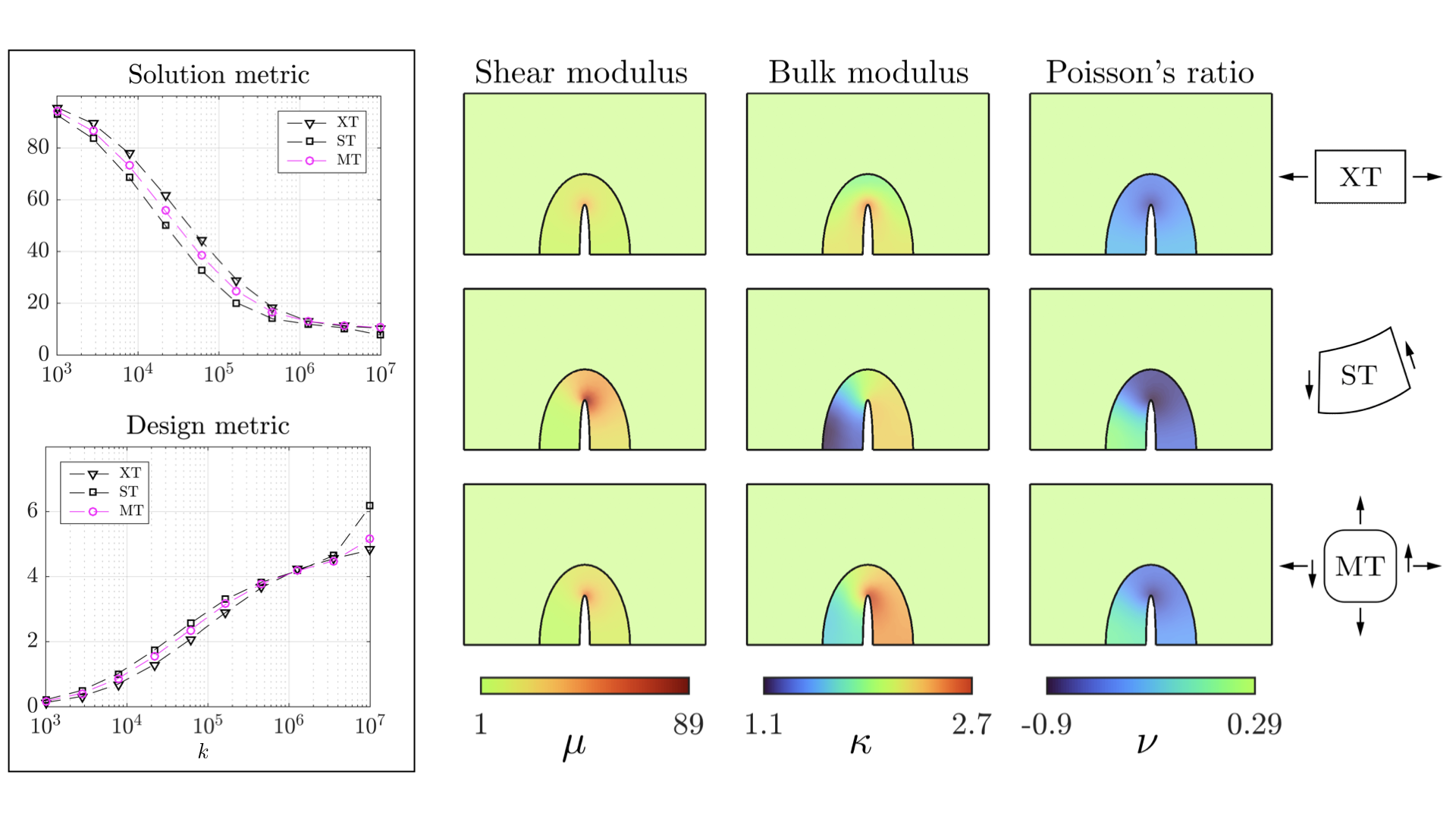}
    \vspace{-0.1in}
    \caption{Design of a carpet cloak surrounding an elliptic cut.
    Left: Convergence in terms of the solution metric~\eqref{performance},~\eqref{performance-multi}, and of the design $H^1$ metric~\eqref{metric-H2}. The solution metric is normalized with respect to the no-cloak case and is expressed in percentages.
    Right: Distribution of the elastic moduli in the body for each design load.
    Each row represents a design, under different optimization loads (XT, ST, MT).}
    \label{fig:carpet-design}
\end{figure}
\begin{figure}[tp]
\centering
\includegraphics[width=.4\textwidth]{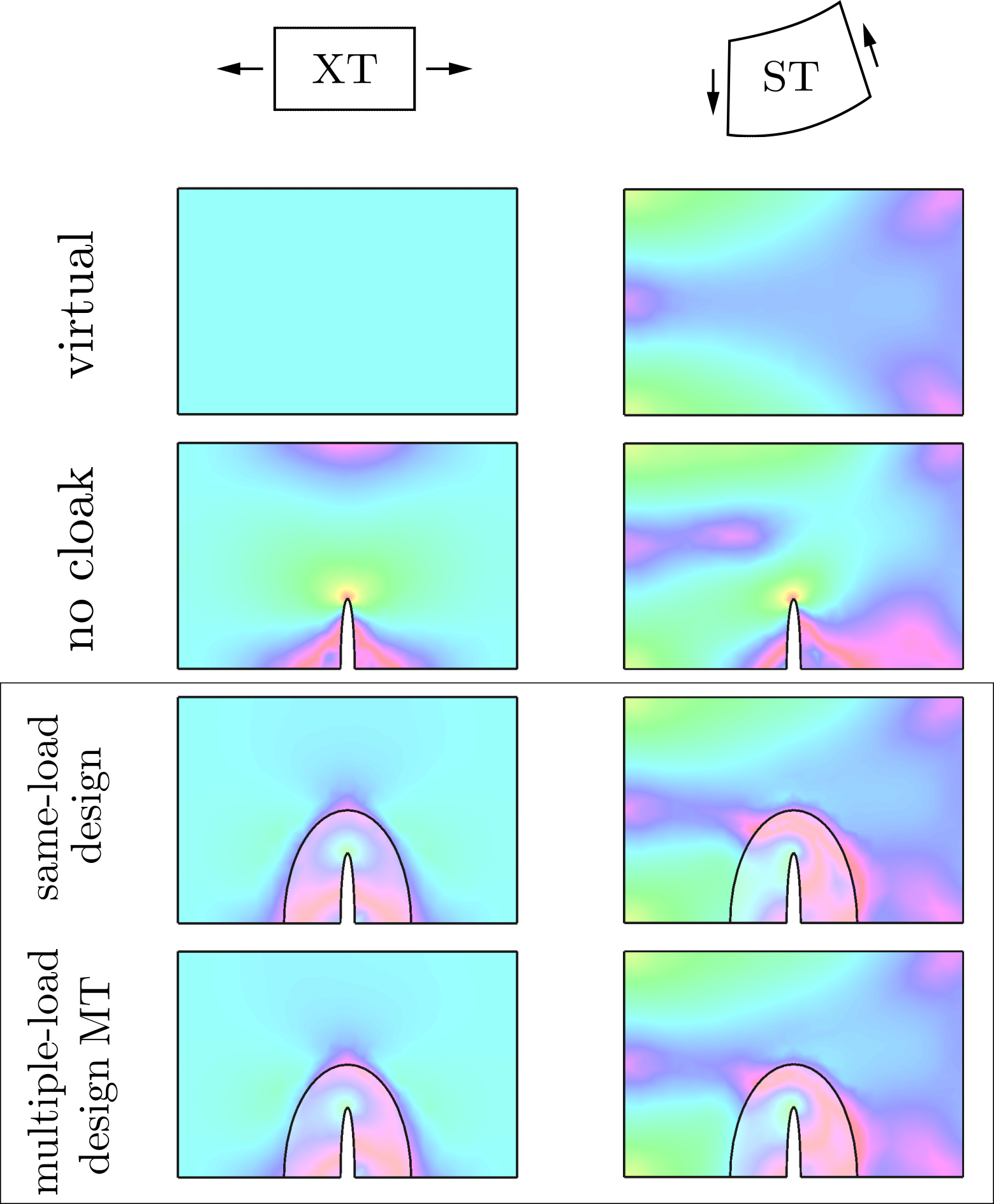}
\vspace{0.1in}
\caption{Performance of the optimal cloaks under different service loads they are optimized for, in the case of a carpet cloak surrounding an elliptic cut.
The performance is given in terms of the Frobenius norm of the stress.
Each column represents a loading condition: XT, ST.
Each row refers to virtual body, body with no cloak, same-load design, and multiple-load design MT.}
\label{fig:carpet-stress}
\end{figure}

We consider cloaking of a cut with the shape of a sharp semi-elliptic hole, see Fig.~\ref{fig:configurations}(b).
In this case the object to hide is not fully embedded in the considered medium, in the sense that it shares part of its boundary.
In particular, the elliptic cut has major semi-axis $\frac{3}{2}L_o$ and is centered on the middle point of the bottom boundary of the same rectangular medium considered in the previous examples.
The cut is surrounded by a half-elliptic cloak with major semi-axis of $2 L_o$.
We only consider the combination MT, with weights $w_{\mathrm X}=w_{\mathrm S}=\frac{1}{2}$, and $w_{\mathrm Y} = 0$.
The coefficients in the augmented objective functions are $k = 10^7$, $m_1 = m_2 = 2$,  $\alpha_1 = \alpha_2 = 3$.
In Fig.~\ref{fig:carpet-design} the designs for each load combination are shown.
The efficacy of each design with respect to each service load is reported in Table~\ref{table:carpet} and Fig.~\ref{fig:carpet-stress}.

\begin{table}[h!]
\centering
{\footnotesize
\begin{tabular}{c | c c | c}
  \toprule
    & XT & ST & average\\
    \midrule
NC & 141.0 & 38.1 &  89.5 \\
\midrule
XT &  \underline{14.6} &  4.9 &  9.8 \\
ST &  16.7 &   \underline{2.9} &9.8 \\
 \midrule
 MT & 15.1 &  3.6  & 9.4 \\
  \bottomrule
\end{tabular}}
\caption{Efficacy of a carpet cloak.
Each row corresponds to one of the two optimization loads (XT, ST) plus the no-cloak case (NC) at the top.
Each column corresponds to a service load, with the last column showing the average.
Excluding the first row and the last column, the main diagonal of the table corresponds to cases in which the optimization and service loads are the same.}
\label{table:carpet}
\end{table}

\subsubsection{Example 3: Design of an elastic cloak for a rectangular inhomogeneity}
\begin{figure}[tp]
    \centering
    \includegraphics[width=\textwidth]{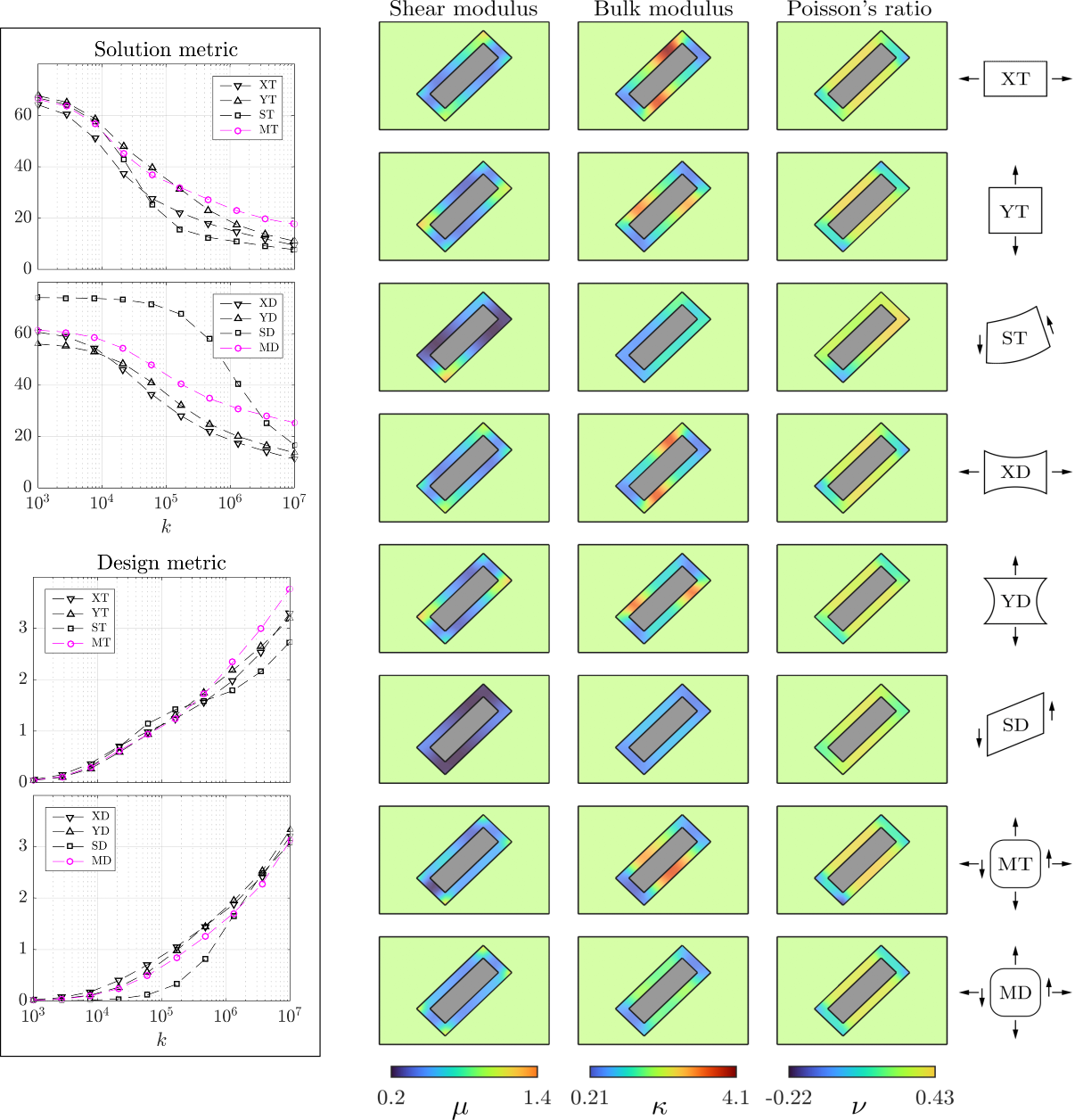}
    \vspace{-0.1in}
    \caption{Design of a cloak surrounding a rectangular hard inhomogeneity (rotated by $45^{\circ}$ with respect to the $x$ and $y$ axis).
    Left: Convergence in terms of the solution metric~\eqref{performance},~\eqref{performance-multi}, and of the design $H^1$ metric~\eqref{metric-H2}. The solution metric is normalized with respect to the no-cloak case and is expressed in percentages.
    Right: Distribution of the elastic moduli in the body for each design load.
    Each row represents a design, under different optimization loads (XT, ST, MT).}
\label{fig:rectangle-design}
\end{figure}
\begin{figure}[tp]
\centering
\includegraphics[width=\textwidth]{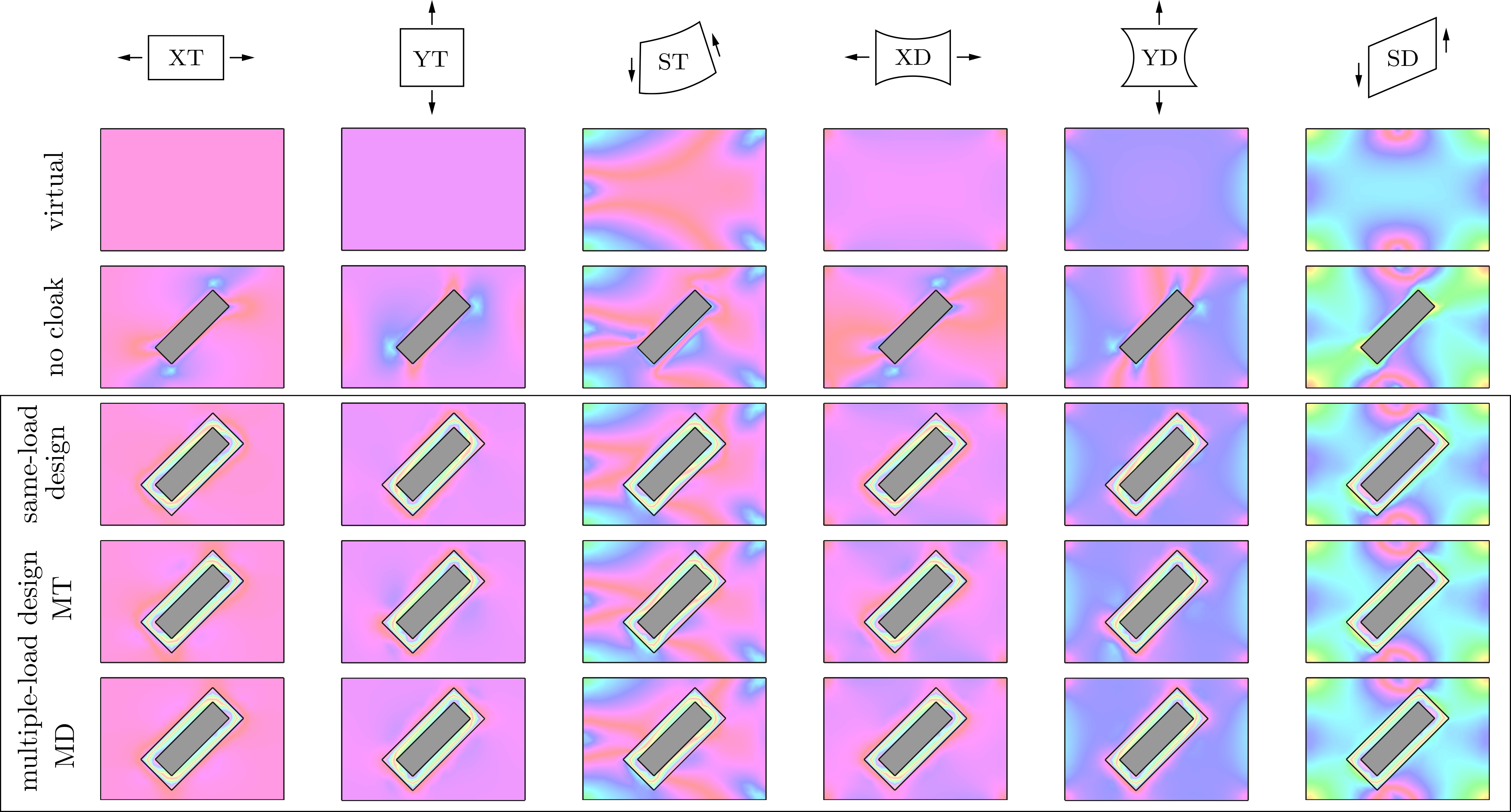}
\vspace{-0.15in}
\caption{Performance of the optimal cloaks under different service loads they are optimized for, in the case of a rectangular hard inhomogeneity.
The performance is given in terms of the Frobenius norm of the stress.
Each column represents a loading condition: XT, YT, ST, XD, YD, SD.
Each row refers to virtual body, body with no cloak, same-load design, multiple-load design MT, and multiple-load design MD.}
\label{fig:rectangle-stress}
\end{figure}

We consider a rectangular hard inhomogeneity inside the same rectangular region of sides $a=6$ and $b=4$ as in the previous example, see Fig.~\ref{fig:configurations}(c).
The inhomogeneity is assumed to have a much higher stiffness than the matrix $\mathring{\mathcal{B}}$.
The rectangular inhomogeneity is rotated $45^{\circ}$ with respect to the $x$ and $y$ axes, and has sides of $\frac{4}{3}L_o$ and $\frac{8}{21}L_o$, while the cloak is a rectangular annulus of length $\frac{5}{3}L_o$ and width $\frac{5}{7}L_o$. The thickness of the rectangular annulus is $\frac{1}{3}L_o$ along both sides.
Both combinations MT and MD have weights $w_{\mathrm X}=w_{\mathrm Y}=w_{\mathrm S}=\frac{1}{3}$.
The coefficients in the augmented objective functions are $k = 10^7$, $m_1 = m_2 = \alpha_1 = \alpha_2 = 1$.
The eight different designs are shown in Fig.~\ref{fig:rectangle-design}, while their efficacy with respect to each service load is reported in Table~\ref{table:rectangle}.
Fig.~\ref{fig:rectangle-stress} shows the stress distribution for some of the design-service load combinations.

\begin{table}[h!]
\centering
{\footnotesize
\begin{tabular}{c | c c c c c c | c}
\toprule
  & XT & YT & ST & XD & YD & SD & average\\
\midrule  
NC & 27.5 & 28.2 & 15.5 & 15.9 & 14.0 & 6.4 & 17.9 \\
\midrule
 XT &  \underline{2.6} & 13.6 &  4.9 &  2.4 &  6.4 & 3.1 &  5.5 \\
 YT& 10.9 &  \underline{3.1}  &  5.5 &  5.8 &  2.6 & 3.2 &  5.2 \\
 ST& 16.0 & 25.3 &  \underline{1.2}  &  6.9 & 12.0 & 2.4 & 10.6 \\
 XD&  3.0 & 13.2 &  4.8 &  \underline{1.8}  &  6.0 & 3.0 &  5.3 \\
 YD& 12.2 &  3.2 &  5.1 &  6.5 &  \underline{1.9}  & 3.1 &  5.3 \\
 SD& 28.9 & 35.0 &  6.8 & 12.7 &  9.7 & \underline{1.1}  & 15.7 \\
\midrule
 MT &  4.4 &  5.9 &  2.2 &  3.8 &  4.7 & 2.7 &  4.0 \\
 MD &  5.0 &  6.6 &  4.4 &  3.0 &  3.4 & 2.8 &  4.2 \\
\bottomrule
\end{tabular}
}
\caption{Efficacy of a cloak in the case of a rectangular inhomogeneity.
Each row corresponds to one of the six optimization loads (XT, YT, ST, XD, YD, SD) plus the no-cloak case (NC) at the top.
Each column corresponds to a service load, with the last column showing the average.
Excluding the first row and the last column, the main diagonal of the table corresponds to cases in which the optimization and service loads are the same.}
\label{table:rectangle}
\end{table}

\subsubsection{Example 4: Design of an elastic cloak for a random distribution of inhomogeneities}

\begin{figure}[tp]
    \centering
    \includegraphics[width=\textwidth]{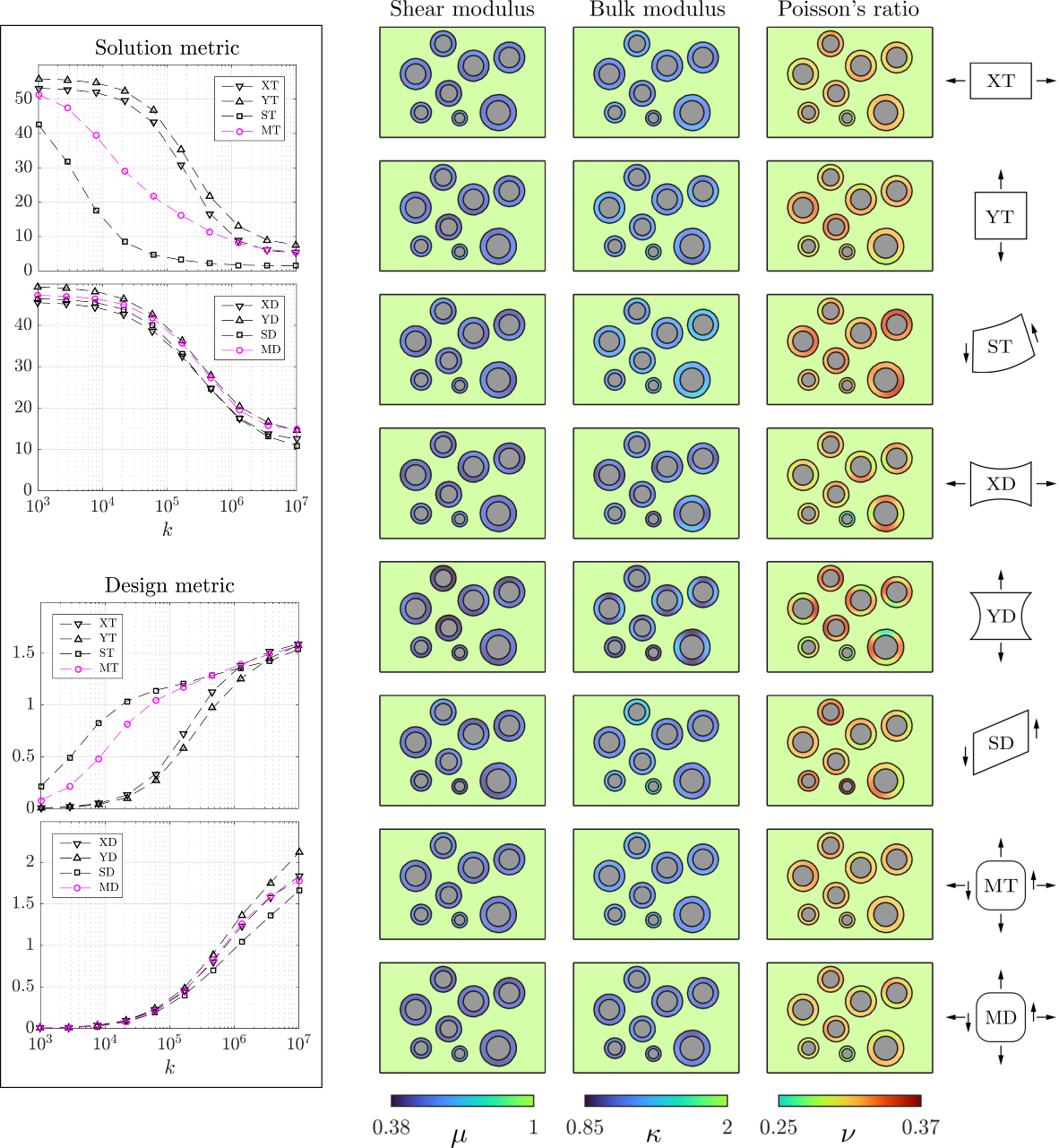}
    \vspace{-0.1in}
    \caption{Design of optimal cloaks surrounding randomly distributed multiple hard inhomogeneities.
    Left: Convergence in terms of the solution metric~\eqref{performance},~\eqref{performance-multi}, and of the design $H^1$ metric~\eqref{metric-H2}. The solution metric is normalized with respect to the no-cloak case and is expressed in percentages.
    Right: Distribution of the elastic moduli in the body for each design load.
    Each row represents a design, under different optimization loads (XT, YT, ST, XD, YD, SD, MT, MD).}
    \label{fig:inhom-design}
\end{figure}
\begin{figure}[tp]
\centering
\includegraphics[width=\textwidth]{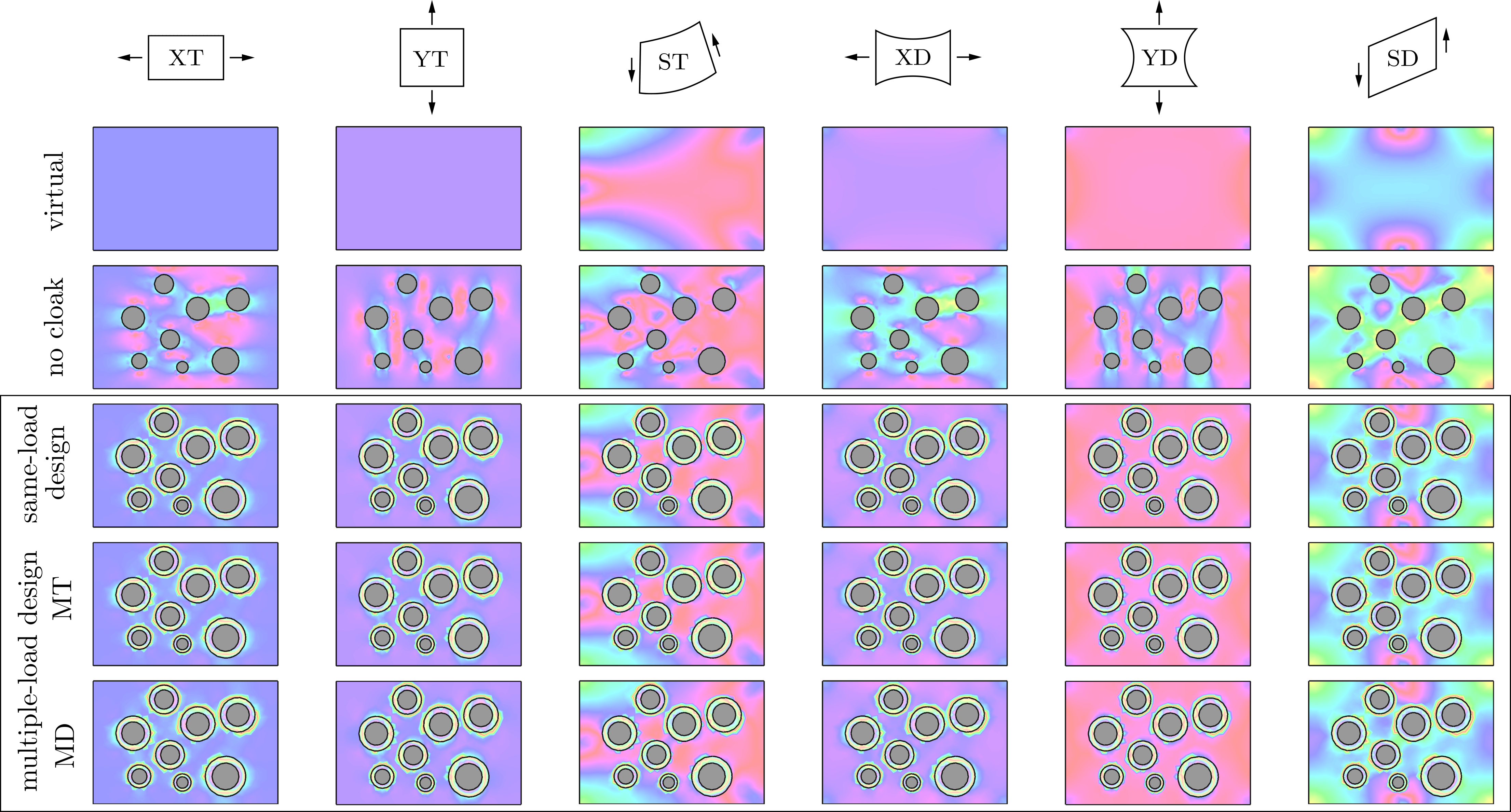}
\vspace{-0.15in}
\caption{Performance of the optimal cloaks under different service loads they are optimized for, in the case of randomly distributed multiple hard inhomogeneities.
The performance is given in terms of the Frobenius norm of the stress.
Each column represents a loading condition: XT, YT, ST, XD, YD, SD.
Each row refers to virtual body, body with no cloak, same-load design, multiple-load design MT, and multiple-load design MD.}
\label{fig:inhom-stress}
\end{figure}

We consider multiple circular inhomogeneities inside the same rectangular region of sides $a=6$ and $b=4$ as in the previous example, see Fig.~\ref{fig:configurations}(d).
The inhomogeneities are assumed to have a much higher stiffness than the matrix $\mathring{\mathcal{B}}$.
In particular, we consider eight disks of different radii between $0.15 L_o$ and $0.45 L_o$.
Each inhomogeneity is surrounded by an annular cloak of external radius $1.5$ times the inner radius.
Both combinations MT and MD have weights $w_{\mathrm X}=w_{\mathrm Y}=w_{\mathrm S}=\frac{1}{3}$.
The coefficients in the augmented objective functions are $k = 10^7$, $m_1 = m_2 = \alpha_1 = \alpha_2 = 1$.
The efficacy of each design shown in Fig.~\ref{fig:inhom-design}  is reported in Table~\ref{table:inhom} with respect to each service load, and is shown in Fig.~\ref{fig:inhom-stress} in terms of the stress distribution.

\begin{table}[h!]
\centering
{\footnotesize
\begin{tabular}{c | c c c c c c | c}
\toprule
  & XT & YT & ST & XD & YD & SD & average\\
\midrule  
NC & 37.4 & 35.7 & 22.7 & 13.8 & 13.3 & 4.5 & 21.2 \\
\midrule
XT &  \underline{2.0} &  3.0 &  0.4 &  1.9 &  2.7 & 0.6 &  1.8 \\
YT &  2.4 &  \underline{2.7} &  0.4 &  2.2 &  2.5 & 0.6 &  1.8 \\
ST &  2.4 &  3.4 &  \underline{0.3} &  2.2 &  3.2 & 0.6 &  2.0 \\
XD &  2.8 &  3.2 &  1.4 &  \underline{1.7} &  2.7 & 0.6 &  2.0 \\
YD &  3.5 &  5.0 &  3.0 &  2.4 &  \underline{1.9} & 0.7 &  2.8 \\
SD &  2.6 &  3.5 &  0.6 &  2.2 &  2.8 & \underline{0.5} &  2.0 \\
\midrule
MT &  2.1 &  2.8 &  0.3 &  2.0 &  2.6 & 0.6 &  1.7 \\
MD &  2.7 &  3.4 &  1.6 &  1.9 &  2.3 & 0.5 &  2.1 \\
\bottomrule
\end{tabular}
}
\caption{Efficacy of a cloak in the case of randomly distributed hard inhomogeneities.
Each row corresponds to one of the six optimization loads (XT, YT, ST, XD, YD, SD) plus the no-cloak case (NC) at the top.
Each column corresponds to a service load, with the last column showing the average.
Excluding the first row and the last column, the main diagonal of the table corresponds to cases in which the optimization and service loads are the same.}
\label{table:inhom}
\end{table}

\subsection{Discussion}

The performance or efficacy of a cloak, intended as the ability to hide the object it is designed for, is shown in Tables~\ref{table:hole}-\ref{table:inhom} in terms of the quantity defined in~\eqref{performance} and expressed in percentages.
It should be noticed that, compared to the optimization conditions, some designs are more effective under service loads that they are not optimized for.
This means that the minimum value of each row in the tables is not necessarily located under the corresponding column.
This occurs because the distance between the no-cloak and the virtual solutions is lower for some loading modes than others, i.e., some values in the first row---showing the no-cloak data---are much lower than others.
Since we use a displacement-based objective function, this is the case for the displacement-controlled optimization loads XD, YD, SD.
However, the minimum value of each column is located in the corresponding row---the boxed values in Tables~\ref{table:hole}, \ref{table:inhom}, \ref{table:carpet}.
This means that the best performance for each service load is achieved by optimizing for the same load as expected.
This is consistent with what is expected from the optimization problem.

Assuming isotropic solids allows us to only work with two design variables $\mu$ and $\kappa$.
In spite of this restriction, our results are promising, as in many cases we were able to reduce the initial difference between the physical and the virtual elastic fields to less than 10\%.
The iterations were stopped once small improvements in the performance corresponded to large changes in the design, as is shown in the convergence plots in Figs.~\ref{fig:hole-design},~\ref{fig:inhom-design},~\ref{fig:carpet-design} (where the solution metrics~\eqref{performance} and~\eqref{performance-multi} are expressed in terms of percentage of the no-cloak one).
The example of the cloaking of a sharp elliptic cut (carpet cloak) is more challenging, as the system tends to choose less regular distributions. For this reason, we worked with higher penalty coefficients in order to obtain more reasonable elastic moduli distributions.
The effectiveness of the designs can also be seen in the stress plots in Figs.~\ref{fig:hole-stress},~\ref{fig:inhom-stress},~\ref{fig:carpet-stress}; when compared to the stress distributions in the virtual cases (first rows in the stress plots), the response of the designs (third, fourth, fifth rows) are much more similar than the uncloaked case (second row).
It should be emphasized that, regardless of the efficacy of a cloak, its presence prevents the stress concentration that one observes in the no-cloak case, suggesting optimal cloaking approach as a way to enhance the toughness of materials.
This is noticeable especially in Fig.~\ref{fig:inhom-stress} for the random distribution of inhomogeneities.
Particular attention should be given to the multiple-load design.
The last columns of Tables~\ref{table:hole} and~\ref{table:inhom} show good average drop of the cloaking metric, and hence an average improvement of the performance of the cloak, with respect to all the service loads.
However, in Table~\ref{table:carpet} there is no significant improvement, which is due to the fact that the average of performance is a meaningful measure only when the normalized distances~\eqref{performance} corresponding to the no-cloak solutions of different load cases are similar.

The continuum approach allows us to span the whole design space, without any restrictions due to a particular choice of the class of lattice materials for the cloak. 
In other words, in designing the optimal elastic cloaks we consider all the admissible values of $\mu>0$ and $\kappa>0$.
For example, in some of the cloak designs Poisson's ratio is negative, see the right columns in Figs.~\ref{fig:hole-design} and~\ref{fig:carpet-design}. This suggests the use of auxetic materials for the purposes of cloaking holes, cavities, and cracks. 
For engineering applications of elastic cloaks, the proposed optimal cloaks in this paper can be additively manufactured following data-driven methods \citep{wilt2020accelerating, wang2022mechanical}, and using functionally graded auxetic lattice materials \citep{ren2018auxetic}.

\section{Conclusions} \label{Sec:Conclusions}

In this paper we formulated elastostatic cloaking as an optimal design problem.
Similar to electromagnetic cloaking, the goal in elastic cloaking is to make an object---a cavity, a through hole, or any type of inhomogeneity or inclusion---invisible to elastic fields. This concealment can be achieved by surrounding the object with a cloaking device with the goal of controlling the elastic field outside of it. Because of the unattainability of exact transformation cloaking, we propose an original formulation of optimal elastic cloaking
based on the adjoint state method, in which the balance of linear momentum is enforced as a constraint.
The objective function measures the distance between the solutions in the physical and in the virtual homogeneous elastic bodies.
The cloak is assumed to be made of isotropic  inhomogeneous linear elastic materials.
Hence, the design parameters are the two elastic moduli in the cloak, namely the bulk modulus $\kappa$ and the shear modulus $\mu$.
In order to guarantee positive definiteness of the elasticity tensor of the cloak, we used a change of variables.
Relatively smooth variations of the elastic moduli within the cloak are enforced via a penalization term, based on an $H^1$ metric defined on the design space.

Variations of the objective function with respect to $\boldsymbol{\gamma}$ and $\mathbf{u}$ give the standard and the adjoint balance equations, respectively. Although the elastic problem is linear, variations of the design parameters $\eta$ and $\xi$ yield nonlinear equations with associated Neumann boundary conditions. The optimization formulation is extended to multiple-load design.
We used mixed finite elements to discretize the weak formulation of the governing equations, and considered several numerical examples of optimal cloaks designed for single and multiple loads. In spite of the restrictive isotropic assumption, the results are promising as in many cases we were able to reduce the initial difference between the physical and the virtual elastic fields to less than 10\%. Moreover, the general continuum approach allows us to span the entire design space, and find results such as the use of auxetic materials for the cloaking of holes and cavities.

A future extension of this work will be to include anisotropic and non-centrosymmetric solids.
We believe that being able to operate within a much larger design space would allow one to get closer to exact cloaking.
Moreover, extending the present formulation to elastodynamics will be the subject of a future communication. When considering elastic waves, not only is it necessary to take into account combinations of loads, but it becomes fundamental to optimize with respect to multiple frequencies.
Other future extensions of our work would consist of investigating the effect of different objective functions (e.g., energy or stress-based) on the design, and extending the present formulation to nonlinear elastostatic cloaking. Extension of the present work to cloaking in elastic plates will also be the subject of another future communication.

\section*{Acknowledgement}

This research was supported by ARO W911NF-18-1-0003 (Dr. Daniel P. Cole).

\bibliographystyle{abbrvnat}
\bibliography{ref}

\appendix

\section{Lagrange Multipliers in the Optimization Problem} \label{App:Lag-mult}

In this appendix we show that a single Lagrange multiplier $\boldsymbol\gamma$ can be associated with both the equilibrium constraint and the traction boundary condition, as was assumed in~\eqref{modified-objective}.
Assuming two different Lagrange multipliers, $\boldsymbol\gamma$ for $\operatorname{div}(\boldsymbol{\mathsf{C}}\nabla\mathbf{u}) +\mathbf{b}$ and $\boldsymbol{\chi}$ for $\bar{\mathbf{t}}-(\boldsymbol{\mathsf{C}}\nabla\mathbf{u})\hat{\mathbf{n}}$,
\eqref{u-variation2} is rewritten as
\begin{equation} \label{u-variation-twolag}
	\delta_{\mathbf{u}} \mathsf{f} =
	\int_{\mathcal{B}} \operatorname{div}(\boldsymbol{\mathsf{C}}\nabla\boldsymbol{\gamma})
	\cdot\delta\mathbf{u}\,\mathrm{d}v
	-\int_{\partial_N\mathcal{B}} (\boldsymbol{\mathsf{C}}\nabla\boldsymbol{\gamma}) \hat{\mathbf{n}}
	\cdot\delta\mathbf{u} \,\mathrm{d}a
	+\int_{\partial_N\mathcal{B}} (\boldsymbol{\gamma} - \boldsymbol{\chi}) \cdot (\boldsymbol{\mathsf{C}} 
	\nabla\delta\mathbf{u})\hat{\mathbf{n}}\,\mathrm{d}a
	+\int_{\partial_o\mathcal{C}} \mathring{\boldsymbol{\mathsf{C}}}(\nabla\mathbf{u}
	-\nabla\tilde{\mathbf{u}})\hat{\mathbf{n}}\cdot\delta\mathbf{u}
	\,\mathrm{d}a
	\,.
\end{equation}
For the sake of simplicity we set $\mathbf{m}=\boldsymbol{\gamma} - \boldsymbol{\chi}$, and $\mathbf{M}=\boldsymbol{\mathsf{C}} (\mathbf{m} \otimes \hat{\mathbf{n}})$, so that the integrand in the third term of~\eqref{u-variation-twolag} reads $\mathbf{M} : \nabla\delta\mathbf{u}$.
We want to show that $\mathbf{m}=\mathbf{0}$.
We also denote with $\mathbf{P}=\mathbf{I}-\hat{\mathbf{n}} \otimes \hat{\mathbf{n}}$ the orthogonal projection to the tangent space of $\partial_N\mathcal B$.\footnote{%
As an orthogonal projection, $\mathbf{P}$ satisfies $\mathbf{P}\mathbf{v}\cdot\mathbf{w}=\mathbf{v}\cdot\mathbf{P}\mathbf{w}$.
Moreover, it can be used to decompose dot products as $\mathbf{v}\cdot\mathbf{w}=\mathbf{P}\mathbf{v}\cdot\mathbf{w}+v_n w_n = \mathbf{v}\cdot \mathbf{P}\mathbf{w}+v_n w_n$.
This extends to all contractions, e.g., $\mathbf{A}:\mathbf{B}=\mathbf{A}\mathbf{P}:\mathbf{B}+\mathbf{A}\hat{\mathbf{n}}\cdot\mathbf{B}\hat{\mathbf{n}} =\mathbf{A}:\mathbf{B}\mathbf{P}+\mathbf{A}\hat{\mathbf{n}}\cdot\mathbf{B}\hat{\mathbf{n}}$.}
Then, one can decompose $\mathbf{M} : \nabla\delta\mathbf{u}$ as
\begin{equation}
	\mathbf{M} : \nabla\delta\mathbf{u} =
	\mathbf{M}\mathbf{P} : \nabla\delta\mathbf{u}
	+ ( \mathbf{M} \hat{\mathbf{n}}) \cdot \nabla_{\hat{\mathbf{n}}}\delta\mathbf{u} =
	\mathbf{M} : (\nabla\delta\mathbf{u}) \mathbf{P}
	+ ( \mathbf{M} \hat{\mathbf{n}}) \cdot \nabla_{\hat{\mathbf{n}}}\delta\mathbf{u}
	\,,
\end{equation}
because $(\nabla\delta\mathbf{u})\hat{\mathbf{n}}=\nabla_{\hat{\mathbf{n}}}\delta\mathbf{u}$, and where $(\nabla\delta\mathbf{u}) \mathbf{P}$ represents the derivatives of $\delta\mathbf{u}$ along directions tangent to $\partial_N \mathcal B$.
Therefore, the third term on the right-hand side of~\eqref{u-variation-twolag} reads
\begin{equation} \label{int-decomp}
	\int_{\partial_N\mathcal B} \mathbf{M} : \nabla\delta\mathbf{u} \, \mathrm{d}a
	=
	\int_{\partial_N\mathcal B} \mathbf{M} : (\nabla\delta\mathbf{u}) \mathbf{P} \, \mathrm{d}a
	+ \int_{\partial_N\mathcal B} ( \mathbf{M} \hat{\mathbf{n}}) \cdot \nabla_{\hat{\mathbf{n}}}\delta\mathbf{u}  \, \mathrm{d}a \,.
\end{equation}
Note that on $\partial_N\mathcal{B}$ the field $\nabla_{ \hat{\mathbf{n}} }\delta\mathbf{u}$ is independent from both $\delta\mathbf{u}$ and $(\nabla\delta\mathbf{u}) \mathbf{P}$.
Therefore, by the fundamental lemma of calculus of variations, the vanishing of $\delta_{\mathbf{u}} \mathsf{f}$ in~\eqref{u-variation-twolag} for arbitrary $\nabla_{ \hat{\mathbf{n}} }\delta\mathbf{u}$ implies the vanishing of the integrand of the last term in~\eqref{int-decomp}, i.e., $\mathbf{M} \hat{\mathbf{n}}=\mathbf{0}$.
This means that $\left[ \boldsymbol{\mathsf{C}} (\mathbf{m} \otimes \hat{\mathbf{n}}) \right] \hat{\mathbf{n}}=\mathbf{0}$, which implies that $\mathbf{m}=\mathbf{0}$ because of the invertibility of $\boldsymbol{\mathsf{C}}$.
This can be seen in a chart in which the first two coordinates are tangent and the third one is orthogonal to $\partial_N\mathcal B$.
Then, the constants $\mathsf{C}^{a 3 b 3}$ can be selected by extracting the rows and columns $3,5,6$ from the Voigt representation of $\boldsymbol{\mathsf{C}}$.
Being a principal submatrix of an invertible matrix, it is invertible, and hence, $\mathsf{C}^{a 3 b 3} m_b=0$ implies that $m_b=0$.

\section{Metrics and Norms in the Design Space} \label{App:Metric}

For two isotropic elasticity tensors $\boldsymbol{\mathsf{C}}$ and $\boldsymbol{\mathsf{C}}'$ corresponding to the pairs $(\xi,\eta)$ and $(\xi',\eta')$, respectively, let us define
\begin{equation}
	d^2(\boldsymbol{\mathsf{C}},\boldsymbol{\mathsf{C}}')
	= m_1\left[\xi(x)-\xi'(x)\right]^2+m_2\left[\eta(x)-\eta'(x))\right]^2
	+ \alpha_1\Vert \nabla (\xi(x)-\xi'(x)) \Vert^2+\alpha_2\Vert\nabla(\eta(x)-\eta'(x)) \Vert^2 \,,
\end{equation}
where $m_1$, $m_2$, $\alpha_1$, and $\alpha_2$ are some positive constants.
Our goal is to prove that for fixed $x$, $d(.,.)$ is a metric.
Obviously, $d(\boldsymbol{\mathsf{C}},\boldsymbol{\mathsf{C}}')=d(\boldsymbol{\mathsf{C}}',\boldsymbol{\mathsf{C}})$, and $d(\boldsymbol{\mathsf{C}},\boldsymbol{\mathsf{C}}')=0$ if and only if $\boldsymbol{\mathsf{C}}=\boldsymbol{\mathsf{C}}'$. It remains to show that $d$ satisfies the triangular inequality.
Notice that
\begin{equation}
\begin{aligned}
	d^2(\boldsymbol{\mathsf{C}},\boldsymbol{\mathsf{C}}')
	& = m_1\left[\xi-\xi'\right]^2+m_2\left[\eta-\eta'\right]^2
	+ \alpha_1\Vert \nabla (\xi-\xi') \Vert^2+\alpha_2\Vert\nabla(\eta-\eta') \Vert^2 \\
	& = m_1\left[(\xi-\xi'')+(\xi''-\xi')\right]^2+m_2\left[(\eta-\eta'')+(\eta''-\eta')\right]^2 
	\\
	& \quad + \alpha_1\Vert \nabla (\xi-\xi'')+\nabla (\xi''-\xi') \Vert^2
	+\alpha_2 \Vert \nabla(\eta-\eta'')+\nabla(\eta''-\eta') \Vert^2  \\
	&= m_1(\xi-\xi'')^2+m_2(\eta-\eta'')^2+m_1(\xi''-\xi')^2+m_2(\eta''-\eta')^2\\
	& \quad +2m_1(\xi-\xi'')(\xi''-\xi')+2m_2(\eta-\eta'')(\eta''-\eta') \\
	& \quad + \alpha_1\Vert \nabla (\xi-\xi'') \Vert^2+ \alpha_1\Vert \nabla (\xi''-\xi') \Vert^2
	+2\alpha_1 \nabla (\xi-\xi'')\cdot\nabla (\xi''-\xi') \\
	& \quad +\alpha_2 \Vert \nabla(\eta-\eta'')\Vert^2
	+\alpha_2 \Vert \nabla(\eta''-\eta') \Vert^2				
	+ 2\alpha_2 \nabla(\eta-\eta'')\cdot\nabla(\eta''-\eta') \\
	& = d^2(\boldsymbol{\mathsf{C}},\boldsymbol{\mathsf{C}}'')
	+d^2(\boldsymbol{\mathsf{C}}'',\boldsymbol{\mathsf{C}}')
	+2m_1(\xi-\xi'')(\xi''-\xi')+2m_2(\eta-\eta'')(\eta''-\eta') \\
	& \quad +2\alpha_1 \nabla (\xi-\xi'')\cdot\nabla (\xi''-\xi') 
	+ 2\alpha_2 \nabla(\eta-\eta'')\cdot\nabla(\eta''-\eta')
	\,.
\end{aligned}
\end{equation}
For vectors $\mathbf{a}=(a_1,a_2,a_3,a_4,a_5,a_6)$ and $\mathbf{b}=(b_1,b_2,b_3,b_4,b_5,b_6)$ in $\mathbb{R}^6$, it is straightforward to show that $\langle\mathbf{a},\mathbf{b}\rangle=m_1a_1b_1+m_2a_2b_2+\alpha_1(a_3b_3+a_4b_4)+\alpha_2(a_5b_5+a_6b_6)$ is an inner product.\footnote{This is for 2D. In 3D, for vectors $\mathbf{a}=(a_1,a_2,a_3,a_4,a_5,a_6,a_7,a_8)$ and $\mathbf{b}=(b_1,b_2,b_3,b_4,b_5,b_6,b_7,b_8)$ in $\mathbb{R}^8$, $\langle\mathbf{a},\mathbf{b}\rangle=m_1a_1b_1+m_2a_2b_2+\alpha_1(a_3b_3+a_4b_4+a_5b_5)+\alpha_2(a_6b_6+a_7b_7+a_8b_8)$ is the inner product.}
The Cauchy-Schwarz inequality states that $\langle\mathbf{a},\mathbf{b}\rangle^2\leq \langle\mathbf{a},\mathbf{a}\rangle \langle\mathbf{b},\mathbf{b}\rangle$.
Let us define $\mathbf{a}=(\xi-\xi'',\eta-\eta'',(\xi-\xi'')_{,1},(\xi-\xi'')_{,2},(\eta-\eta'')_{,1},(\eta-\eta'')_{,2})$, and $\mathbf{b}=(\xi''-\xi',\eta''-\eta',(\xi''-\xi')_{,1},(\xi''-\xi')_{,2},(\eta''-\eta')_{,1},(\eta''-\eta')_{,2})$.\footnote{In 3D, $\mathbf{a}=(\xi-\xi'',\eta-\eta'',(\xi-\xi'')_{,1},(\xi-\xi'')_{,2},(\xi-\xi'')_{,3},(\eta-\eta'')_{,1},(\eta-\eta'')_{,2},(\eta-\eta'')_{,3})$, and $\mathbf{b}=(\xi''-\xi',\eta''-\eta',(\xi''-\xi')_{,1},(\xi''-\xi')_{,2},(\xi''-\xi')_{,3},(\eta''-\eta')_{,1},(\eta''-\eta')_{,2},(\eta''-\eta')_{,3})$.} 
Notice that
\begin{equation}
\begin{aligned}
	\langle\mathbf{a},\mathbf{b}\rangle &=
	m_1(\xi-\xi'')(\xi''-\xi')+m_2(\eta-\eta'')(\eta''-\eta') 
	+\alpha_1 \nabla (\xi-\xi'')\cdot\nabla (\xi''-\xi') 
	+ \alpha_2 \nabla(\eta-\eta'')\cdot\nabla(\eta''-\eta')\,, \\
	\langle\mathbf{a},\mathbf{a}\rangle &=
	m_1(\xi-\xi'')^2+m_2(\eta-\eta'')^2 
	+\alpha_1 \Vert \nabla (\xi-\xi'') \Vert^2 + \alpha_2 \Vert\nabla(\eta-\eta'') \Vert^2\,, \\
	\langle\mathbf{b},\mathbf{b}\rangle &=
	m_1(\xi''-\xi')^2+m_2(\eta''-\eta')^2 
	+\alpha_1 \Vert \nabla (\xi''-\xi') \Vert^2 + \alpha_2 \Vert\nabla(\eta''-\eta') \Vert^2\,.
\end{aligned}
\end{equation}
The Cauchy-Schwarz inequality implies that
\begin{equation}
\begin{aligned}
	& \left[m_1(\xi-\xi'')(\xi''-\xi')+m_2(\eta-\eta'')(\eta''-\eta') 
	+\alpha_1 \nabla (\xi-\xi'')\cdot\nabla (\xi''-\xi') 
	+ \alpha_2 \nabla(\eta-\eta'')\cdot\nabla(\eta''-\eta')\right]^2 \\
	& \quad \leq \left[ m_1(\xi-\xi'')^2+m_2(\eta-\eta'')^2 
	+\alpha_1 \Vert \nabla (\xi-\xi'') \Vert^2 + \alpha_2 \Vert\nabla(\eta-\eta'') \Vert^2 \right] \\
	& \quad \times \left[ m_1(\xi''-\xi')^2+m_2(\eta''-\eta')^2 
	+\alpha_1 \Vert \nabla (\xi''-\xi') \Vert^2 + \alpha_2 \Vert\nabla(\eta''-\eta') \Vert^2 \right]
	=d^2(\boldsymbol{\mathsf{C}},\boldsymbol{\mathsf{C}}'') 
	d^2(\boldsymbol{\mathsf{C}}'',\boldsymbol{\mathsf{C}}')\,.
\end{aligned}
\end{equation}
Thus
\begin{equation}
\begin{aligned}
	& m_1(\xi-\xi'')(\xi''-\xi')+m_2(\eta-\eta'')(\eta''-\eta') 
	+\alpha_1 \nabla (\xi-\xi'')\cdot\nabla (\xi''-\xi') + \alpha_2 \nabla(\eta-\eta'')\cdot\nabla(\eta''-\eta') \\
	& \quad \leq d(\boldsymbol{\mathsf{C}},\boldsymbol{\mathsf{C}}'') 
	d(\boldsymbol{\mathsf{C}}'',\boldsymbol{\mathsf{C}}')
	\,.
\end{aligned}
\end{equation}
Therefore 
\begin{equation}
\begin{aligned}
	d^2(\boldsymbol{\mathsf{C}},\boldsymbol{\mathsf{C}}')
	& = d^2(\boldsymbol{\mathsf{C}},\boldsymbol{\mathsf{C}}'')
	+d^2(\boldsymbol{\mathsf{C}}'',\boldsymbol{\mathsf{C}}')
	+2m_1(\xi-\xi'')(\xi''-\xi')+2m_2(\eta-\eta'')(\eta''-\eta') \\
	& \quad +2\alpha_1 \nabla (\xi-\xi'')\cdot\nabla (\xi''-\xi') + 2\alpha_2 \nabla(\eta-\eta'')\cdot\nabla(\eta''-\eta')\\
	& \leq d^2(\boldsymbol{\mathsf{C}},\boldsymbol{\mathsf{C}}'')
	+d^2(\boldsymbol{\mathsf{C}}'',\boldsymbol{\mathsf{C}}')
	+2d(\boldsymbol{\mathsf{C}},\boldsymbol{\mathsf{C}}'') 
	d(\boldsymbol{\mathsf{C}}'',\boldsymbol{\mathsf{C}}')
	\,.
\end{aligned}
\end{equation}
This implies that $d(\boldsymbol{\mathsf{C}},\boldsymbol{\mathsf{C}}')\leq d(\boldsymbol{\mathsf{C}},\boldsymbol{\mathsf{C}}'')+d(\boldsymbol{\mathsf{C}}'',\boldsymbol{\mathsf{C}}')$, and hence $d(.,.)$ defines a metric for a fixed $x$.
From \eqref{metric-H2}, notice that
\begin{equation} \label{metric}
\begin{aligned}
	d^2_{\boldsymbol{\mathsf{C}}}(\boldsymbol{\mathsf{C}},\boldsymbol{\mathsf{C}}')
	& = \int_{\mathcal{C}} d^2(\boldsymbol{\mathsf{C}}(x),\boldsymbol{\mathsf{C}}'(x)) \, \mathrm d v \\
	& \leq \int_{\mathcal{C}} d^2(\boldsymbol{\mathsf{C}}(x),\boldsymbol{\mathsf{C}}''(x)) \, \mathrm d v
	+\int_{\mathcal{C}} d^2(\boldsymbol{\mathsf{C}}''(x),\boldsymbol{\mathsf{C}}'(x)) \, \mathrm d v 
	+2\int_{\mathcal{C}} d(\boldsymbol{\mathsf{C}}(x),\boldsymbol{\mathsf{C}}''(x))\,
	d(\boldsymbol{\mathsf{C}}''(x),\boldsymbol{\mathsf{C}}'(x)) \, \mathrm d v
		\,.
\end{aligned}
\end{equation}
Using the Cauchy-Schwarz inequality
\begin{equation} 
\begin{aligned}
	\left(\int_{\mathcal{C}} d(\boldsymbol{\mathsf{C}}(x),\boldsymbol{\mathsf{C}}''(x))\,
	d(\boldsymbol{\mathsf{C}}''(x),\boldsymbol{\mathsf{C}}'(x)) \, \mathrm d v\right)^2
	& \leq \int_{\mathcal{C}} d^2(\boldsymbol{\mathsf{C}}(x),\boldsymbol{\mathsf{C}}''(x))
	\, \mathrm d v
	\int_{\mathcal{C}} d^2(\boldsymbol{\mathsf{C}}''(x),\boldsymbol{\mathsf{C}}'(x)) 
	\, \mathrm d v \\
	& \quad = d^2_{\boldsymbol{\mathsf{C}}}(\boldsymbol{\mathsf{C}},\boldsymbol{\mathsf{C}}'')
	\,d^2_{\boldsymbol{\mathsf{C}}}(\boldsymbol{\mathsf{C}}',\boldsymbol{\mathsf{C}}')
	\,.
\end{aligned}
\end{equation}
Thus, $\int_{\mathcal{C}} d(\boldsymbol{\mathsf{C}}(x),\boldsymbol{\mathsf{C}}''(x))\,	d(\boldsymbol{\mathsf{C}}''(x),\boldsymbol{\mathsf{C}}'(x)) \, \mathrm d v \leq d_{\boldsymbol{\mathsf{C}}}(\boldsymbol{\mathsf{C}},\boldsymbol{\mathsf{C}}'') \,d_{\boldsymbol{\mathsf{C}}}(\boldsymbol{\mathsf{C}}',\boldsymbol{\mathsf{C}}')$. Substituting this back into \eqref{metric}, one concludes that $d_{\boldsymbol{\mathsf{C}}}(\boldsymbol{\mathsf{C}},\boldsymbol{\mathsf{C}}')\leq d_{\boldsymbol{\mathsf{C}}}(\boldsymbol{\mathsf{C}},\boldsymbol{\mathsf{C}}'')+d_{\boldsymbol{\mathsf{C}}}(\boldsymbol{\mathsf{C}}'',\boldsymbol{\mathsf{C}}')$, and hence $d_{\boldsymbol{\mathsf{C}}}(.,.)$ is a metric.

\end{document}